\documentclass[11pt, a4paper]{article}
\pdfoutput=1

\usepackage{jcappub}
\usepackage{subfigure}
\usepackage{bm}
\usepackage{amssymb}
\usepackage{epsfig}
\usepackage{float}
\usepackage{caption}
\usepackage{color}
\usepackage{upgreek}
\usepackage{amsmath}
\usepackage{graphicx}
\usepackage{float}
\usepackage[utf8]{inputenc}

\begin{document}

\setcounter{tocdepth}{2}

\title{Scale-Dependent Gravitational Couplings in Parameterised Post-Newtonian Cosmology}

\author{Daniel B. Thomas,}
\emailAdd{dan.b.thomas1@gmail.com}
\author{Timothy Clifton and}
\emailAdd{t.clifton@qmul.ac.uk}
\author{Theodore Anton}
\emailAdd{t.j.anton@qmul.ac.uk}

\affiliation{Department of Physics and Astronomy, Queen Mary University of London, UK.}


\abstract{Parameterised Post-Newtonian Cosmology (PPNC) is a theory-agnostic framework for testing gravity in cosmology, which connects gravitational physics on small and large scales in the Universe. It is a direct extension of the Parameterised Post-Newtonian (PPN) approach to testing gravity in isolated astrophysical systems, and therefore allows constraints on gravity from vastly different physical regimes to be compared and combined. We investigate the application of this framework to a class of example scalar-tensor theories of gravity in order to verify theoretical predictions, and to investigate for the first time the scale-dependence of the gravitational couplings that appear within its perturbation equations. In doing so, we evaluate the performance of some simple interpolating functions in the transition region between small and large cosmological scales, as well as the uncertainties that using such functions would introduce into the calculation of observables. We find that all theoretical predictions of the PPNC framework are verified to high accuracy in the relevant regimes, and that simple interpolating functions perform well (but not perfectly) between these regimes. This study is an important step towards being able to use the PPNC framework to analyse cosmological datasets, and to thereby test if/how the gravitational interaction has changed as the Universe has evolved.}

\maketitle
\flushbottom

\bibliographystyle{unsrt}

\vspace{1cm}

\section{Introduction}

Tests of relativistic theories of gravity can be performed in physical environments that range from the Solar System \cite{Will_1993, Poisson_2014}, through to gravitational wave emission from binaries \cite{Wex:2014nva, GW150914}, and cosmological tests \cite{Clifton_2012, Ishak_2018}. The motivation for these tests are varied, covering the basic scientific need to test physical theories in all areas in which they are to be applied, through to the desire to understand the constraints that observations can impose upon new propositions. On these grounds, one could argue that it is particularly important to test gravity in cosmology, due to the vast extrapolations that are required to apply gravitational theory on cosmological scales, and the relative abundance of new theories that appear in this field.

The wide array of spatial and temporal scales that need to be considered in cosmology, as well as the different physical environments that can exist, and the large number of theoretical proposals, also provide complications. These are factors that do not exist in the same way in other areas of gravitational physics, and which therefore present new challenges that need to be overcome. Historically, progress has been made in this field by either performing detailed predictions for a specific theory, before comparing to the data, or by constructing frameworks that cover multiple different theories or classes of theories. The latter of these approaches is more efficient, in the sense that fewer computations need to be done to constrain multiple different cases. It does, however, require us to be able to determine generic behaviour produced by different theories of gravity.

The gold standard in frameworks for testing gravity is the Parameterised Post Newtonian (PPN) formalism, which has seen great success in a variety of weak field systems \cite{Will_2014}. However, it was not designed with cosmology in mind, and is not immediately applicable to observations made over cosmologically interesting scales. Work in producing frameworks for testing gravity in cosmology has therefore often tended to focus on creating new independent frameworks (see e.g. \cite{Clifton_2012, Gleyzes_2017, Martinelli_2021}). Such approaches usually rely on cosmological perturbation theory, and must therefore be extrapolated into the non-linear regime\footnote{See however Refs. \cite{nlpaper,sankar, lucassims}.}. They also separate out the expansion of the ``background'' from other aspects of gravity, which overlooks a vital source of information, as well as obscuring the relationship with frameworks that are used in non-cosmological contexts (such as PPN).

Our approach to this problem is to take as our starting point the PPN framework, and assume this to be an appropriate description of gravitational physics on scales where Newtonian (and post-Newtonian) gravity are valid. We then construct the set of cosmological models that are consistent with this hypothesis, and which are statistically homogeneous and isotropic on large scales, by joining together many regions of space-time that are described in this way \cite{Sanghai_2015, Sanghai_2016}. The result is a cosmology in which the background expansion and the weak gravitational field in the non-linear and mildly non-linear regime are consistent, and described by the same set of PPN parameters (suitably extended for the self-consistency of the cosmological model, and to include dark energy). As a corollary of this approach we also gain information about the behaviour of cosmological perturbations on the very largest scales \cite{Sanghai_2019, Anton_2021}. We call the result `Parameterised Post-Newtonian Cosmology' (PPNC).

The PPNC framework  consistently parameterises gravity in the non-linear regime and at the level of the background, as well as on super-horizon scales. It is constructed from a set of time-dependent parameters that directly reduce to the PPN parameters in the appropriate limits, and that therefore allows the same aspects of the gravitational interaction to be constrained in both cosmological and non-cosmological systems. Furthermore, it provides information about both the small and large-scale limits of linear cosmological perturbations. It is therefore the natural extension of the PPN framework into the cosmological regime. In this paper we study the transition between the small and large-scale cosmological regimes of this approach using a pedagogical set of example theories, and how this transition should be expected to impact the growth of overdensities and influence observables.

This paper is laid out as follows. In section \ref{sec:ppnc} we recap the mathematical formalism of the PPNC approach and the class of theories we will be considering. In section \ref{sec:numerics} we examine the behaviour of the perturbations and examine the quality of some simple interpolation functions. In section \ref{sec:paramevolution} we further examine the use of these interpolation functions to evolve the perturbations and calculate some simple observables. We conclude in section \ref{sec:conc}. {  We choose units such that $c=1$, and use base 10 for the logarithms in all plots}.

\section{Mathematical formalism}
\label{sec:ppnc}

Here we will present the salient details of the PPNC approach, which is based on an extended version of the PPN formalism suitably transformed for use in cosmology, before discussing a class of scalar-tensor theories of gravity as an example. For full details of the PPNC approach the reader is referred to Refs. \cite{Sanghai_2015,Sanghai_2016,Sanghai_2019,Anton_2021}. For the scalar-tensor theories of gravity the reader may refer to, for example, Ref. \cite{Clifton_2012}.

\subsection{Parameterised Post-Newtonian Cosmology}
\label{sec:ppnc2}

The geometry of the constructed space-time in the PPNC approach can be written as
\begin{equation} 
ds^2=a(\tau)^2\left[-(1-2{\Phi})d{\tau}^2+(1+2{\Psi})\delta_{ij}d{x}^id{x}^j \right] \label{eqn_flrw}\text{,}
\end{equation}
where $\Phi\ll 1$ and $\Psi \ll 1$ are both scalar functions of $\tau$ and $x^i$, and $a(\tau)$ is the scale factor as a function of conformal time. The reader will note that vector and tensor gravitational fields have been neglected at leading order, which is consistent with expectations from cosmological perturbation theory, as well as the order at which these objects appear in the post-Newtonian expansion that is used to generate this metric. We have also opted to consider geometries in which the background in spatially flat; this is motivated by CMB observations, but still allows the inclusion of small amounts of spatial curvature through the perturbation $\Psi$.

The metric presented in Eq. (\ref{eqn_flrw}) is written in the longitudinal (conformal Newtonian) gauge, which has the benefit of being well defined in both the cosmological perturbation theory and post-Newtonian sectors of the expansions that we require \cite{Clifton:2020oqx}. It also corresponds to the leading-order part of the `post-Newtonian gauge', in which the PPN approach is typically formulated. We note that the metric in Eq. (\ref{eqn_flrw}) is {\it not} a starting ansatz in this approach, but is constructed from small regions of space-time described by the PPN ``test metric'', which is itself a post-Newtonian expansion about Minkowski space. In this sense, the behaviour of the scale factor $a(\tau)$ and the potentials $\Phi$ and $\Psi$ are emergent quantities, related directly to the gravitational potentials that occur in the test metric. We expect this geometry to be applicable to all astrophysical objects, including everything down to the scale of neutron stars and black holes, as all such bodies are expected to be well modelled by Newtonian (and post-Newtonian) gravity.

The gravitational parameters required for this construction, at the stated level of accuracy, are as follows:
\begin{equation} \label{funcs}
\{ \alpha, \, \gamma , \, \alpha_c , \, \gamma_c \} \, ,
\end{equation}
where each of these should be understood to be a function of conformal time $\tau$, but not spatial coordinates $x^i$. The first of these quantities appears in the Newtonian limit of the field and geodesic equations in exactly the same place as Newton's constant $G$. In the context of weak-field gravity in non-cosmological systems, it is therefore routinely set equal to one. In the cosmological context, we can do this at the present time $\tau_0$, but this does not necessarily mean that $\alpha$ will retain this value at different points in our cosmic history. It must therefore be a function of time, with the boundary condition $\alpha(\tau_0)=1$ at the present time $\tau_0$.

The parameter $\gamma (\tau)$ is the `curvature of space' parameter, which is tightly constrained by solar system observations to lie within one part in $10^5$ of its GR value of one \cite{Bertotti_2003} at the present time (but not necessarily in the past). Finally, $\alpha_c$ and $\gamma_c$ are cosmological parameters, which are required for consistency of the relevant equation when cosmological evolution is permitted, and which must obey the constraint
\begin{equation} \label{int}
4 \pi G\,\bar{\rho} = \frac{\alpha_c + 2 \gamma_c + \hat{\gamma}_c}{\alpha-\gamma+ \hat{\gamma}} \, ,
\end{equation}
where $\bar{\rho}$ is the average cosmological mass density, and the hats denote differentiation with respect to the number of e-foldings, $\ln a$. These two parameters include information about the dark energy (which here is not taken to be a part of $\bar{\rho}$, the mass density of baryons and dark matter).

The Friedmann equations that emerge from averaging the small-scale geometry take the following form:
\begin{eqnarray}
\mathcal{H}^2 &=& \frac{8 \pi G a^2}{3}\, \gamma \, \bar{\rho}-\frac{2a^2}{3} \, \gamma_c \label{eqn_ppncbkgd1}\\
\mathcal{H}' &=& -\frac{4 \pi G a^2}{3} \, \alpha \, \bar{\rho}+\frac{a^2}{3} \, \alpha_c\, \text{,}\label{eqn_ppncbkgd2}
\end{eqnarray}
where $\mathcal{H}= a'/a$ is the conformal Hubble rate, and primes denote differentiation with respect to $\tau$. The parameter which determines the magnitude of the effective Newton's constant $\alpha$ can be seen to enter into the second Friedmann equation, multiplying the average mass density. Conversely, it is the curvature of space parameter $\gamma$ which enters into this position in the first Friedmann equation. It can be seen that Eq. (\ref{int}) acts as an integrability condition on these equations, and that $\alpha_c$ and $\gamma_c$ take the place of dark energy terms.

The equations governing the perturbations $\Phi$ and $\Psi$ are given by \cite{Sanghai_2016} 
\noindent
\begin{eqnarray}
-\mathcal{H}^2 \Phi- \mathcal{H} \Psi'+\frac{1}{3} \nabla^2\Psi &=& -\frac{4\pi G \bar{\rho} a^2}{3} \, \mu \, \delta  \label{pert1}\\
2\mathcal{H}'\Phi+\mathcal{H}\Phi'+\Psi''+\mathcal{H}\Psi'+\frac{1}{3} \nabla^2\Phi &=& -\frac{4\pi G  \bar{\rho} a^2 }{3} \, \xi \, \delta \, , \label{pert2}
\end{eqnarray}
and \cite{Anton_2021}
\begin{equation}
\Psi'_{,i}+\mathcal{H}\Phi_{,i}=4\pi G a^2  \, \mu \, \rho\, v_i +\mathcal{G} \, \mathcal{H} \Psi_{,i}\label{eqn_parammomcon} \, \text{,}
\end{equation}
where $\nabla^2$ denotes the spatial Laplacian constructed from partial derivatives, and where $\mu$ and $\xi$ are generalisations of Newton's constant\footnote{In previous papers we denoted $\xi=\mu (1-\zeta)$ \cite{Sanghai_2019, Anton_2021}, where $\zeta$ was used as a generalisation of the ``slip'' \cite{caldwell2007constraints, amendola2008measuring}. We refrain from doing so here, as the slip will be used more directly later on.}. The density contrast is written in these equations as $\delta= \delta \rho/\bar{\rho}$, and the 3-velocity it written as $v_i$\footnote{The reader should take the factor $\rho v_i$ in Eq. (\ref{eqn_parammomcon}) to mean the scalar part of this quantity, which in the linear regime is equal to $\bar{\rho}\, v_{,i}$, where $v$ is a scalar potential.}. The $\mathcal{G}$ is a further function, which determines the form of the momentum constraint equation (\ref{eqn_parammomcon}) \cite{Anton_2021}.

The PPNC approach tells us the small-scale limit that the effective Newton's constant and momentum constraint parameters must take is given by
\begin{eqnarray}
\lim_{L \to 0} \mu&=& \gamma \label{eqn_musmall} \, , \qquad
\lim_{L \to 0} \xi= {\alpha} \, , \qquad
\lim_{L \to 0}  \mathcal{G}= \frac{\alpha-\gamma}{\gamma}+ \hat{\gamma} \, \text{,}
\end{eqnarray}
while adiabatic perturbations on very large scales require
\begin{eqnarray}
\lim_{L \to \infty} \mu&=&\gamma-\frac{1}{3} \hat{\gamma}+\frac{1}{12\pi G \bar{\rho}} \, \hat{\gamma}_c\label{eqn_lowkgamma}\\
\lim_{L \to \infty} \xi&=&\alpha-\frac{1}{3}\hat{\alpha}+\frac{1}{12\pi G \bar{\rho}} \, \hat{\alpha}_c\label{eqn_lowkalpha} \\[4pt]
\lim_{L \to \infty} \mathcal{G}&=&0\label{eqn_bigglarge} \, .
\end{eqnarray}
The PPNC approach therefore gives us a remarkable set of equations, parameterised by four functions of time (\ref{funcs}) subject to one constraint (\ref{int}), from which we can write both the background equations (\ref{eqn_ppncbkgd1})-(\ref{eqn_ppncbkgd2}) and the small and large-scale limits of the perturbation equations (\ref{pert1})-(\ref{eqn_parammomcon}), including the fully non-linear regime.

The equations presented above are expected to represent the cosmological behaviour of any theory of gravity that fits into the PPN framework, in terms of direct generalizations of the PPN parameters. What remains to be understood is how the functions $\mu$, $\zeta$ and $\mathcal{G}$ interpolate between the large and small-scale limits presented above. In this study we will examine how some simple theories behave in this transition regime, and compare their behaviour to the simplest elementary functions. Knowledge of how these functions behave as they transition between the large and small-scale limits is important for understanding the formalism, and may be significant for the calculation of some cosmological observables, as well as direct integration of the parameterised equations.

{  The PPNC framework does not require density contrasts to be small. However, in this work we are interested in studying the transition regime between large and small scales, which is within the regime we expect to well approximated by linear perturbation theory. This is because the small-scale end of the linear perturbation regime overlaps with the regime in which post-Newtonian expansions are valid. As all non-linear cosmological structure formation happens within the post-Newtonian regime, the transition away from small-scales must occur within the perturbative regime. As a result, for the remainder of this work} we can safely expand the mass density as $\rho = \bar{\rho} + \delta \rho$ so that the gravitational physics is tractable without having to perform N-body simulations (see Refs. \cite{nlpaper, sankar} for a discussion of this in the context of modified gravity). The small-scale end of the range of $k$-values that we will consider (with $L\lesssim 100\,$Mpc) is then assumed to be well-approximated by post-Newtonian expansions so that the results in Eq. (\ref{eqn_musmall}) can be applied \cite{Goldberg_2017a}, while on the largest scales we can use the limits given in Eqs. (\ref{eqn_lowkgamma})-(\ref{eqn_bigglarge}). Linearising in all perturbations then allows us to write the Fourier space versions of Eqs. (\ref{pert1})-(\ref{eqn_parammomcon}) as
\begin{eqnarray}
-\mathcal{H}^2 \Phi- \mathcal{H} \Psi'-\frac{1}{3}k^2  \Psi &=& -\frac{4\pi G \bar{\rho} a^2}{3} \, \mu \, \delta  \\
2\mathcal{H}'\Phi+\mathcal{H}\Phi'+\Psi''+\mathcal{H}\Psi'-\frac{1}{3}k^2 \Phi &=& -\frac{4\pi G  \bar{\rho} a^2 }{3} \, \xi\, \delta\\[4pt]
\Psi'+\mathcal{H}\Phi&=&4\pi G a^2  \, \mu \, \bar{\rho} \, v +\mathcal{G}\, \mathcal{H} \Psi\text{,}
\end{eqnarray}
where we have followed the usual convention of using the same symbols for quantities defined in Fourier space and real space to avoid further cluttering our notation. It is the scale dependence of the Fourier space couplings $\mu$, $\xi$, and $\mathcal{G}$ that will be investigated below\footnote{We note that in the large and small-scale limits described by Eqs. (\ref{eqn_musmall})-(\ref{eqn_bigglarge}) it makes no difference whether $\mu$, $\xi$ and $\mathcal{G}$ are defined in Fourier space or real space, as in both limits they are scale-independent.}.

\subsubsection{Conceptual comparison to other approaches}
\label{sec_eftcomparison}

{  Before continuing, we note that the PPNC approach is different in both concept and execution to the Effective Field Theory (EFT) approaches to modified gravity in cosmology that have recently found prominence in the literature \cite{eft1,eft2,eft3}.

One key difference is that PPNC does not require us to specify the field content of the theory, but instead relies on the specification of the possible relationships between gravitational potentials and matter fields, which are often in the form of Coulomb potentials (or generalisations of Coulomb potentials\footnote{Standard screening mechanisms (chameleon, Vainshtein, symmetron), and long-range massive fields, have not yet been included in this approach, but we hope to remedy this situation in future work.}). This approach is borrowed from the PPN formalism upon which PPNC is constructed, and which has proven so successful in theory-independent tests of gravity in the Solar System. Conversely, in EFT approaches one specifies the field content from the outset, which is often taken to be a single additional scalar field. A second key difference is that EFTs are constructed using cosmological perturbation theory, in which background and perturbation equations are treated separately. On the other hand, the PPNC approach has its foundation in post-Newtonian expansions, from which one can construct both cosmological background and perturbations \cite{Sanghai_2017}.

This differences in starting point have some important consequences. For one, the structure of the EFT approach means that functions that are introduced in parameterisations will appear in either the background or the perturbation equations, but not both (as they are treated separately). This is in contrast to the PPNC approach where parameterizing functions are introduced into the post-Newtonian sector initially, and hence simultaneously appear in both the cosmological background and perturbation equations that emerge. Of course, for any specific theory that fits into both approaches the EFT background and perturbation functions would be found to depend on the underlying parameters of the theory in such a way that they could not be chosen independently. The PPNC approach makes this interdependence explicit, and as a consequence does not (in its current formulation) appear to be consistent with designer theories in which cosmological background and perturbations can be modified independently.

The fact that the PPNC approach is not built explicitly on either cosmological perturbation theory or an EFT expansion has some further advantages. The first is that it allows the time-dependence of the PPN parameters to be constrained with cosmology, and therefore for the constraints that are imposed in the Solar System to be directly extended into a much wider context \cite{Will_2014}. The second is that the equations governing the growth and behaviour of inhomogeneities do not require the density contrast to be small, which means that they can provide a complete description of cosmological structure formation on all scales, such that the non-linear regime does not have to be handled in a different or separate manner. This could prove to be an important benefit for upcoming surveys such as Euclid, and we leave to future work the creation of N-body (or COLA \cite{cola}) simulations within this framework. Lastly, the fact that the PPNC formalism does not require the specification of the field content of the theory from the outset gives it the potential to have considerably more theory independence.}

\subsection{Scalar-Tensor Theories of Gravity}

The example theories we will use to investigate the transition regime from small to large scales will be the Bergmann-Wagoner class of canonical scalar-tensor theories \cite{bergmann1968comments, wagoner1970scalar}. These have the action
\begin{equation} \label{action}
S=\frac{1}{16 \pi G}\int d^4 x \sqrt{-g}\left[\phi R -\frac{\omega}{\phi}\nabla^\mu \phi \nabla_\mu \phi-2\Lambda\right]\text{,}
\end{equation}
where $\omega=\omega(\phi)$ and $\Lambda=\Lambda(\phi)$ are functions that specify the coupling between the scalar and tensor degrees of freedom in the theory, and the scalar field potential, respectively. This class of theories contains within it the very well studied Brans-Dicke theory ($\omega=$constant and $\Lambda=0$) \cite{jbd}, as well as GR with a cosmological constant ($\omega \rightarrow \infty$ and $\Lambda =$constant). These are fully conservative theories, in that they do not have any preferred frame or preferred location effects, while they possess the full set of conservation laws (energy, momentum angular momentum and centre-of-mass motion). They are the sub-class of Horndeski theories with $G_2 =G_2 (\phi)$ and $G_4=G_4(\phi)$, and with all other $G_i=0$ \cite{horndeski1974second}.

In the perturbed Robertson-Walker geometry given by Eq. (\ref{eqn_flrw}), the Friedmann equation in these theories can be written as
\begin{eqnarray} \label{stf}
&&\mathcal{H}^2=\frac{8\pi G a^2}{3\phi} \, \bar{\rho}-\mathcal{H} \frac{{\phi}'}{\phi}+\frac{\omega {\phi'}^2}{6\phi^2} +\frac{\Lambda a^2}{3\phi } \, ,
\end{eqnarray}
and the Klein-Gordon equation for the scalar field as
\begin{eqnarray} \label{sts}
&&\frac{{\phi''}}{\phi}=\frac{8 \pi G a^2}{\left(3+2\omega \right)\phi} \, \bar{\rho} - 2\mathcal{H}\frac{{\phi'}}{\phi}-\frac{d\omega}{d\phi} \frac{{\phi'}^2}{\left(3+ 2\omega \right)\phi}  +\frac{4\Lambda a^2}{\left(3+ 2\omega \right)\phi } - \frac{2a^2}{(3+2 \omega)} \frac{d\Lambda}{d\phi} \, ,
\end{eqnarray}
where $\bar{\rho}$ is again the mass density and radiation has been neglected. The second Friedmann equation follows from differentiating Eq. (\ref{stf}), and eliminating the second derivative of the scalar field using Eq. (\ref{sts}).

The perturbation equations around a Robertson-Walker background can be split up into constraint and evolution equations; we present these equations up to linear order in Fourier space. The Hamiltonian and momentum constraints can be written as
\vspace{-0.25cm}
\begin{eqnarray} \nonumber
&&\hspace{-1.2cm} 3 \mathcal{H}^2 \delta \phi + \frac{\omega}{2}\left( \frac{\phi'}{\phi}\right)^2 \delta \phi + 3 \mathcal{H} \delta \phi' - \omega \frac{{\phi}'}{{\phi}} \delta \phi' + 6 \mathcal{H} {\phi} \Psi' + 3 {\phi}' \Psi' +k^2 \delta \phi  \\ &&\hspace{-0.5cm} + 2 \phi k^2 \Psi
+6\mathcal{H}^2\phi \Phi +6\mathcal{H}\phi'\Phi- \omega \frac{{\phi}'^2}{{\phi}} \Phi
- \frac{d \omega}{d\phi} \frac{\phi'^2}{\phi} \frac{\delta \phi}{2} - a^2 \frac{d\Lambda}{d\phi} \delta \phi = 8 \pi G a^2 \, \delta \rho \label{ham}
\end{eqnarray}
\vspace{-0.5cm}
\begin{flalign}
&{\rm and} \hspace{2cm} \mathcal{H} \delta \phi - \omega \frac{\phi'}{\phi} \delta \phi - \delta \phi'-2 \mathcal{H} \phi \Phi -\phi' \Phi-2 \phi \Psi' = -8 \pi G a^2 \, \bar{\rho}\, v \, ,&
\end{flalign}
where $v$ is the velocity potential. We have also abused notation to write $\phi$ and $\delta \phi$ as the background and perturbation to the scalar field. As is usual in longitudinal gauge, the shear evolution equation also reduces to a constraint, which in this case can be written in the particularly useful form
\vspace{-0.25cm}
\begin{equation}
\Phi-\Psi=\frac{\delta \phi}{\phi} \, .
\end{equation}
The evolution equations for the matter variables can now be written as
\vspace{-0.25cm}
\begin{eqnarray}
v'+\mathcal{H}v-\Phi &=& 0\\
\delta\rho'+3\mathcal{H}\delta\rho+3\bar{\rho}\Psi' - k^2 \bar{\rho}\, v &=& 0 \, ,
\end{eqnarray}
while the perturbed Klein-Gordon equation gives
\vspace{-0.25cm}
\begin{eqnarray}
&&\hspace{-1cm}\left[4 \mathcal{H} \phi' \delta \phi  + 2 \phi'' \delta \phi + 2 \phi' \delta \phi'  + 2 \phi'^2 \Phi  \right] \frac{d \omega}{d\phi} +\phi'^2 \delta \phi \frac{d^2\omega}{d\phi^2} - 2 a^2 \delta \phi \frac{d \Lambda}{d\phi} +2 a^2 \phi \delta \phi'  \frac{d^2 \Lambda}{d\phi^2} \nonumber \\
&&\hspace{-1cm}\quad+(3  +2 \omega) \left[ \delta \phi'' + 2 \mathcal{H} \delta \phi' + 4 \mathcal{H} \phi' \Phi + 2 \phi'' \Phi +\phi' \Phi' + 3  \phi' \Psi' +k^2 \delta \phi\right]= 8 \pi G a^2 \, \delta \rho \, ,
\end{eqnarray}
and the Raychaudhuri equation gives
\vspace{-0.25cm}
\begin{eqnarray}
&&-2\phi\mathcal{H}\Phi'
-4\mathcal{H}'\phi\Phi-2\mathcal{H}^2\phi\Phi+\frac{2}{3}\phi k^2\Phi
-2\phi\Psi''-4\phi\mathcal{H}\Psi'-\frac{2}{3}\phi k^2\Psi \nonumber \\
=&&2\mathcal{H}'\delta\phi+\mathcal{H}^2\delta\phi+2\phi''\Phi+\delta\phi''+\mathcal{H}\delta\phi'+\frac{2}{3} k^2\delta\phi
-\omega\frac{\phi'^2}{2\phi^2}\delta\phi+\omega\frac{\phi'}{\phi}\delta\phi' \nonumber\\
&&+\phi'\Phi'+2\phi'\mathcal{H}\Phi+2\phi'\Psi' -a^2 \frac{d\Lambda}{d\phi}  \delta \phi+\frac{d\omega}{d\phi}\frac{\phi'^2}{2\phi}\delta\phi+\omega\frac{\phi'^2}{\phi}\Phi \text{.}
\label{ray}
\end{eqnarray}
This provides the complete set of linearized scalar equations in these theories.

The relevant PPN parameters $\alpha$ and $\gamma$ are obtained by taking a post-Newtonian expansion about Minkowski space and comparing the resultant geometry to the test metric (see Ref. \cite{Will_1993} for details). The result is
\begin{eqnarray} \label{ppn}
\alpha=\frac{(4+2\omega)}{(3+2\omega)}\frac{1}{\phi} \qquad {\rm and} \qquad
\gamma=\frac{(2+2\omega)}{(3+2\omega)}\frac{1}{\phi} \, .
\end{eqnarray}
Usual practise, within the literature on weak-field gravity, would be to set $\phi$ (by a choice of units) so that $\alpha=1$. Such a choice means that Newton's constant is recovered in the proper place in the Newton-Poisson equation and the geodesic equation. The appropriate choice in cosmology is to make this choice at $\tau=\tau_0$, so that Newton's constant takes the value $G$ at the present time, but is allowed to take different values at other points in cosmic history. This naturally incorporates the idea of a ``varying Newton's constant'' \cite{uzan2011varying}.

In addition to the usual PPN parameters, we are required to introduce the following additional two parameters in order to construct the cosmological equations \cite{Sanghai_2017}:
\begin{eqnarray}
&&\hspace{-1cm}a^2\,\alpha_c=-\omega \frac{\phi'^2}{\phi^2}-\frac{\phi''}{\phi}+\mathcal{H}\frac{\phi'}{\phi}+\frac{\phi'^2}{2\phi \left(3+2\omega \right)}\frac{d\omega}{d\phi}  +\frac{\left(1+2\omega \right)}{\left(3+2\omega \right)}\frac{\Lambda \,a^2}{\phi}+ \frac{1}{\left(3+2\omega \right)}\frac{d \Lambda}{d\phi}\label{eqn_alphac}\\
&&\hspace{-1cm}a^2 \, \gamma_c=-\frac{\omega}{4}\frac{\phi'^2}{\phi^2}-\frac{1}{2} \frac{\phi''}{\phi}+\frac{\mathcal{H}}{2}\frac{\phi'}{\phi}-\frac{\phi'^2}{2\phi \left(3+ 2\omega \right)}\frac{d\omega}{d\phi} +\frac{\left(1-2\omega \right)}{\left(3+2\omega \right)}\frac{\Lambda \, a^2}{2 \phi}+ \frac{2}{\left(3+2\omega \right)}\frac{d \Lambda}{d\phi}\label{eqn_gammac}\, .
\end{eqnarray}
One can verify that these parameters satisfy the constraint equation (\ref{int}). In the limits $\omega\rightarrow \infty$ and $\Lambda \rightarrow \,$constant,  we can see that these equations reduce to $\alpha_c = -2 \gamma_c = \Lambda$ and $\alpha = \gamma=1$, thereby recovering GR with a cosmological constant, as expected\footnote{The background equations can be used to verify that $\omega \phi'^2/\phi^2 \rightarrow 0$ in this limit.}.

In what follows we will take $\omega$ and $\Lambda$ to be constants (unless otherwise specified), and integrate the background and perturbation equations of these theories directly in order to verify the theoretical predictions of the large and small-scale limits that are given by the PPNC formalism. We will further use our numerical solutions to investigate how the effective Newton's constants $\mu$ and $\xi$ vary between these two limits within this class of theories. The results will then be used to determine how well a simple interpolating function would have performed in the PPNC equations, when they themselves are directly integrated. We will use the resulting density contrasts, and some simple observables, to determine the accuracy that the PPNC equations would have achieved in comparison to the direct integration of the example theories outlined above. {  Due to theories that introduce new scales not being contained in the set currently covered by the PPNC approach (see comments in Section \ref{sec_eftcomparison}), we expect this transition behaviour to be broadly representative of the general case. }\\

\section{Application in Example Theories}
\label{sec:numerics}

In this section we will solve the differential equations that describe the background cosmology and all relevant perturbed quantities in the example theories described above. We first solve for the background cosmology\footnote{Note that we neglect the effects of a radiation component throughout our calculations.} to redshifts larger than $1\,100$, and verify that the evolution we obtain obeys the background PPNC equations (\ref{eqn_ppncbkgd1})-(\ref{eqn_ppncbkgd2}). We then solve for the perturbations, using suitable initial conditions, by evolving $\Psi$ and $\phi$ using the Raychaudhuri and Klein-Gordon equations, while calculating $\Phi$, $\delta\rho$ and $v$ at each time step using the anisotropic stress, Hamiltonian and momentum constraint equations. We work in code units where $c=1$, and such that the conformal time today is unity, $\tau_0=1$. 

\subsection{Background Cosmology and Parameter Evolution}
\label{sec:bg}

The background scalar-tensor equations (\ref{stf})-(\ref{sts}) can be integrated once values for $\omega$ and $\Lambda$ have been provided, along with suitable boundary conditions for $a(\tau)$ and $\phi(\tau)$. For illustrative purposes, the integrations we present in this section will be for several different combinations of constant values of $\omega$ and $\Lambda$. We will not concern ourselves with whether or not the values we choose are compatible with observations (some of them will not be), as at this stage we are concerned only with exemplifying how the PPNC formalism works in example theories that differ from GR, and not with modelling any particular physical phenomena. For these purposes it is useful to consider theories that are allowed to differ substantially from GR. The boundary conditions we will use are that $a(\tau_0)=1$ and that $\phi(\tau_0)=(4+2 \omega)/(3+2 \omega)$ in the attractor solution for $\phi(\tau)$. This last condition is chosen so that $\alpha(\tau_0)=1$ from Eq. (\ref{ppn}). Both choices can be thought of as specifications of units.

\begin{figure}[t!]
\centering
\includegraphics[height =5cm]{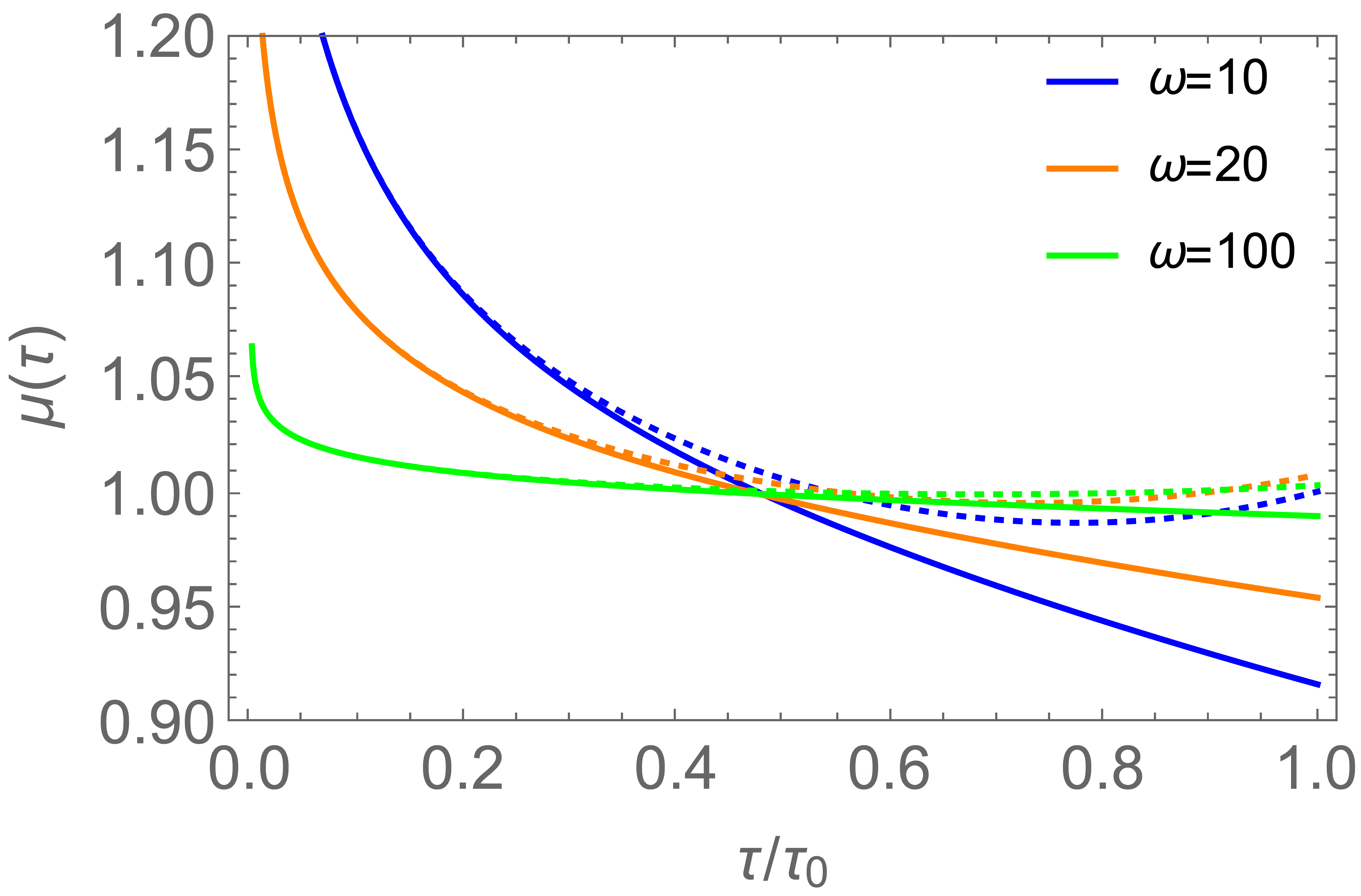}
\includegraphics[height =5cm]{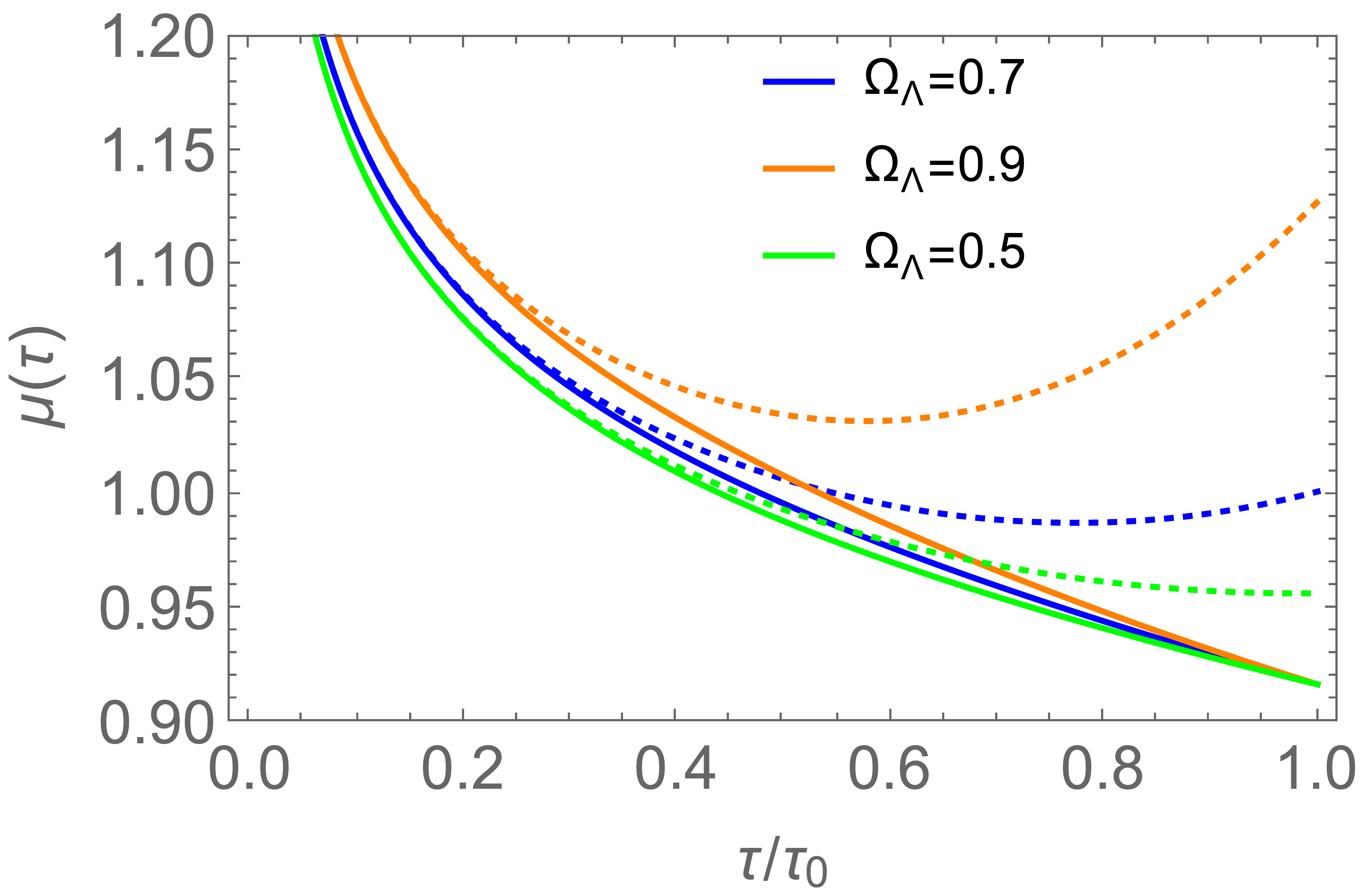}
\caption{Large (dotted) and small (solid) scale limits of $\mu(\tau)$. Left: differing values of $\omega$, with $\Omega_\Lambda=0.7$. Right: differing values of $\Omega_\Lambda$, with $\omega=10$. 
}
\label{fig_gamma}
\vspace{0.5cm}

\centering
\includegraphics[height =4.9cm]{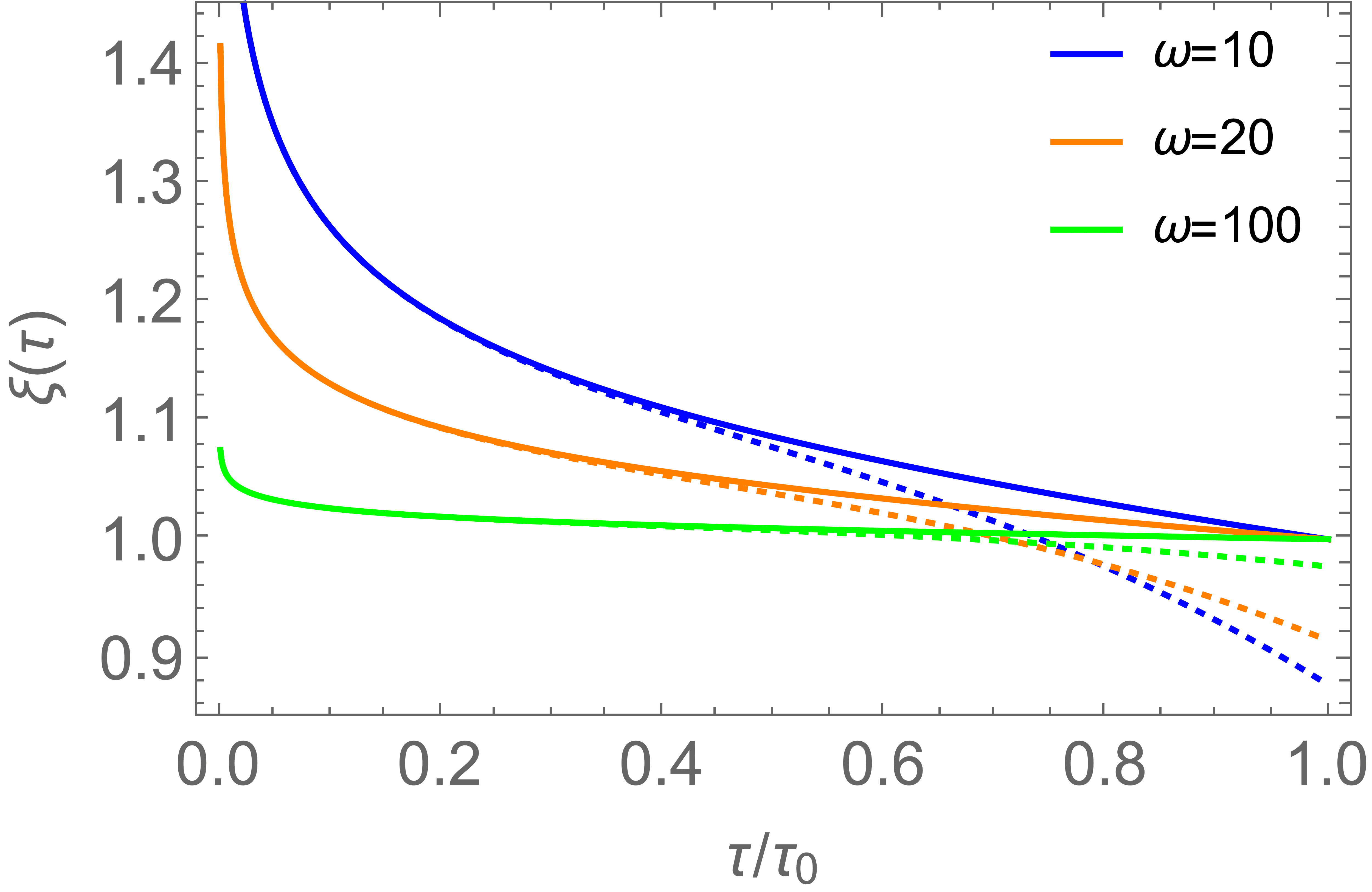}
\includegraphics[height =4.9cm]{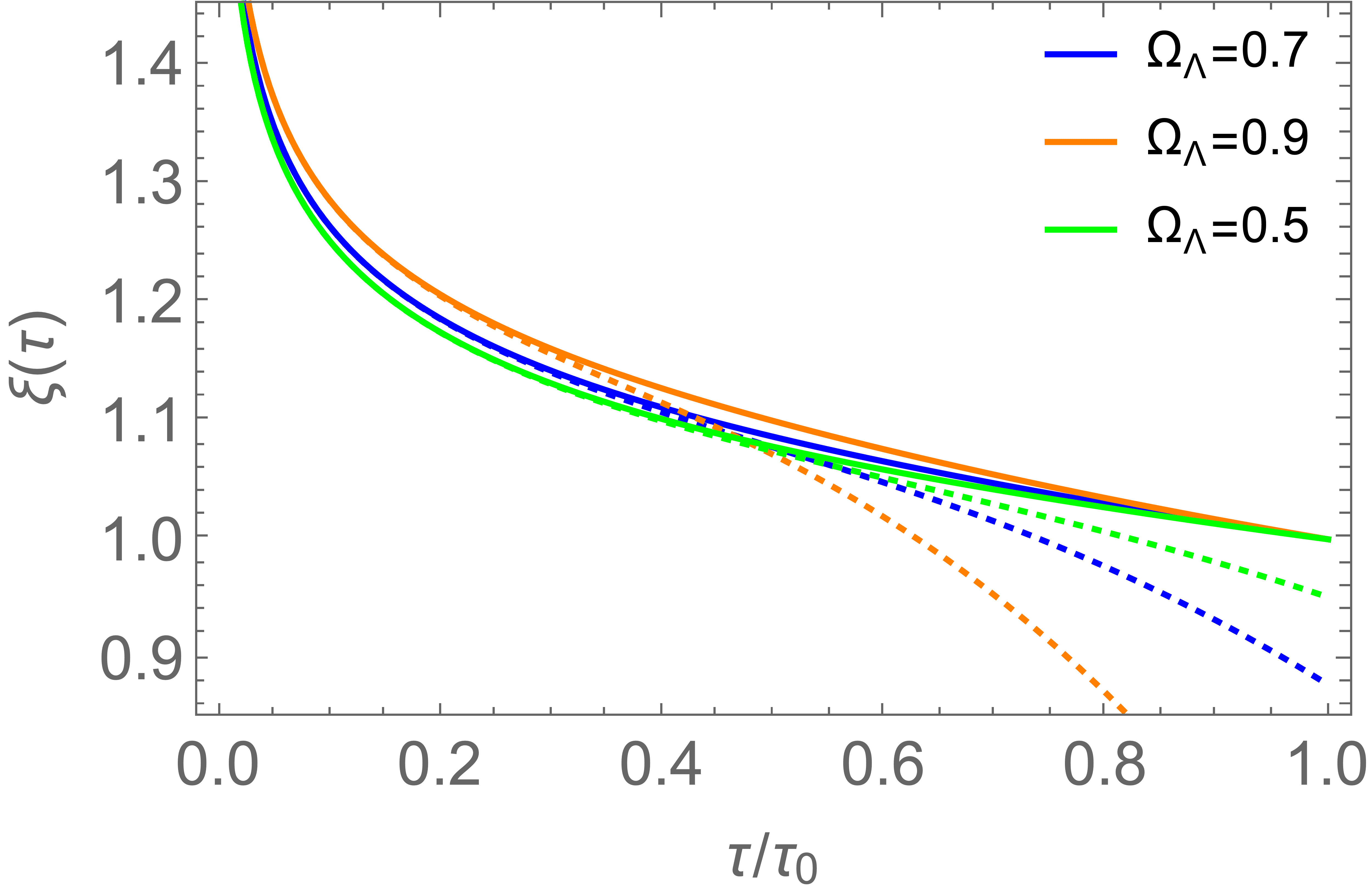}
\caption{Large (dotted) and small (solid) scale limits of $\xi(\tau)$. Left: differing values of $\omega$, with $\Omega_\Lambda=0.7$. Right: differing values of $\Omega_\Lambda$, with $\omega=10$. 
}
\label{fig_alpha}
\end{figure}

The behaviour of $a(\tau)$ and $\phi(\tau)$ have been well studied in the literature on these theories\footnote{During dust domination the attractor solution is $a(\tau) \propto \tau^{\frac{2 (1+\omega)}{2+\omega}}$ and $\phi(\tau) \propto \tau^{\frac{2}{2+\omega}}$ \cite{nariai1968green}.}, and we will not present them here beyond commenting that $\phi$ increases with time for positive values of $\omega$. Once these solutions are known, however, they can be used to derive values for the PPNC parameters specified in Eqs. (\ref{ppn})-(\ref{eqn_gammac}). In turn, these can be used to derive the values of the parameters specified in Eqs. (\ref{eqn_musmall})-(\ref{eqn_bigglarge}). We plot the large and small-scale values of $\mu(\tau)$ and $\xi(\tau)$ in Figures \ref{fig_gamma} and \ref{fig_alpha}, for several different choices of $\omega$ and $\Omega_\Lambda$. The reader will note that the small-scale values of these two quantities (i.e. the solid lines) correspond directly to the values of the PPN parameters $\gamma$ and $\alpha$ at different times in cosmic history. We consider how such time dependence could be fitted in Appendix \ref{app:params}.

In both Figures \ref{fig_gamma} and \ref{fig_alpha} we can see that at early times the behaviour of $\mu$ and $\xi$ is very similar on both large and small spatial scales. This is because the $\gamma$ and $\alpha$ terms dominate the expressions for these quantities on large scales. In fact, it is only in the presence of a non-zero $\Lambda$ that the large and small-scale limits are different. This appears to be a feature of the particular theories we are considering, and is not expected to be a generic property of theories of gravity in general. The curves that depict the small-scale limit of $\xi$ in Figure \ref{fig_alpha} can be seen to converge at $\tau=\tau_0$ (today), which is due to our choice of boundary condition on $\phi$. This also explains why the solid curves in the right-hand plot in Figure \ref{fig_gamma} converge today; as the value of $\gamma$ depends only on $\omega$, once $\phi_0$ has been fixed. 

Going back in time, $\alpha$ and $\gamma$ increase for all of our theories. The differences between small and large scales, however, increase into the future. Increasing the value of  $\Omega_\Lambda$ can be seen to exacerbate this situation, while increasing $\omega$ has the opposite effect. In general, larger values of $\omega$ can be seen to result in a slower evolution in the values of $\mu$ and $\xi$, as expected (as this is the limit in which GR is recovered). While $\alpha$ is always greater than the GR value of unity, it can be seen that $\gamma$ crosses the GR value of unity during its evolution (starting above it and finishing below it, in the cases we consider). It will be the behaviour between the large and small-scale limits depicted in these plots that will concern us for much of the rest of this study. For this we will also need to understand the evolution of perturbations.

\subsection{Perturbations and Scale-Dependent Couplings}
\label{sec_numpert}

In this section we will evolve the perturbations in our example theories, and use them to determine the scale dependence of the couplings that appear in the PPNC parameterisation. This will help to provide guidance of how the interpolation between different spatial scales works within a mathematically well-defined alternative to GR.

\subsubsection{Initial Conditions}

To begin, let us specify the initital conditions for our numerical integration. We wish to specify these as adiabatic fluctuations that satisfy \cite{Malik_2009}
\begin{equation} \label{ad}
\frac{\delta \phi}{\phi'}=\frac{\delta \rho}{\rho'} \, .
\end{equation}
We will set initial conditions for our perturbations by starting the evolution at an early enough time that all Fourier modes of interest start off outside the horizon. Isocurvature modes are neglected, as they are generally expected to be small on these scales \cite{gordon2000adiabatic}.

Under the adiabatic condition in Eq. (\ref{ad}), it can be seen that the large-scale limit of Eqs. (\ref{ham})-(\ref{ray}) admit the solution
\begin{eqnarray}
\Phi = \frac{\zeta'}{\mathcal{H}} -\frac{\mathcal{H}'}{\mathcal{H}^2} \zeta + \zeta
\qquad {\rm and} \qquad
\Psi = c_1 - \zeta
\end{eqnarray}
with the momentum constraint giving $v=\mathcal{H}^{-1} \zeta$ and where
\begin{equation}
\zeta = -\mathcal{H} \frac{\delta \phi}{\phi'}= -\mathcal{H} \frac{\delta \rho}{\rho'} = c_1 \mathcal{H} \frac{\int^{\tau} a(\tilde{\tau})^2  \phi (\tilde{\tau}) \, d\tilde{\tau}}{a^2 \phi} + c_3 \frac{\mathcal{H}}{a^2 \phi}
\end{equation}
is the comoving curvature perturbation on uniform density hypersurfaces. The $c_1$ and $c_3$ in these expressions are constants, and taking $\zeta (\tau_i) = c_2$ to be another constant at the initial time $\tau_i$ allows us to write initial conditions for $\delta \phi$ and $\delta \phi'$ as
\begin{eqnarray}
\delta \phi \vert_{\tau_i} = -\frac{ {\phi}'}{\mathcal{H}} \Bigg\vert_{\tau_i} \, c_2
\qquad {\rm and} \qquad
\delta {\phi}' \vert_{\tau_i} = -  {\phi'}\vert_{\tau_i} \, c_1-  \left[ \left( \frac{{\phi''}}{\phi}- 2 \mathcal{H} \frac{{\phi'}}{\phi} - \frac{{\phi'}^2}{\phi^2} \right)\frac{\phi}{\mathcal{H}} \right] \Bigg\vert_{\tau_i} \,  c_2 
\end{eqnarray}
and the initial conditions for $\Psi$ and $\Psi'$ as
\begin{equation}
\Psi\vert_{\tau_i} =  c_1 - c_2
\qquad {\rm and} \qquad
{\Psi}'\vert_{\tau_i} = - \mathcal{H}\vert_{\tau_i} \, c_1+ \left[\frac{{\phi'}}{\phi} + 2 \mathcal{H} - \frac{{\mathcal{H}'}}{\mathcal{H}}  \right]  \Bigg\vert_{\tau_i} \, c_2  \, .
\end{equation}
Together with the constraint equations, these conditions are sufficient to integrate the Raychaudhuri and Klein-Gordon equations, and therefore serve as initial conditions for super-horizon adiabatic perturbations. The two constants $c_1$ and $c_2$ in the equations above can be thought of as specifying the initial amplitude of the growing and decaying modes within this setup. To ensure that the decaying mode is under control, we impose on these constants the supplementary condition\footnote{This condition should be expected to be satisfied at early times in these theories, as $\delta \phi'  \rightarrow 0$ during radiation domination, when the source term in the scalar field evolution equation vanishes.} $\delta \phi' \vert_{\tau_i} =0$, which leaves us with a single constant that specifies the amplitude of our perturbations at the initial time $\tau_i$.

\subsubsection{Present-Day Values of Gravitational Couplings}

\begin{figure}
\centering
\includegraphics[height =4.9cm]{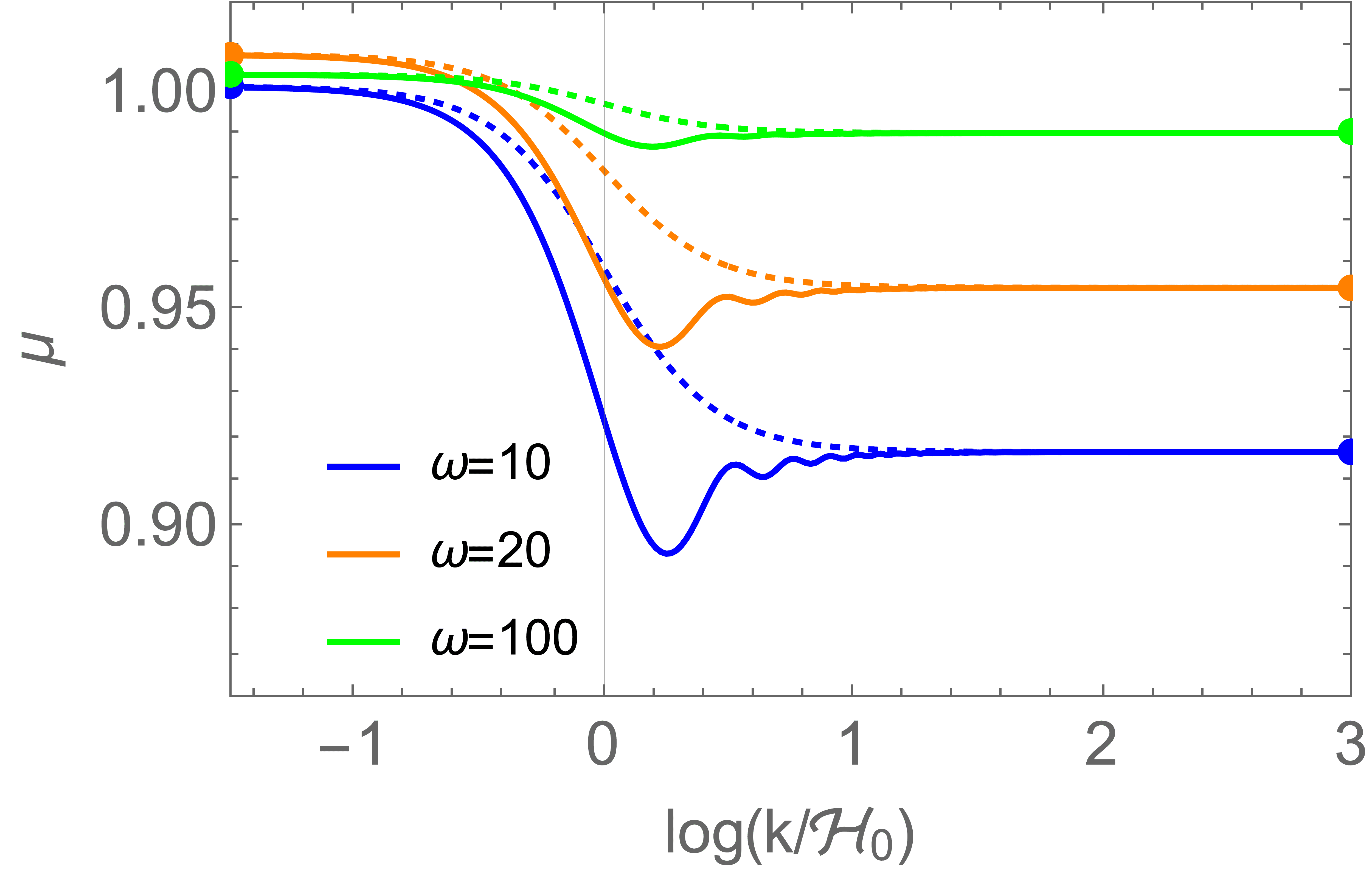}
\includegraphics[height =4.9cm]{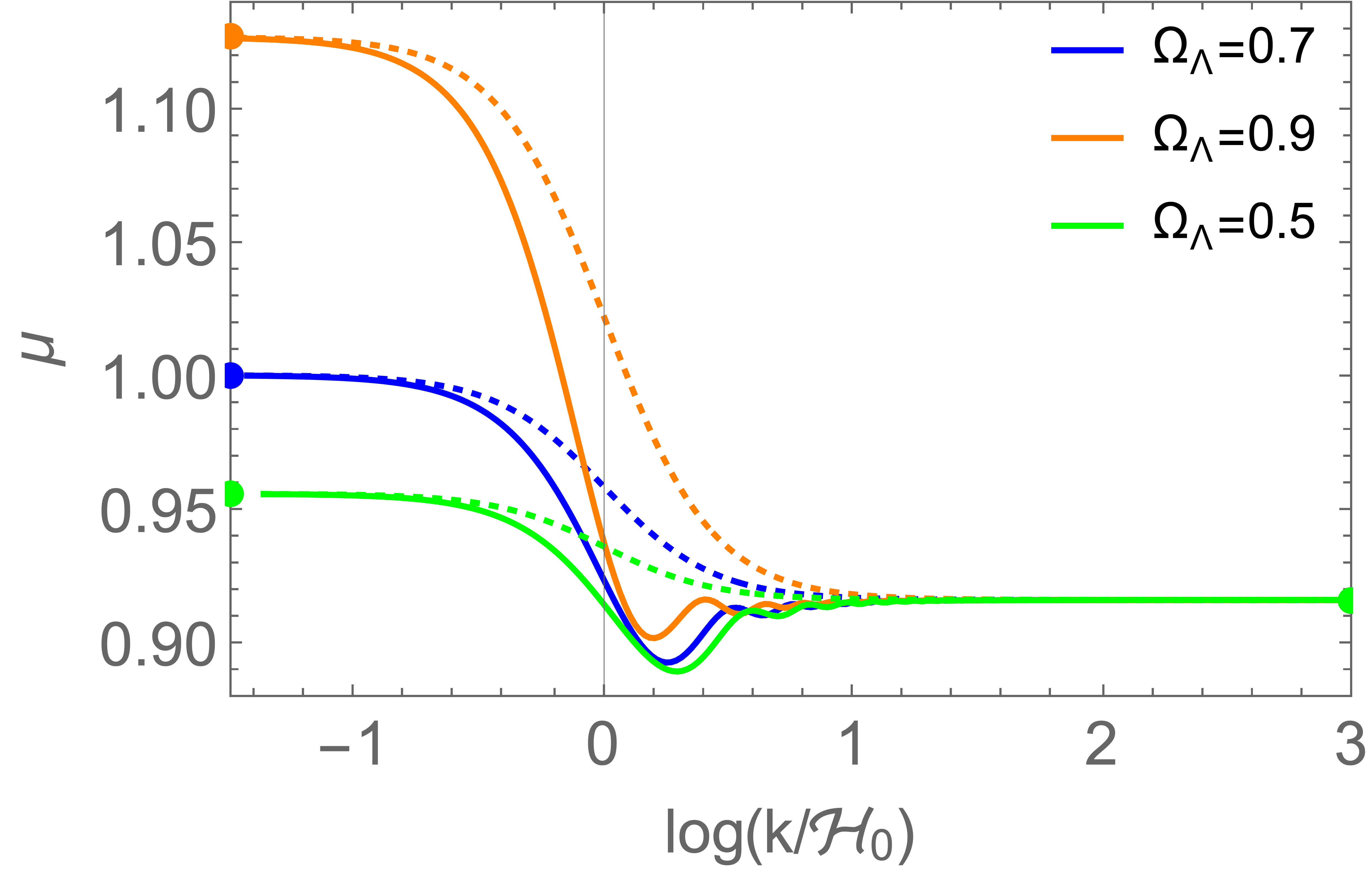}
\caption{Solid lines depict the function $\mu(k)$ evaluated at $z=0$ and constructed as per Eq. (\ref{eqn_mudefn}). Dotted lines show the simple interpolating function from Eq. (\ref{eqn_timinterp}). Dots denote the asymptotic values expected from Eqs. (\ref{eqn_musmall})-(\ref{eqn_bigglarge}). Left: differing values of $\omega$, with $\Omega_\Lambda=0.7$. Right: differing values of $\Omega_\Lambda$, with $\omega=10$. 
}
\label{fig_hamcon}
\vspace{0.5cm}

\centering
\includegraphics[height =4.9cm]{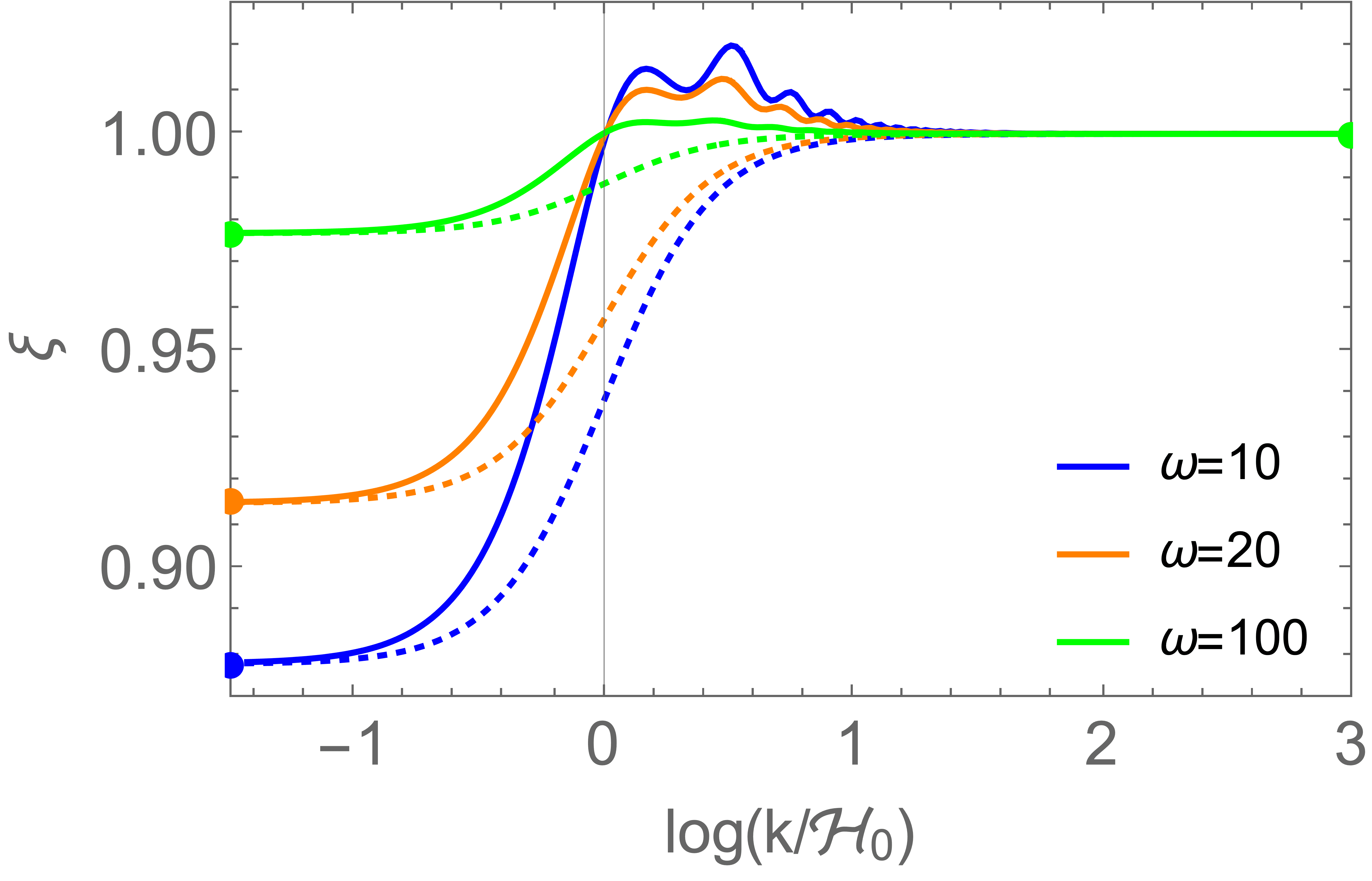}
\includegraphics[height =4.9cm]{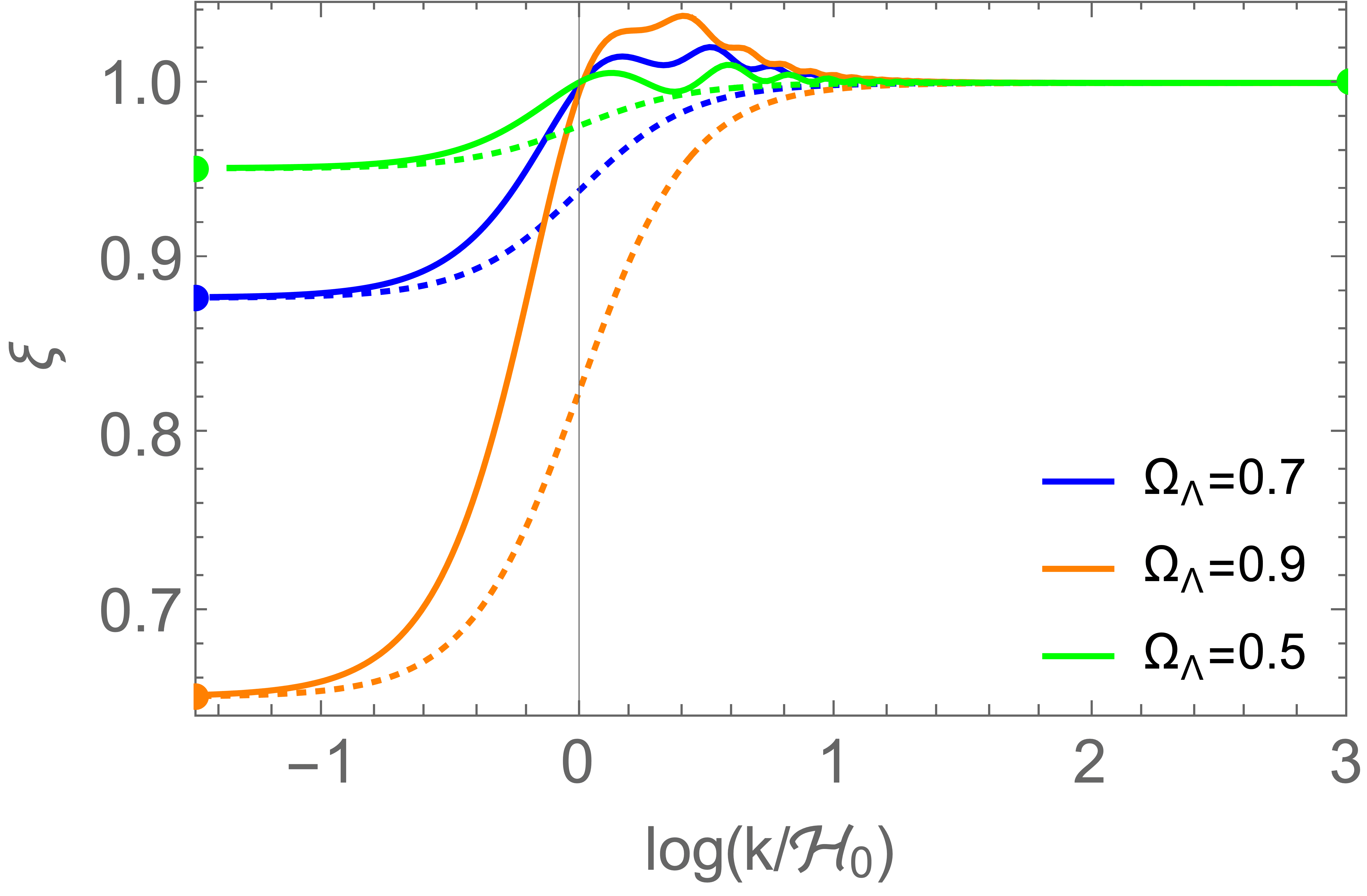}
\caption{Solid lines depict the function $\xi(k)$ evaluated at $z=0$ and constructed as per Eq. (\ref{eqn_mudefn2}). Dotted lines show the simple interpolating function from Eq. (\ref{eqn_timinterp}). Dots denote the asymptotic values expected from Eqs. (\ref{eqn_musmall})-(\ref{eqn_bigglarge}). Left: differing values of $\omega$, with $\Omega_\Lambda=0.7$. Right: differing values of $\Omega_\Lambda$, with $\omega=10$. 
}
\label{fig_raynow}
\vspace{0.5cm}

\centering
\includegraphics[height =4.8cm]{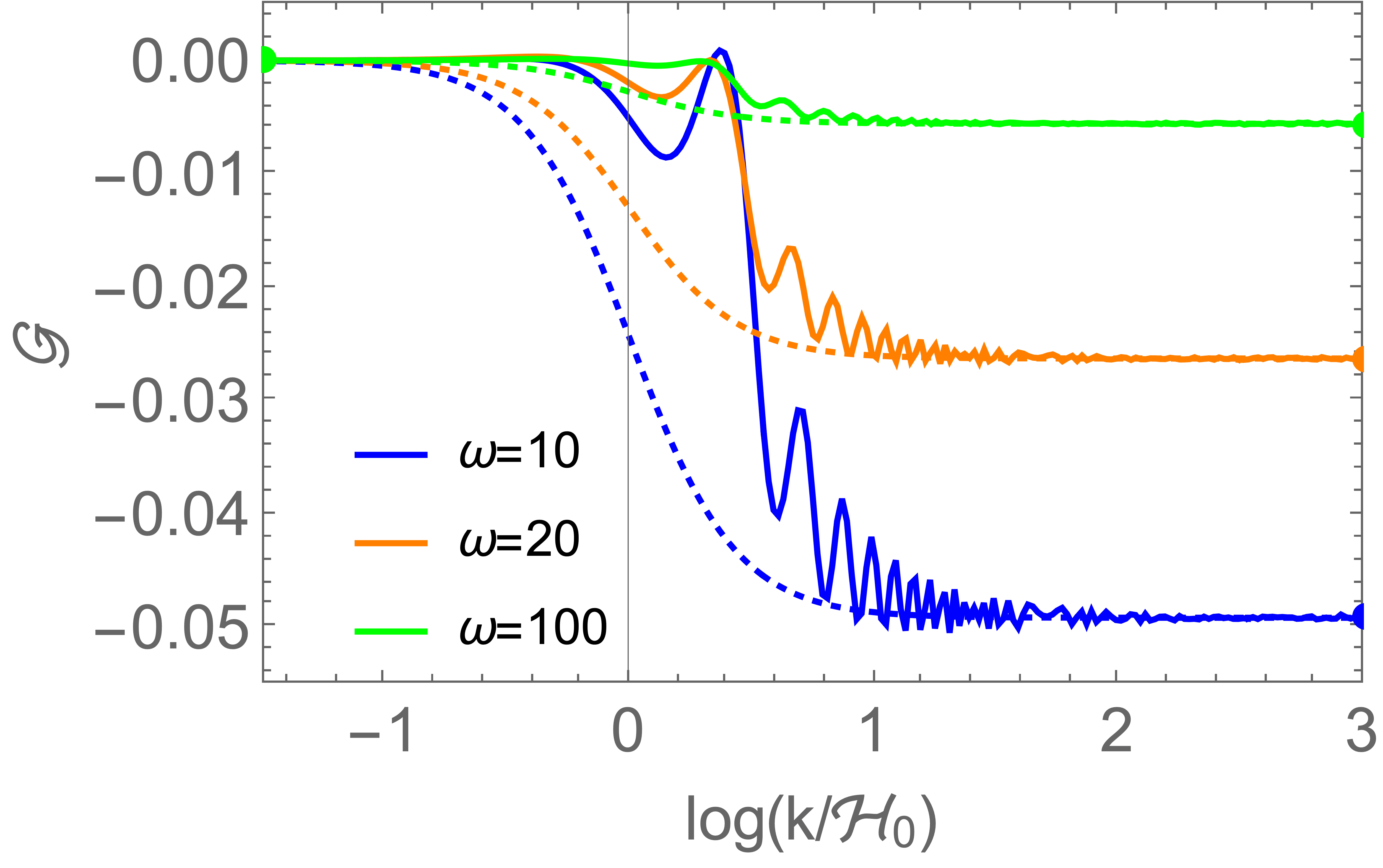}
\includegraphics[height =4.8cm]{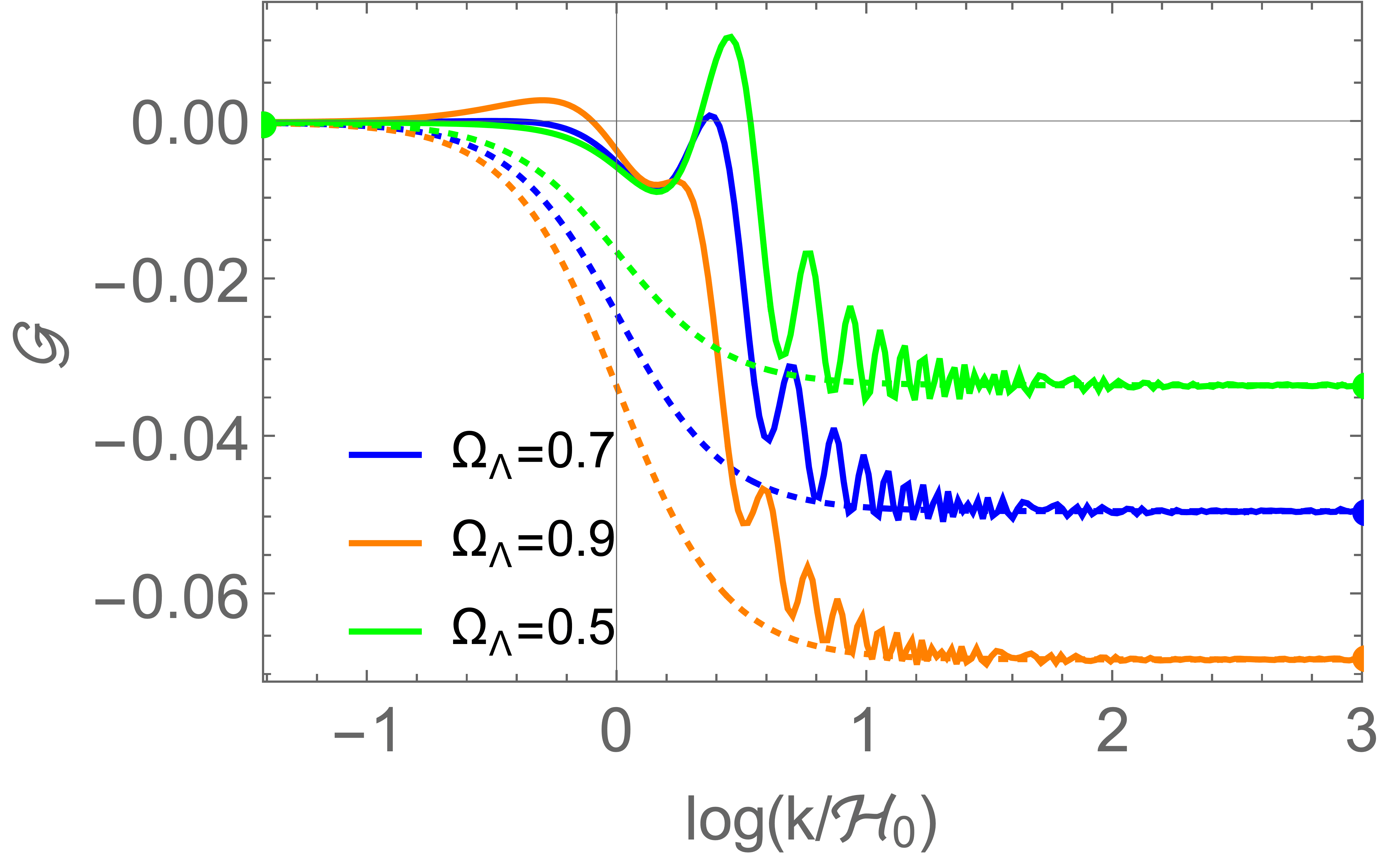}
\caption{Solid lines depict the function $\mathcal{G}(k)$ evaluated at $z=0$ and constructed as per Eq. (\ref{eqn_gdefn}). Dotted lines show the simple interpolating function from Eq. (\ref{eqn_timinterp}). Dots denote the asymptotic values expected from Eqs. (\ref{eqn_musmall})-(\ref{eqn_bigglarge}). Left: differing values of $\omega$, with $\Omega_\Lambda=0.7$. Right: differing values of $\Omega_\Lambda$, with $\omega=10$. 
 }
\label{fig_momcon}
\end{figure}

In this section we examine the gravitational couplings $\mu$, $\xi$ and $\mathcal{G}$ at the present time $\tau=\tau_0$, by numerically evolving our perturbation equations from the initial conditions specified above. This is achieved by taking the values of the $\Phi$, $\Psi$ and $\delta \rho$ from our integration, and substituting them into the following re-arranged versions of Eqs. (\ref{pert1})-(\ref{eqn_parammomcon}), obtained after linearisation and Fourier transformation:
\begin{eqnarray}
\mu&=&\frac{3\mathcal{H}^2 \Phi+3\mathcal{H} \Psi'+k^2\Psi}{4\pi G a^2 \delta\rho} \label{eqn_mudefn} \\
\xi&=&-\frac{6\mathcal{H}'\Phi+3 \mathcal{H}\Phi'+3 \Psi''+3 \mathcal{H}\Psi'-k^2\Phi}{{4\pi G a^2 \delta\rho}} \label{eqn_mudefn2} \\
\mathcal{G}&=&\frac{\Psi'+\mathcal{H}\Phi-4\pi G a^2  \mu \bar{\rho} v}{\mathcal{H}\Psi} \, \text{.} \label{eqn_gdefn}
\end{eqnarray}
The results, plotted in Figures \ref{fig_hamcon} to \ref{fig_momcon}, show the values of the coupling functions $\mu$, $\xi$ and $\mathcal{G}$ at $z=0$ in these theories.

In each of these figures the wavenumber $k$ of the perturbative modes is plotted on the x-axis in units of the Hubble rate, $\mathcal{H}$. We can clearly see the scale-independent limits predicted by Eqs. (\ref{eqn_musmall})-(\ref{eqn_bigglarge}) are being approached as $k \gg \mathcal{H}$ on the right of each plot, and $k \ll \mathcal{H}$ on the left. Each line in each plot represents a particular combination of constants in the theory, but in each case we can see a smooth transition between these two limits. The vertical black lines in each plots correspond to the Hubble scale $k = \mathcal{H}$, which can be seen to be the scale about which the transition takes place. In each figure the left-hand plot shows the effect of changing the value of the coupling parameter $\omega$, while the right-hand plot shows the effect of changing the amount of dark energy $\Omega_\Lambda$.

In Figures \ref{fig_hamcon} to \ref{fig_momcon} we have included simple interpolation functions, to explore how well they fit through the transition scale. We considered several possibilities for interpolating between the large and small-scale limits, but found the simplest to be a $\tanh$ function\footnote{More complicated interpolations are considered in Appendix \ref{app:func}.}, as suggested in \cite{Sanghai_2019}:
\begin{equation}
\label{eqn_timinterp0}
f(k)=\bar{f}+ A_f \tanh\left( \ln \frac{k}{\mathcal{H}}  \right)\, ,
\end{equation}
where $\bar{f}= (S+L)/2$ is the mean of the small (S) and large (L) scale limits of the coupling, and $A_f= (S-L)/2$ is its amplitude. Using the definition of the $\tanh$ function, in terms of exponentials, allows us to write this as
\begin{equation}
\label{eqn_timinterp}
f(k) \propto \frac{1-\frac{k^2}{\mathcal{H}^2}}{1+\frac{k^2}{\mathcal{H}^2}} \, ,
\end{equation}
which is similar to previous attempts to parameterise the scale dependence of gravitational couplings using Pad\'{e} approximants \cite{amin2008subhorizon, bakerbull}. This interpolation function is fixed to reach a value halfway between the two endpoints at the Hubble scale, but otherwise to require no additional parameters (i.e. they are a `zero-parameter' fit). While they give a reasonable first approximation, these simple functions clearly miss the detailed (oscillating) features. We will examine these features in more detail in later sections, but note that they appear to be similar to the Gravity Acoustic Oscillations (GAO) discussed in Ref. \cite{gao}.

In all cases, we can see that deviations from GR reduce to zero in the limit $\omega \rightarrow \infty$ (as expected), and that $\Omega_\Lambda$ acts to increase the difference between the values of couplings at small and large scales (with increasing values of $\Omega_\Lambda$ causing a bigger difference).  For the values of $\xi$ plotted in Figure \ref{fig_raynow} we can see that all curves approach unity as $k \rightarrow \infty$, as required from the present-day value of $\alpha$ and Eq. (\ref{eqn_musmall}). Correspondingly, the values of $\mu$ plotted in Figure \ref{fig_hamcon} approach the values of $\gamma$ given in Eq. (\ref{ppn}), again as anticipated from Eq. (\ref{eqn_musmall}). Of course, when using this formalism to constrain gravity using cosmological observations one may wish to use observations performed in nearby astrophysical systems to constrain the present day value of $\gamma$, which requires it to be very close to unity \cite{Bertotti_2003}.

\begin{figure}
\centering
\includegraphics[height =5cm]{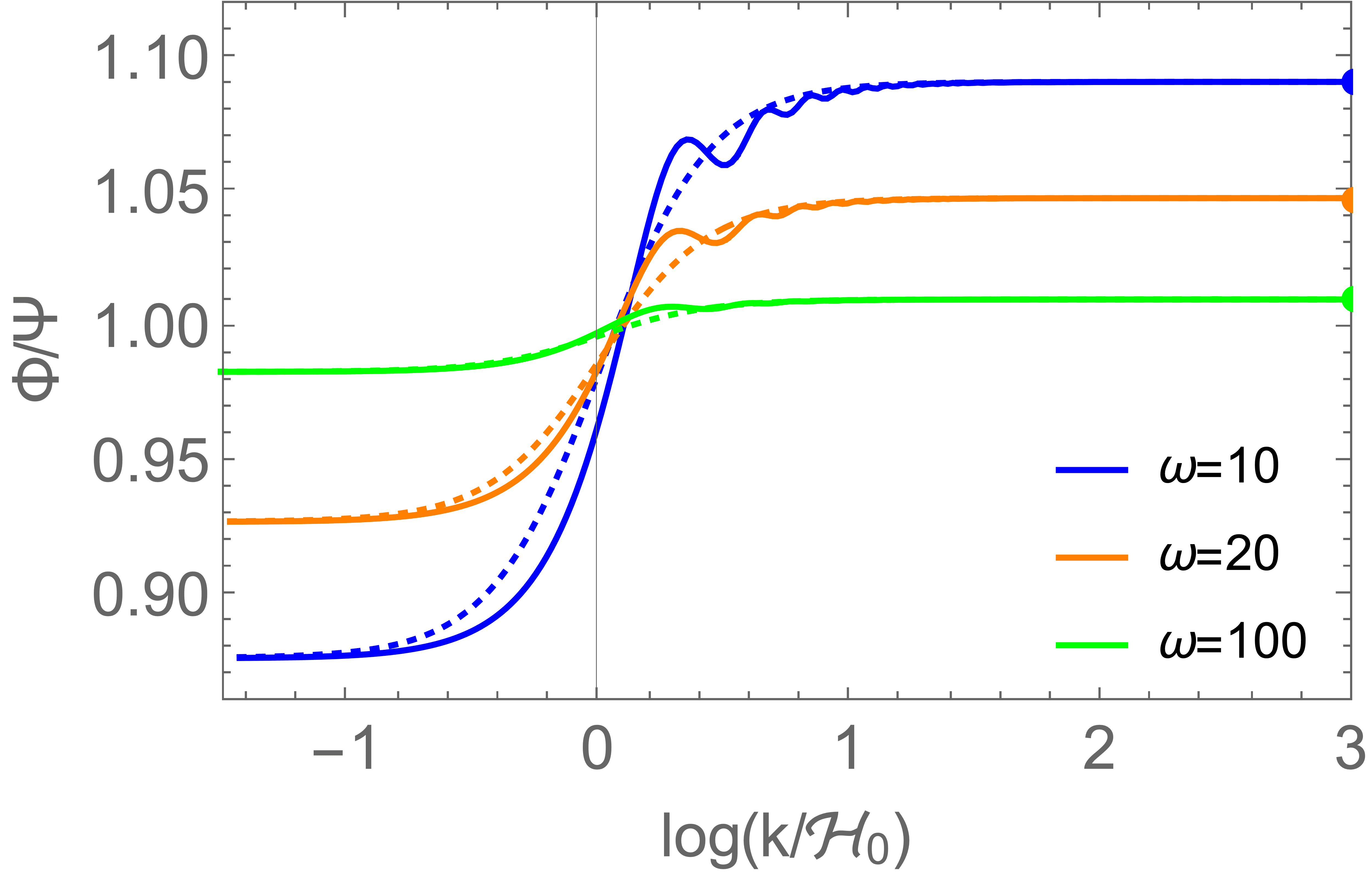}
\includegraphics[height =5cm]{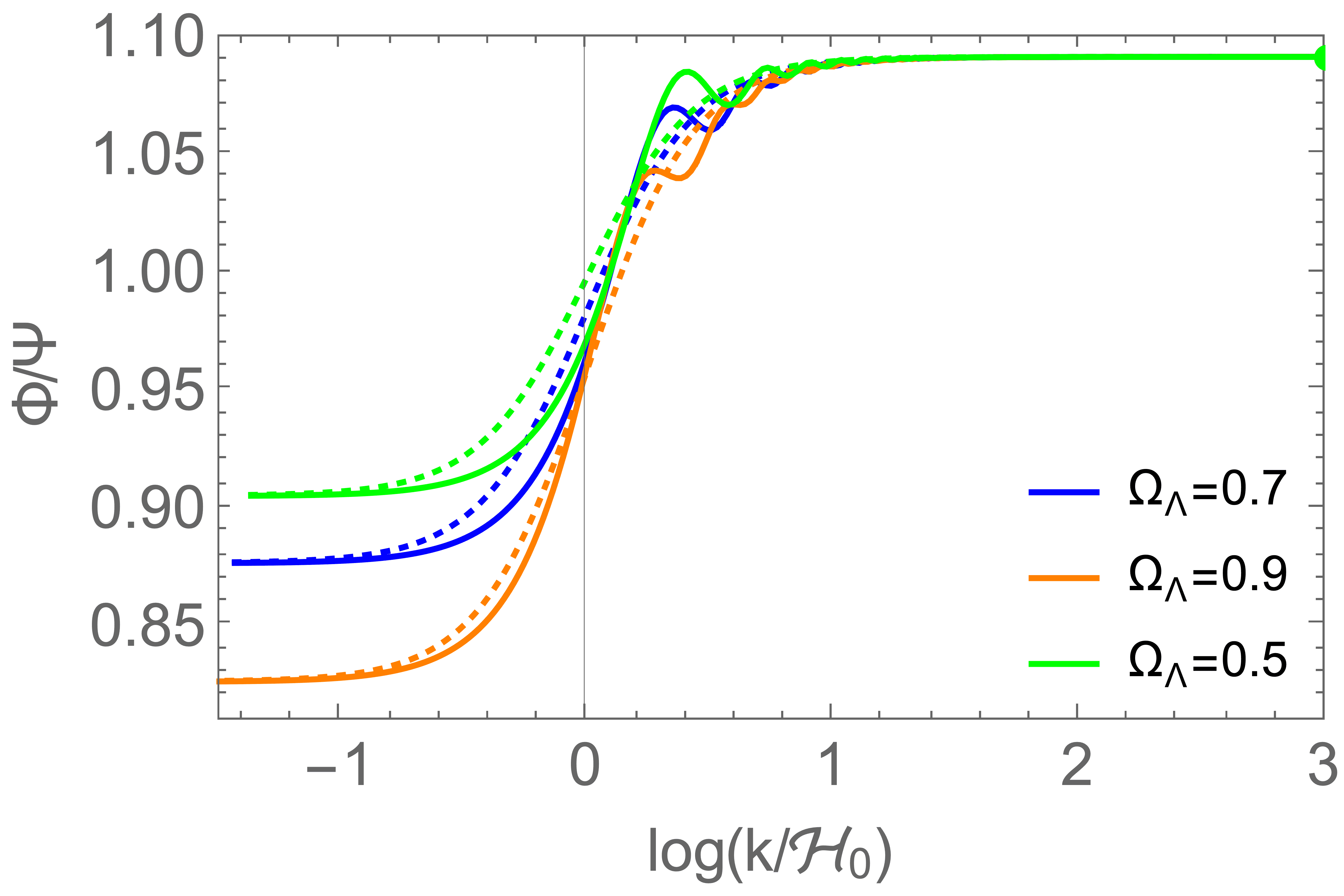}
\caption{Solid lines depict the function $\eta=\Phi/\Psi$ at $z=0$. Dotted lines show the simple interpolating function from Eq. (\ref{eqn_timinterp}). Dots denote the expected asymptotic value $\alpha/\gamma$. Left: differing values of $\omega$, with $\Omega_\Lambda=0.7$. Right: differing values of $\Omega_\Lambda$, with $\omega=10$. 
}
\label{fig_slip}
\end{figure}

The final quantity we examine over this transition region is the gravitational `slip', which is the ratio of the two gravitational potentials $\eta = {\Phi}/{\Psi}$, and is shown in Figure \ref{fig_slip}. The value this quantity should take on small scales can be readily calculated from the PPN approach, from which we obtain $\eta={\alpha}/{\gamma}$. This value can be seen to be obtained from the large-$k$ limit of our numerical evolution of the perturbations. On large scales, however, it is so far not clear how to obtain an analytic expression for $\eta$ (such an expression was missing from section \ref{sec:ppnc2}). At our present level of understanding, we therefore take the value of the slip on large scales to be an additional parameter of the framework, which can be discarded if and when an analytic expression in terms of other PPNC parameters is obtained. We can see from Figure \ref{fig_slip} that the small-scale value of $\eta$ varies with $\omega$, but not $\Omega_\Lambda$ (as expected). On the other hand, the large-scale limit varies with both parameters, and is less than unity for all the parameter values we have examined within our class of example theories. Interestingly, the slip is the only one of our coupling functions that has quantitatively different large and small-scale limits even in the matter-dominated era\footnote{It is possible that this is related to the difficulty of deriving a large-scale limit for the slip from the PPNC parameters given in Eq. (\ref{funcs}).}.

\subsubsection{Evolution of Perturbations and Couplings}

\begin{figure}
\centering
\includegraphics[height =5.0cm]{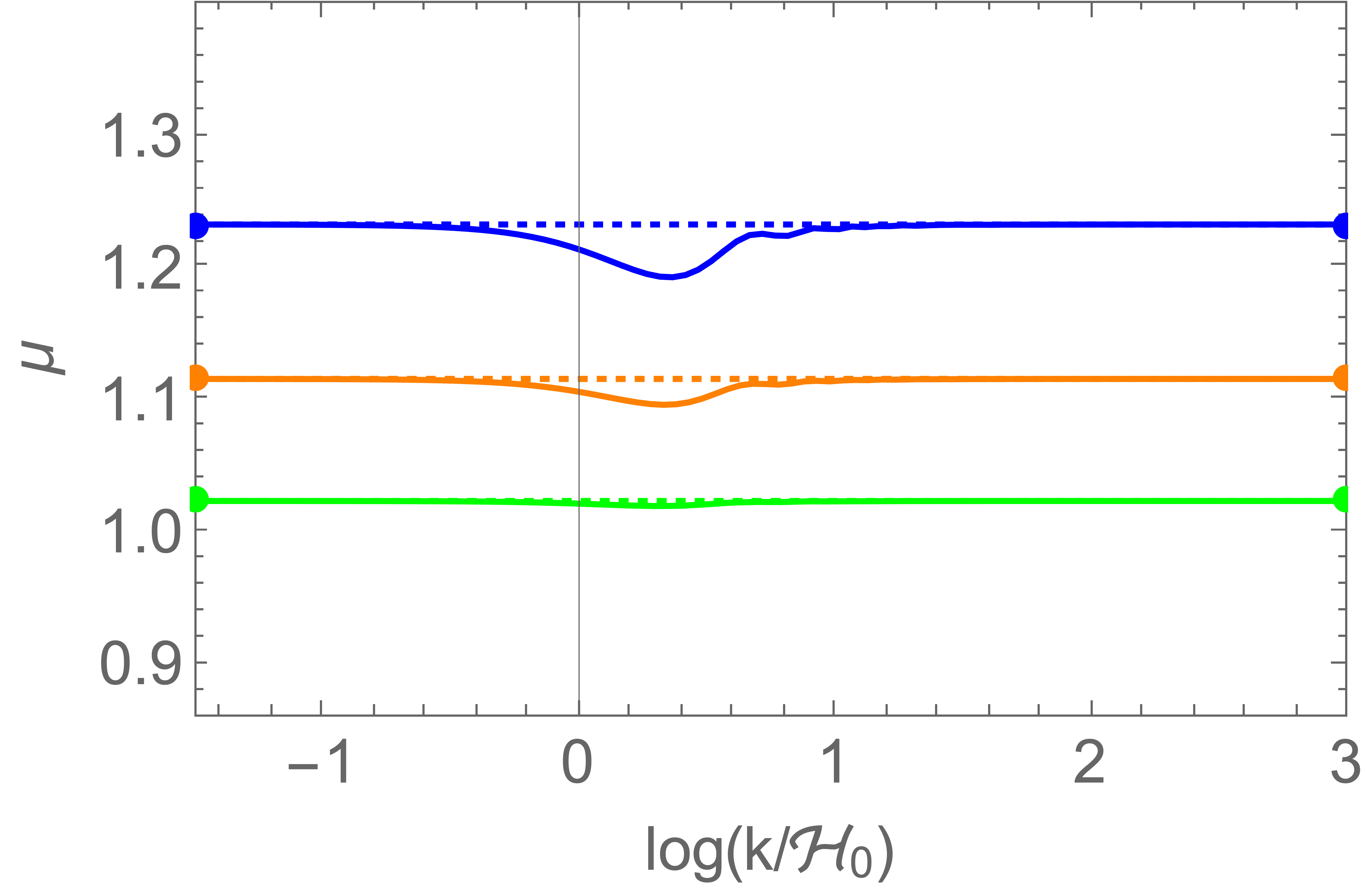}
\includegraphics[height =5.0cm]{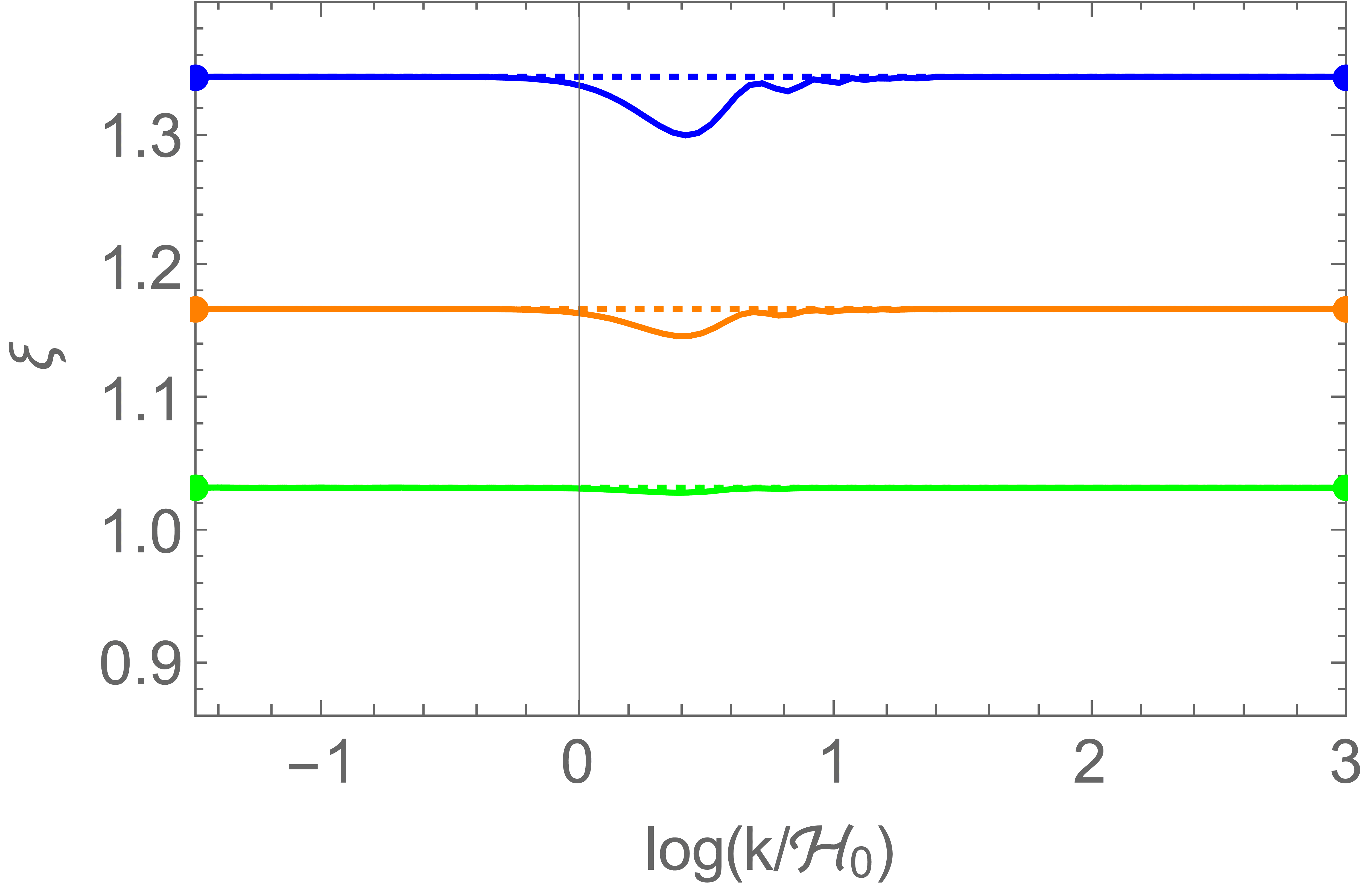}\\
\includegraphics[height =5.0cm]{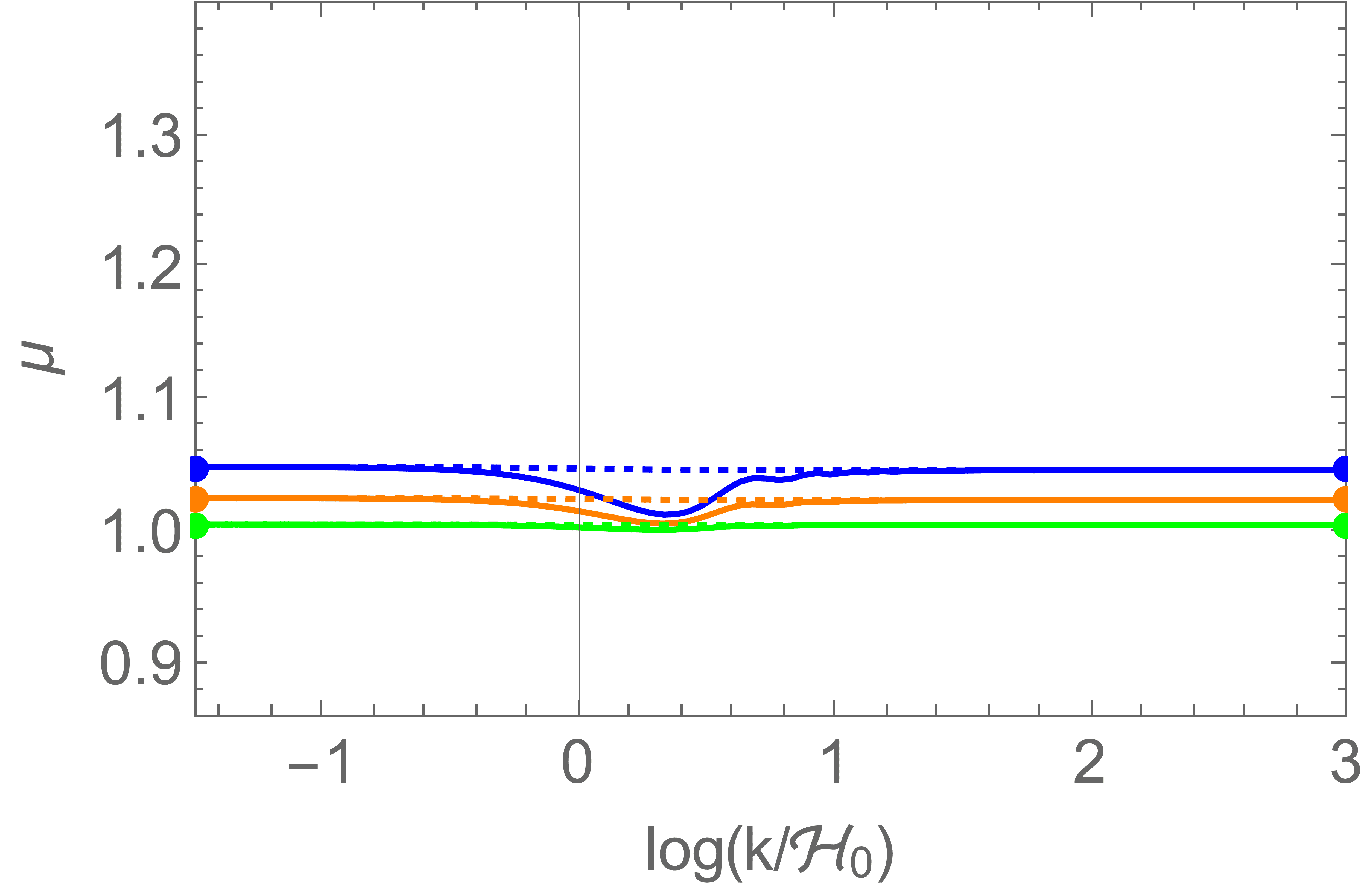}
\includegraphics[height =5.0cm]{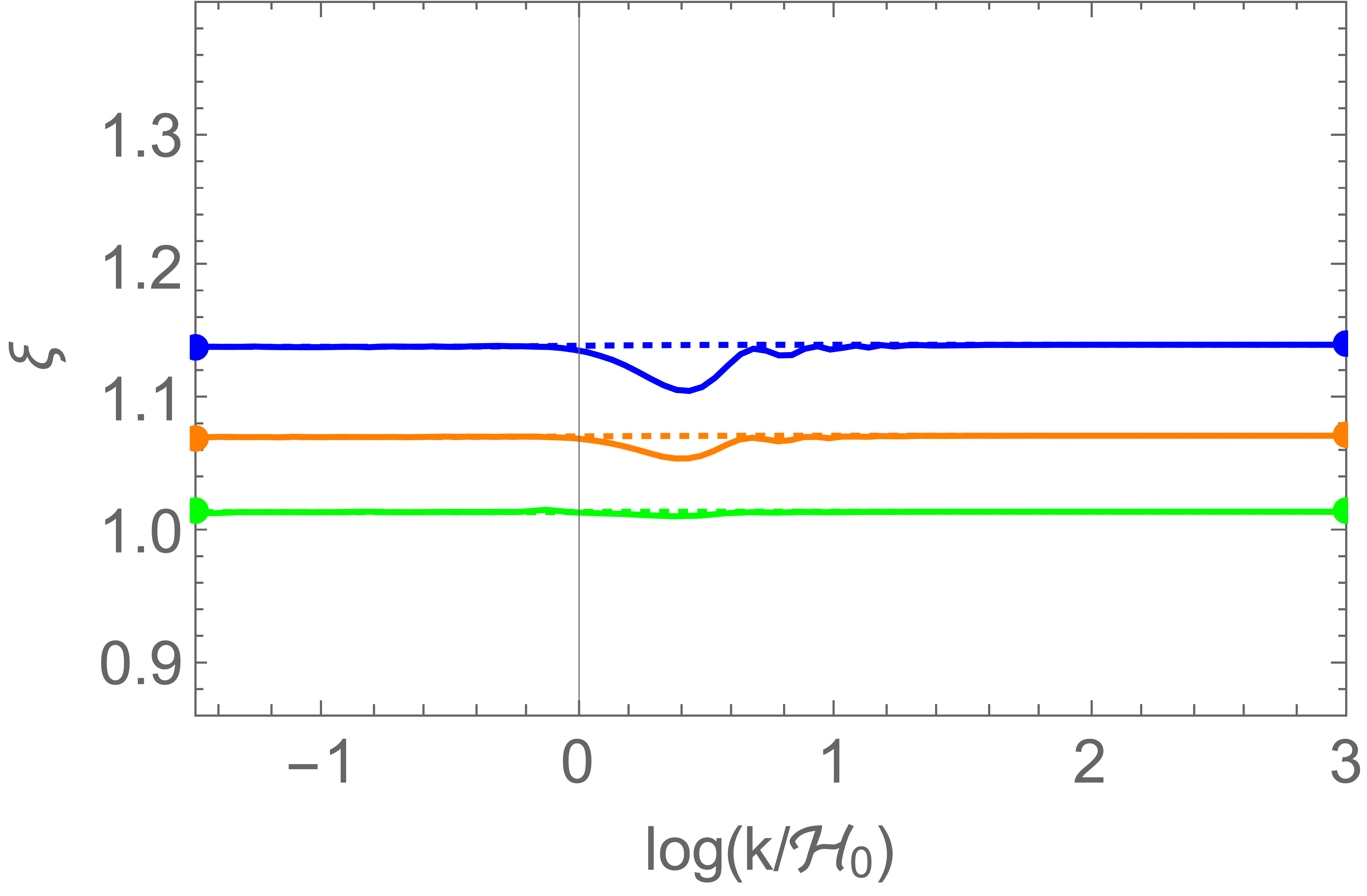}\\
\includegraphics[height =5.0cm]{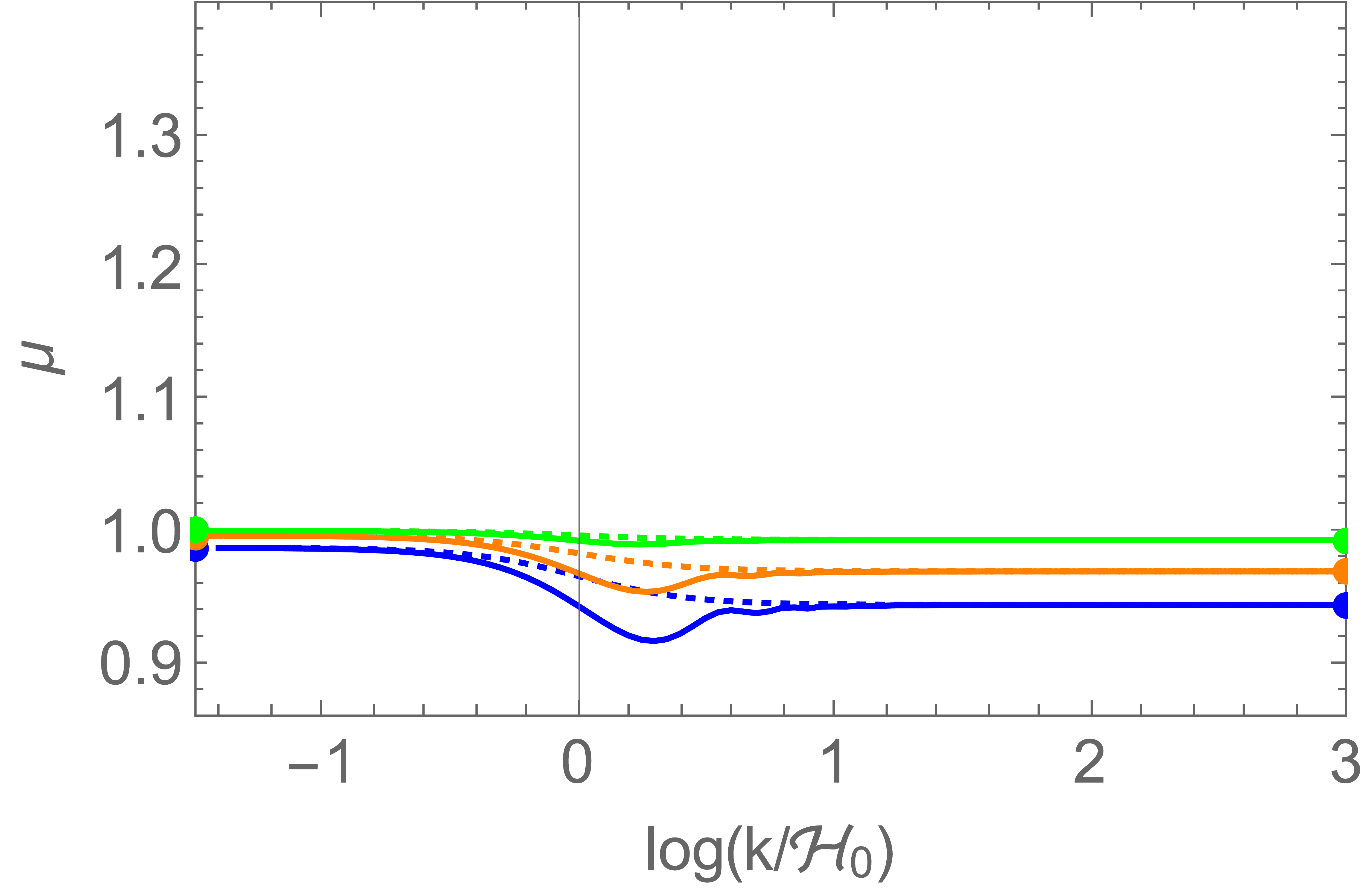}
\includegraphics[height =5.0cm]{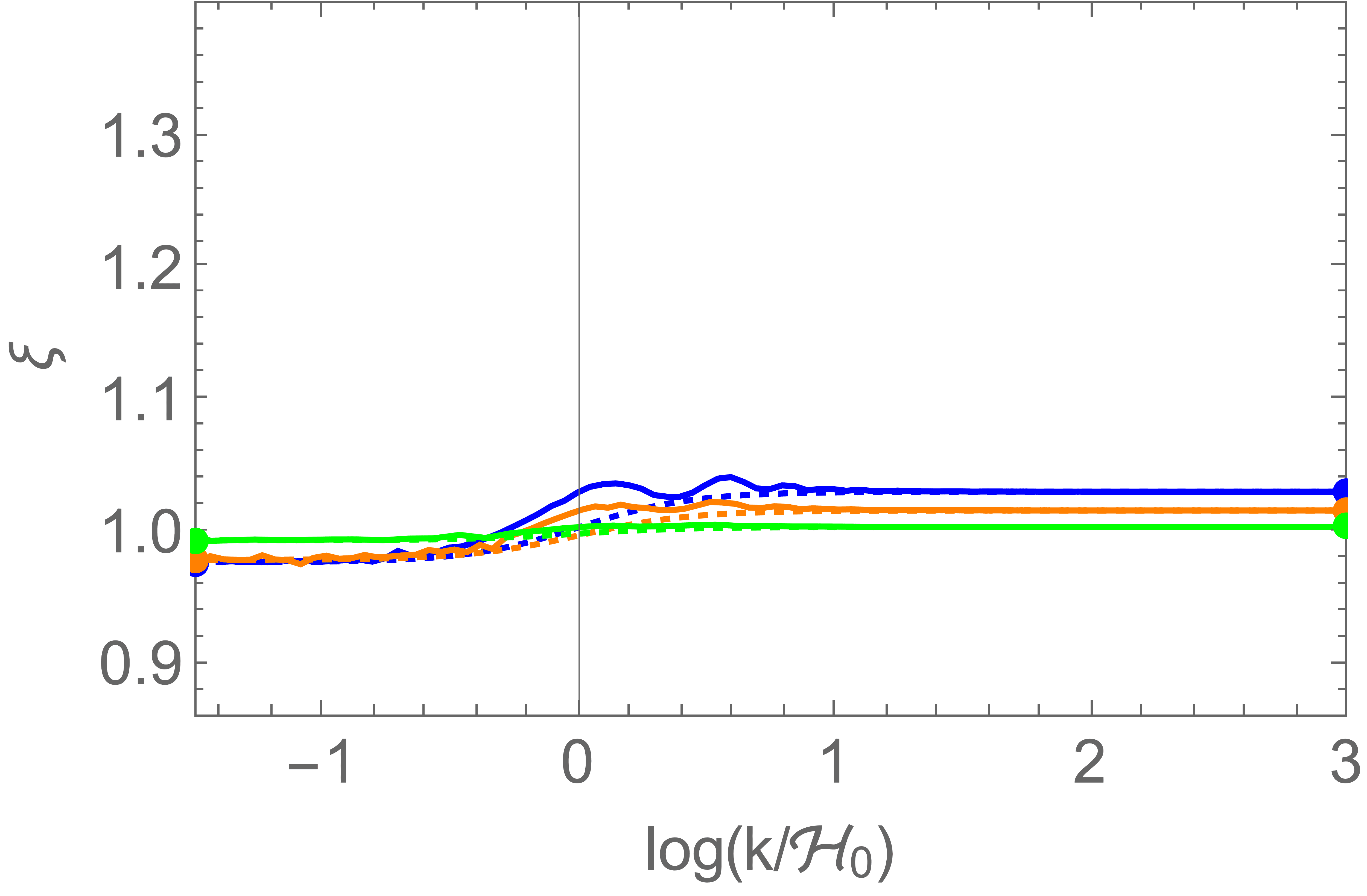}\\
\includegraphics[height =5.0cm]{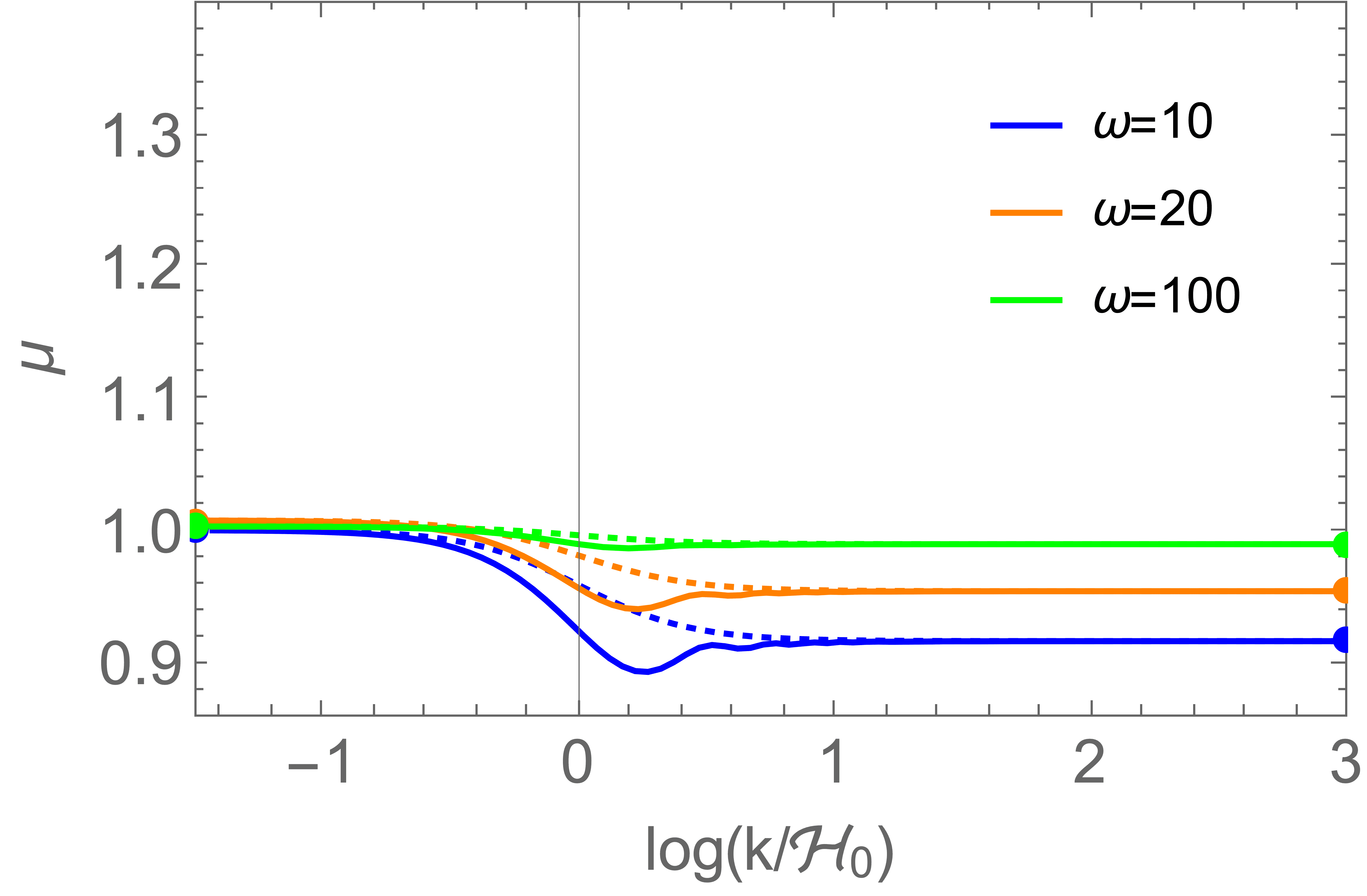}
\includegraphics[height =5.0cm]{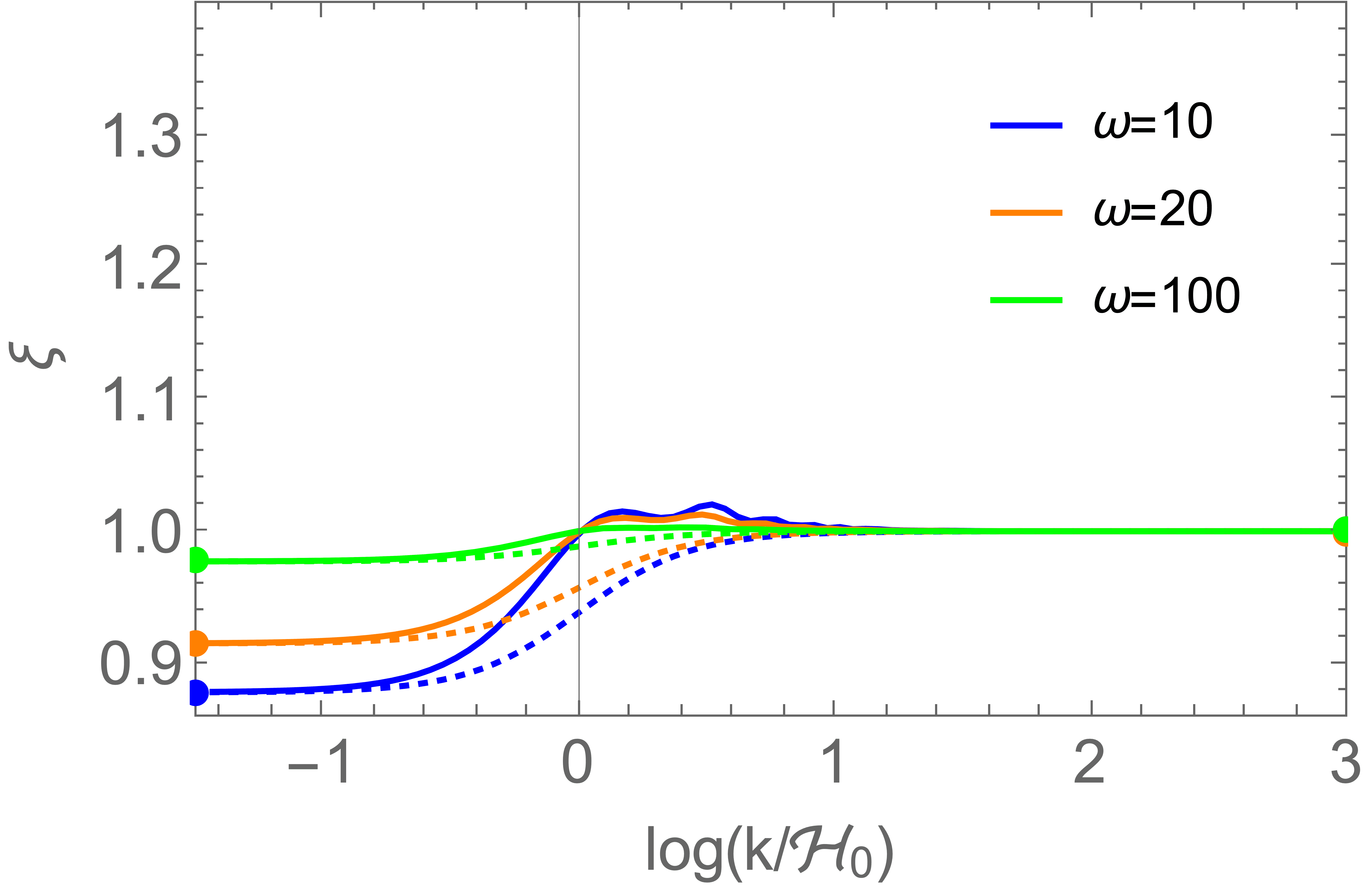}\\
\caption{Solid lines depict the function $\mu(k)$ (left) and $\xi(k)$ (right), at $z=0, \, 0.25, \, 2.33$ and $19$ (from bottom to top). The colours correspond to differing values of $\omega$, with $\Omega_\Lambda=0.7$ in each case. Dotted lines show the interpolating function from Eq. (\ref{eqn_timinterp}). 
}
\label{fig_manytimes2}
\end{figure}

\begin{figure}
\centering
\includegraphics[height =5.0cm]{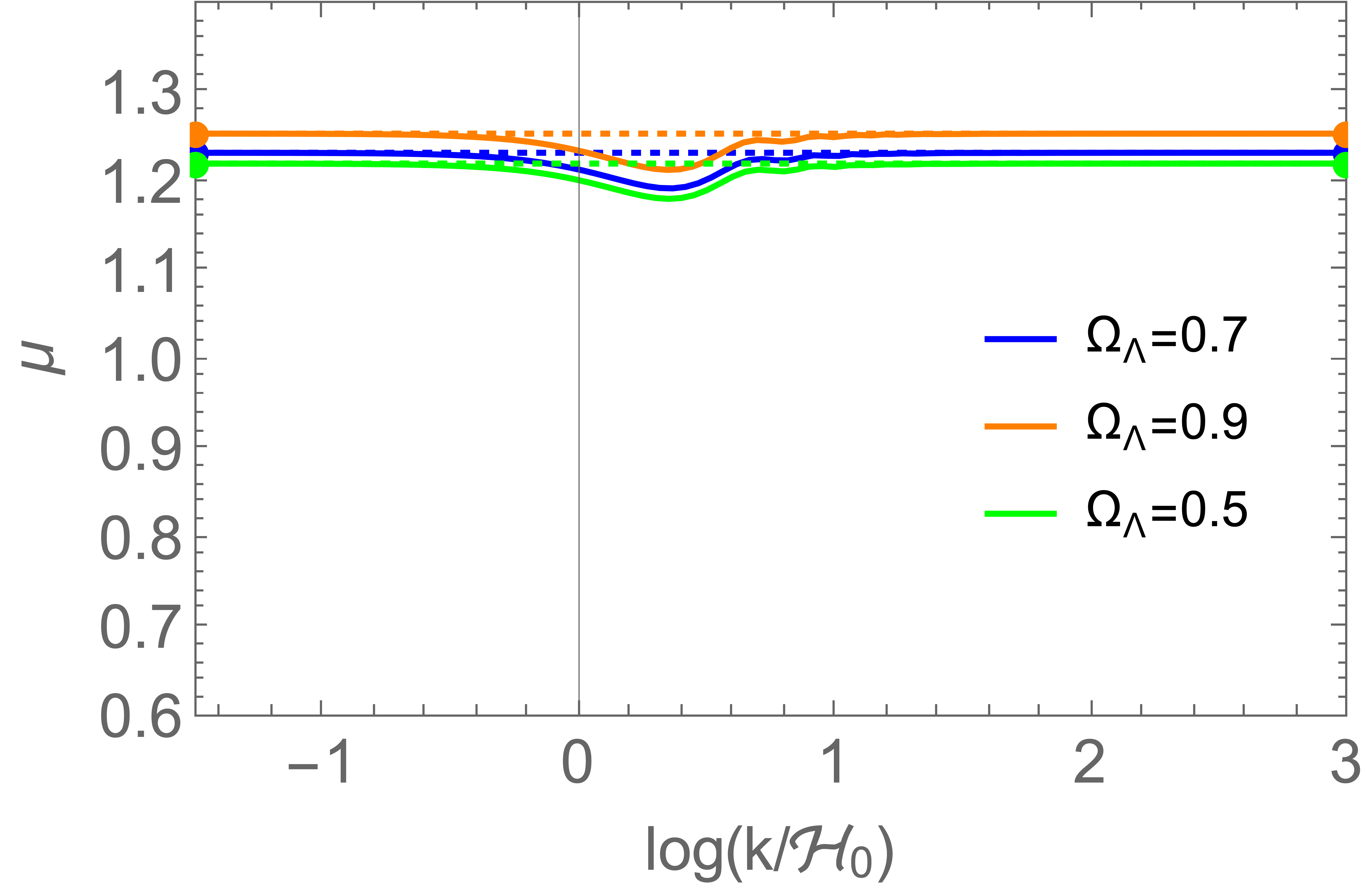}
\includegraphics[height =5.0cm]{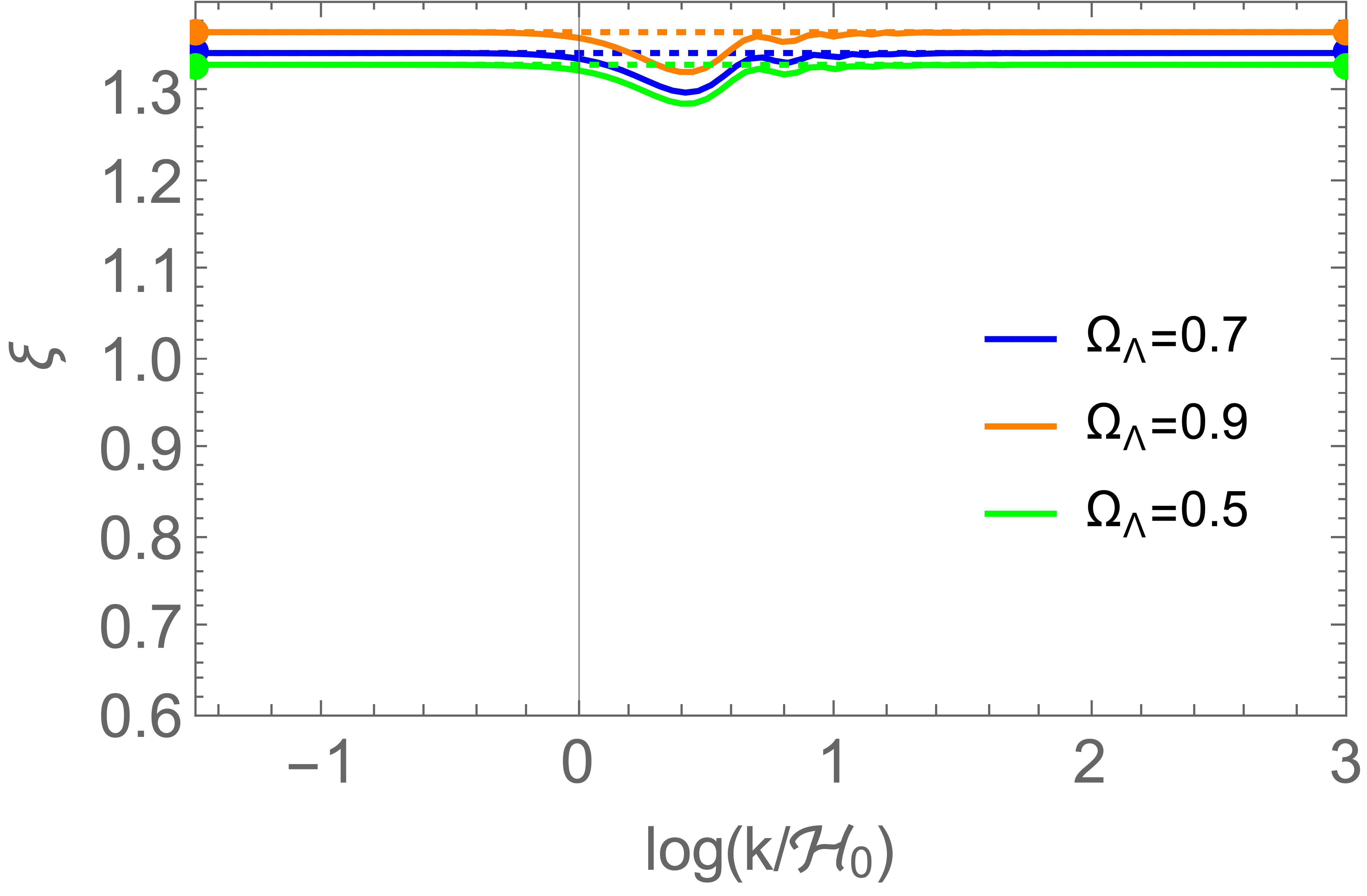}\\
\includegraphics[height =5.0cm]{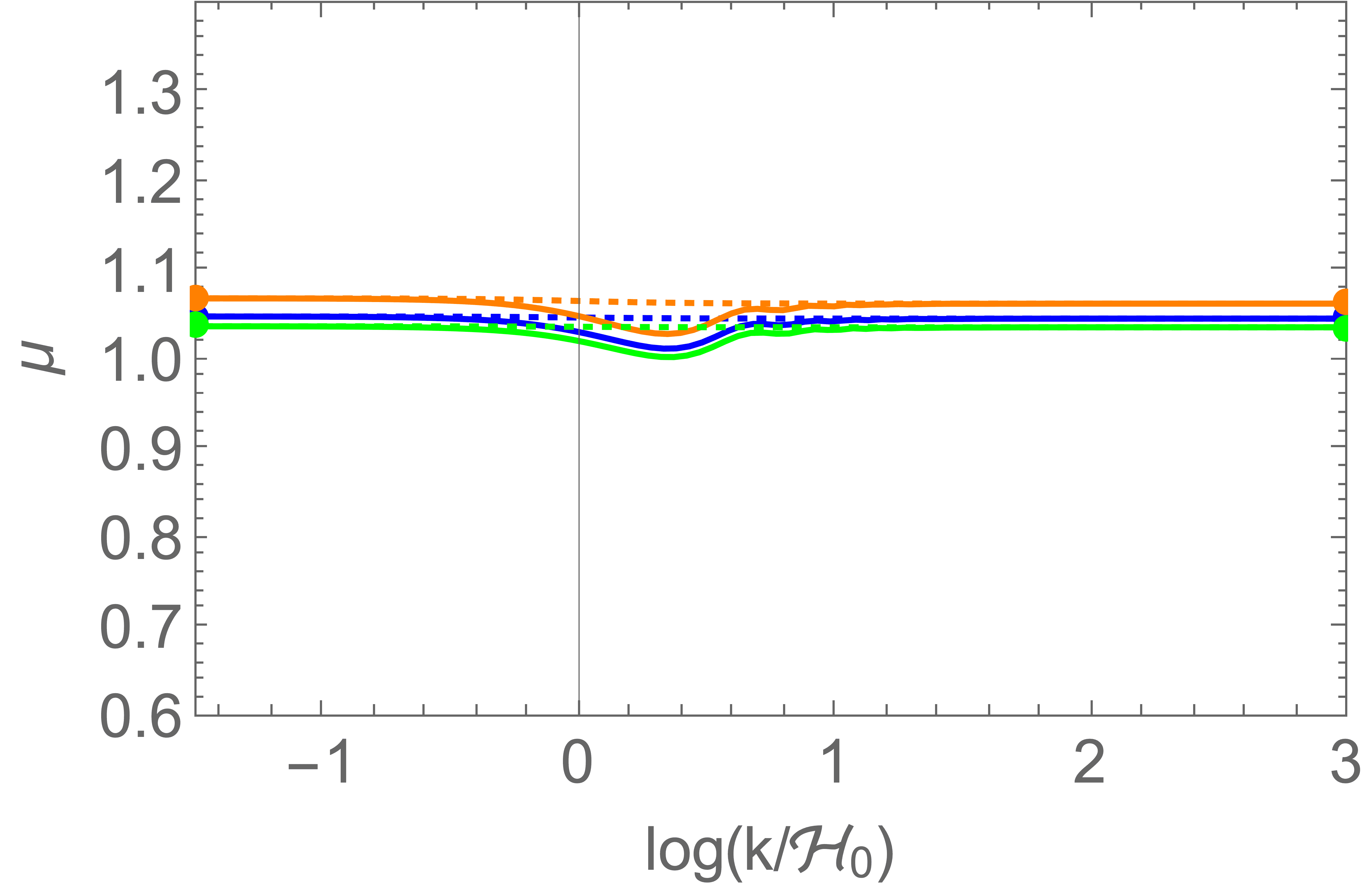}
\includegraphics[height =5.0cm]{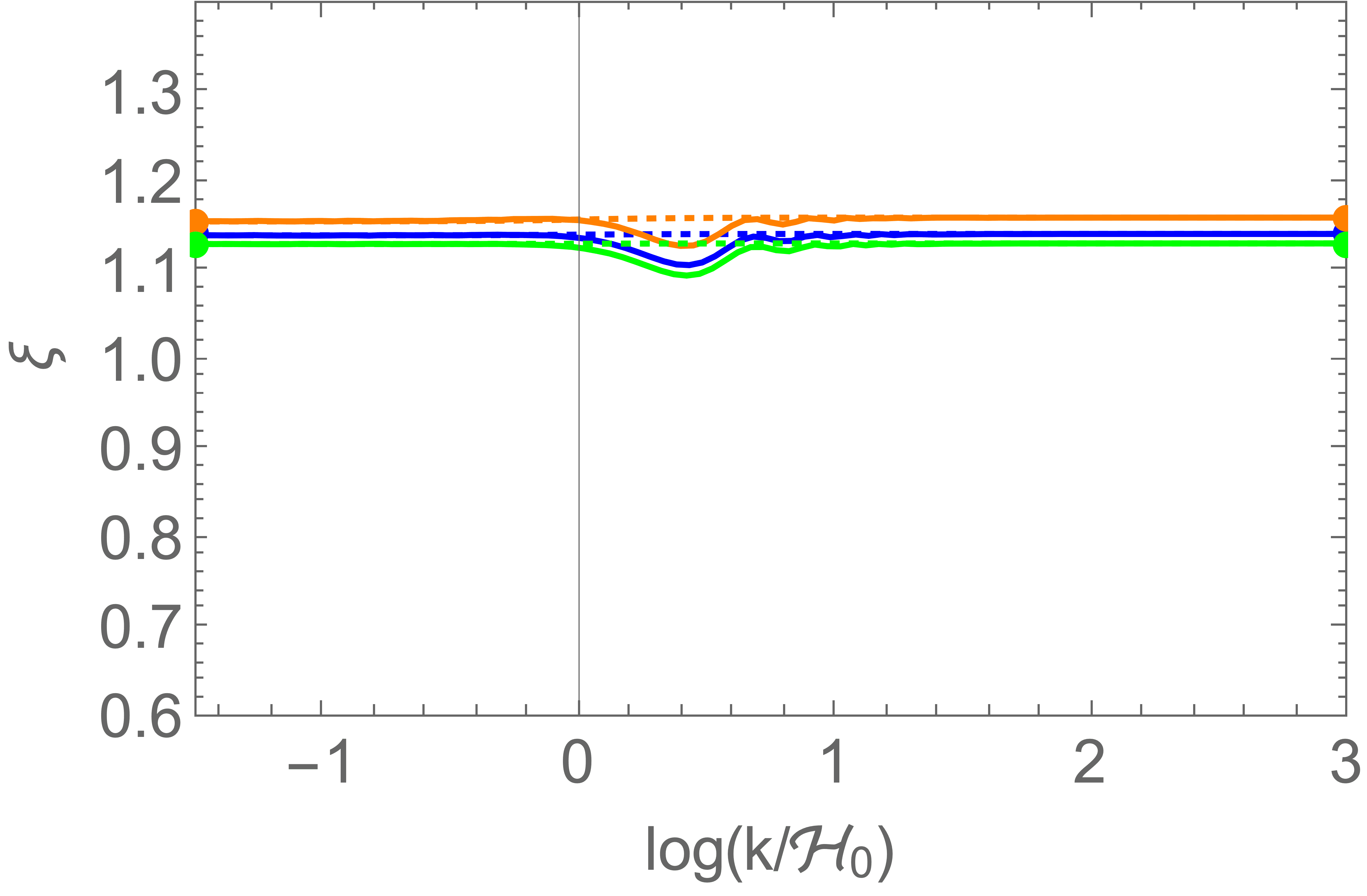}\\
\includegraphics[height =5.0cm]{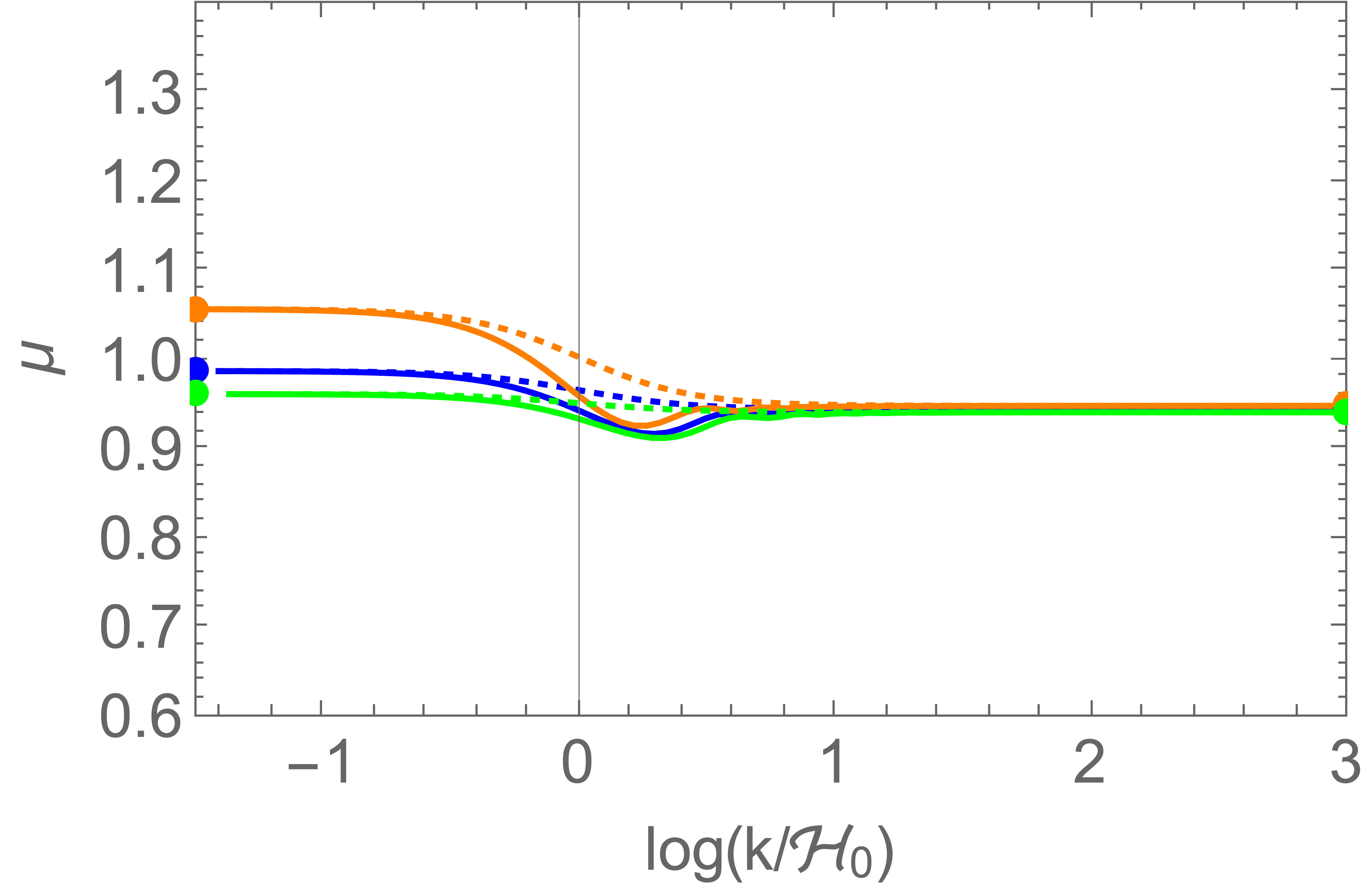}
\includegraphics[height =5.0cm]{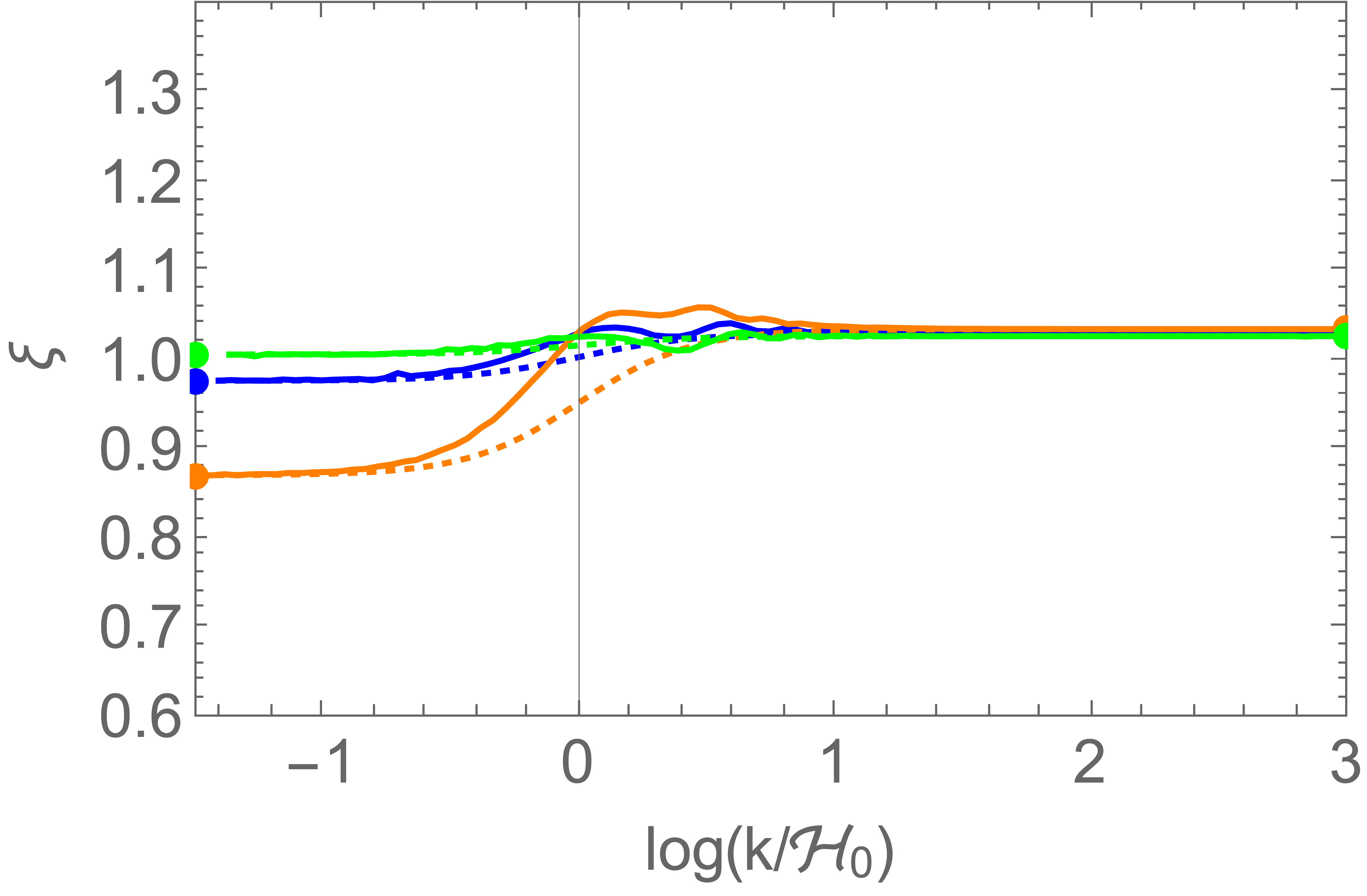}\\
\includegraphics[height =5.0cm]{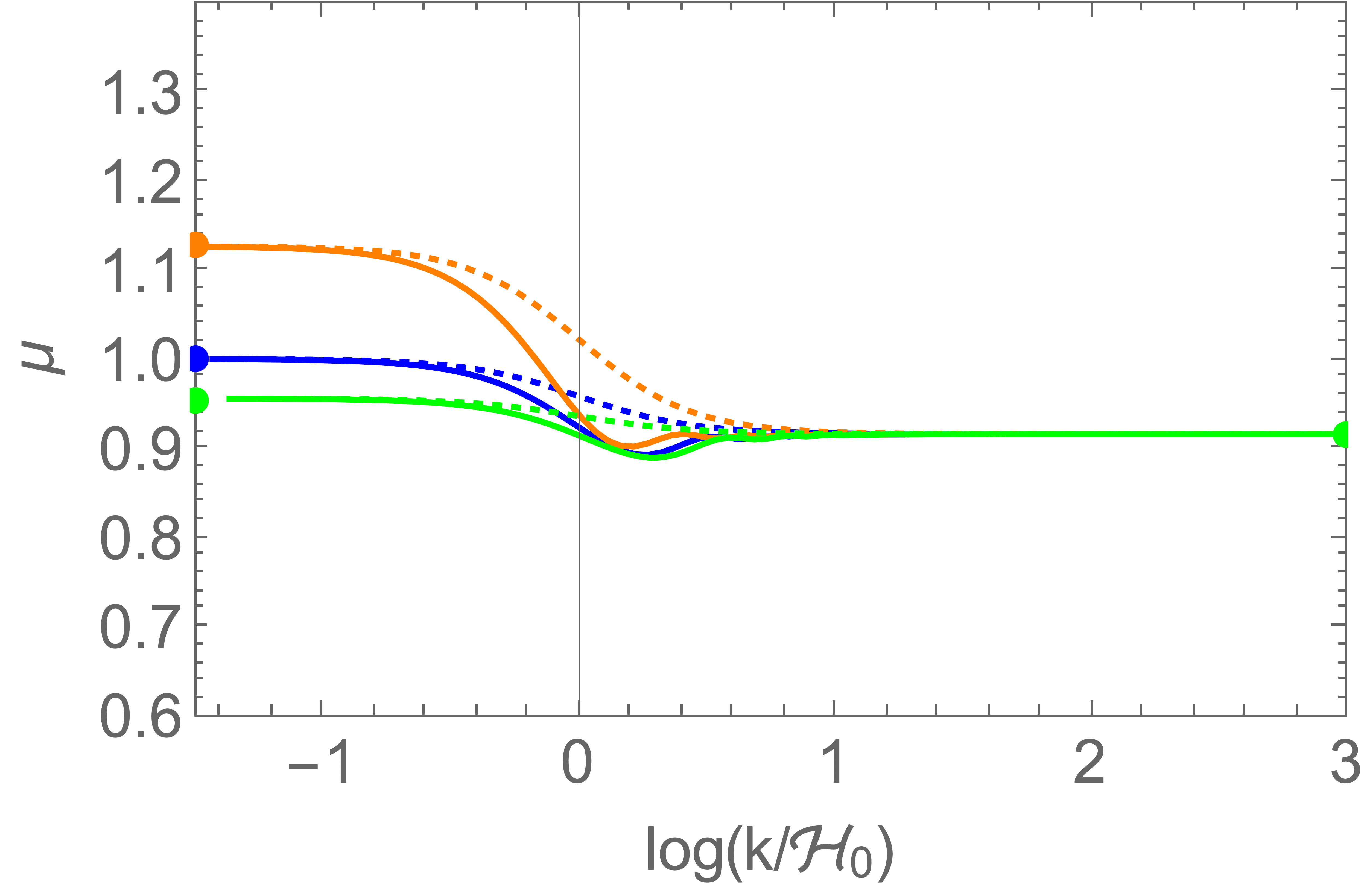}
\includegraphics[height =5.0cm]{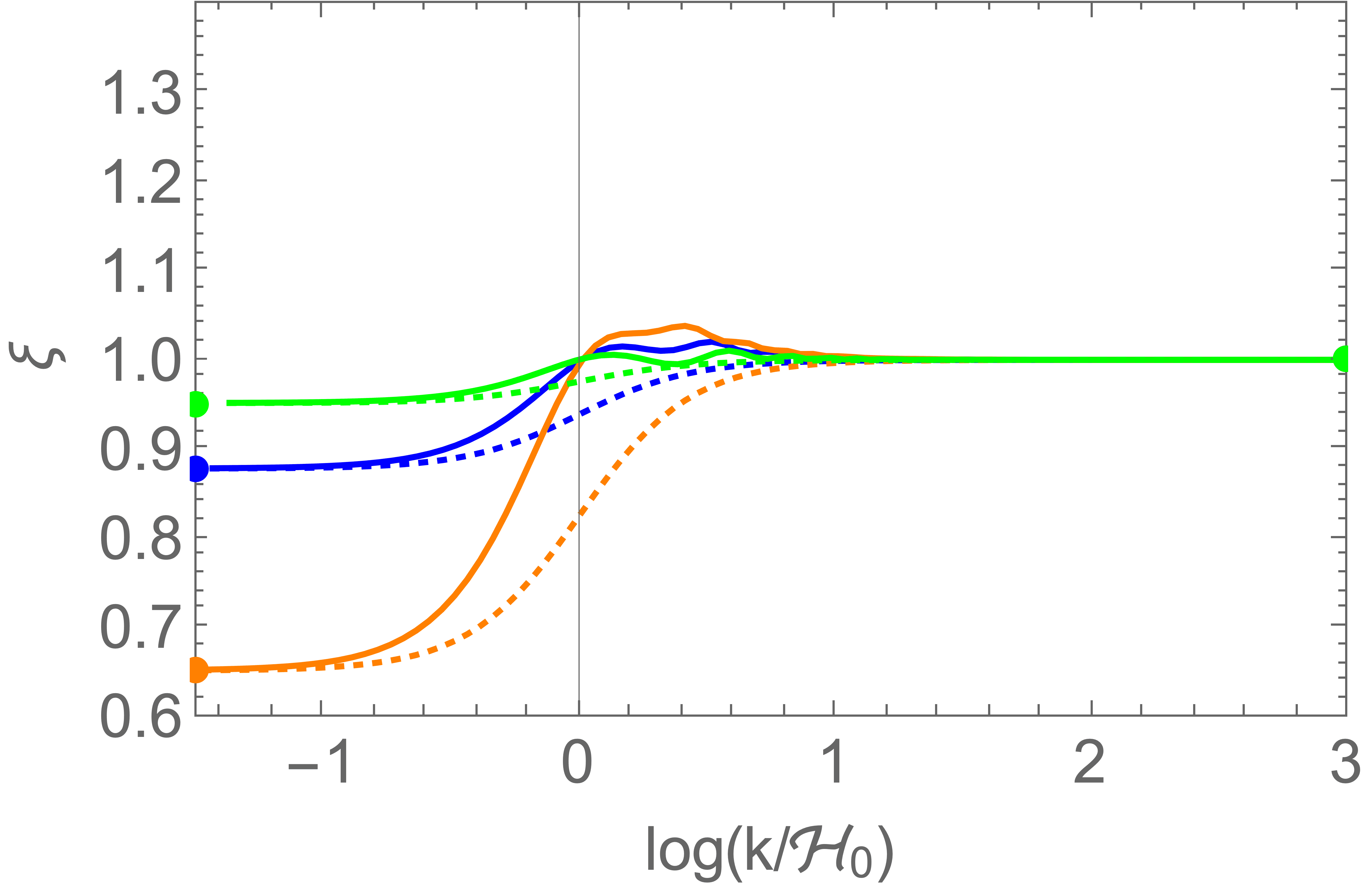}\\
\caption{Solid lines depict the function $\mu(k)$ (left) and $\xi(k)$ (right), at $z=0, \, 0.25, \, 2.33$ and $19$ (from bottom to top). The colours correspond to differing values of $\Omega_\Lambda$, with $\omega=10$ in each case. Dotted lines show the interpolating function from Eq. (\ref{eqn_timinterp}). 
}
\label{fig_manytimes2varylambda}
\end{figure}

Figures \ref{fig_manytimes2} and \ref{fig_manytimes2varylambda} show the time evolution of the couplings $\mu$ and $\xi$, for varyious values of $\omega$ and $\Omega_\Lambda$. From the bottom row of plots upwards, we show the spatial dependence of these functions at $z =0,\, 0.25, \, 2.33$ and $19$. As expected, it can be seen that varying $\Omega_\Lambda$ only makes a difference at late times, where it acts to increase (decrease) the large-scale value of $\mu$ ($\xi$) compared to the small scales. The effect of changing $\omega$ is a little more complicated. During the matter-dominated era the large-scale behaviour of $\mu$ and $\xi$ is dominated by parameters $\gamma$ and $\alpha$, so that there is no difference between the behaviour on large and small scales (note that interesting features around the horizon scale still exist). During $\Lambda$ domination, however, we see that decreasing $\omega$ results in increasing deviations from GR\footnote{The behaviour in the matter-dominated era persists to $z=0$ in a universe with $\Omega_\Lambda=0$.}.

As the universe transitions to a $\Lambda$-dominated phase, we can see that the value of $\mu$ becomes close (but not identical) to its GR value on large scales regardless of the value of $\omega$, while on small scales decreasing $\omega$ still results in substantial deviations from the GR value of unity. This behaviour is reversed for $\xi$: the large-scale behaviour can be seen to deviate substantially from the GR value, while on small scales it approaches one (exactly, by construction, as described above). We note for the reader that the bottom two panels of Figures \ref{fig_manytimes2} and \ref{fig_manytimes2varylambda} correspond precisely to the left-hand plots of Figures \ref{fig_hamcon} and \ref{fig_raynow}, reproduced here for ease of comparison.

In the interests of brevity, we have not plotted the time dependence of the slip $\eta$ or momentum constraint parameter $\mathcal{G}$ here. Both are relatively simple. The large-scale value of $\mathcal{G}$ is zero at all times, and the small-scale value varies monotonically from being close to zero in the matter-dominated era, to around $-0.05$ today (for $\omega=10$ and $\Omega_\Lambda=0.7$). The oscillations close to the horizon scale shown in Figure \ref{fig_momcon} persist at a similar amplitude over time, but are shifted vertically as the small-scale value of $\mathcal{G}$ evolves. The small-scale value of $\eta$ depends only on $\omega$, with no time dependence (because the time dependence in $\alpha$ and $\gamma$ is the same, therefore cancelling-out when the ratio is taken). The large-scale value  of $\eta$ decreases monotonically from about $0.95$ deep in the matter-dominated era to around $0.9$ today (for $\omega=10$ and $\Omega_\Lambda=0.7$). Again, the oscillations around the Hubble scale do not change significantly over time.

\subsection{Accuracy of Interpolating Functions}
\label{sec:acc}

\begin{figure}
\centering
\hspace{-0.5cm}
\includegraphics[width =0.5\textwidth]{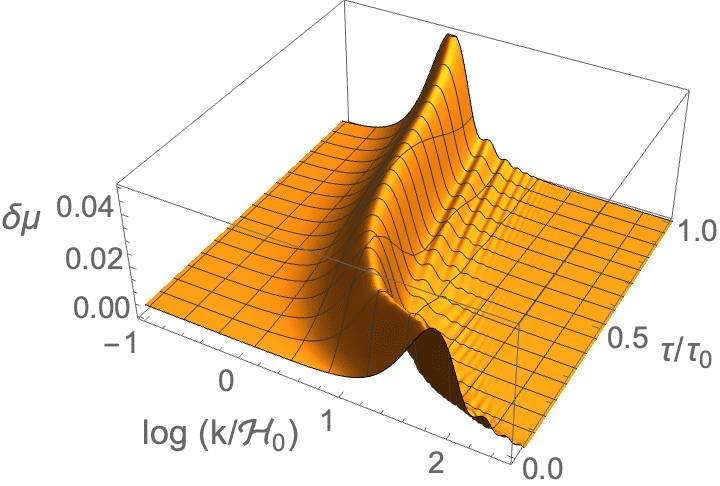} \,\,
\includegraphics[width =0.5\textwidth]{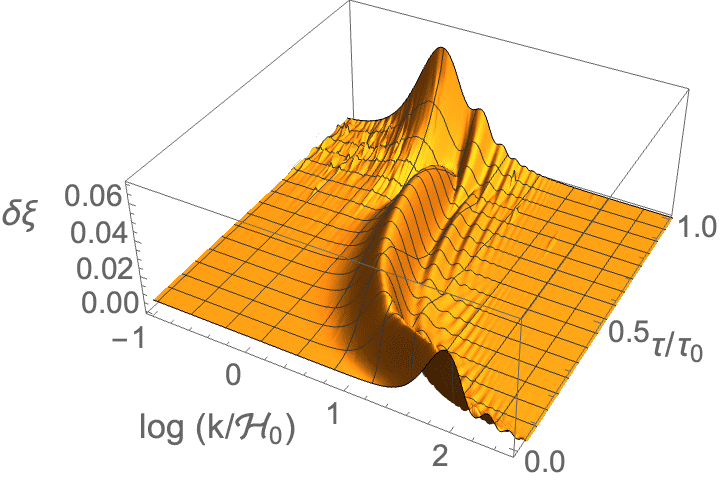}
\caption{The fractional error defined in Eq. (\ref{eqn_muaccuracy}), for the interpolation function from Eq. (\ref{eqn_timinterp0}). Plots show $\delta \mu$ (left) and $\delta \xi$ (right), as a function of time $\tau$ and scale $k$.}
\label{fig_kfail3d_ham}
\end{figure}

Let us now examine the performance of the interpolation function from Eq. (\ref{eqn_timinterp0}), in reproducing the coupling functions from our set of example theories. For this, we will use as a specific example theory given by $\omega=10$ and $\Omega_\Lambda=0.7$ (i.e. the common case in Figs. \ref{fig_manytimes2} and \ref{fig_manytimes2varylambda}). We specify the fractional difference between the simple interpolating function and the true value for this theory using the notation
\begin{equation}
\label{eqn_muaccuracy}
\delta f = \left\vert \frac{f_\text{true}-f_{\text{interp}}}{f_\text{true}} \right\vert \,\text{,}
\end{equation}
where $f$ here is being used to denote any one of the gravitational couplings (i.e. $\mu$, $\xi$ or $\mathcal{G}$). Figure \ref{fig_kfail3d_ham} shows the accuracy of the interpolation as a function of time and scale for $\mu$ and $\xi$, as quantified by this fractional error. The range of times being shown in this figure extends from deep in the matter-dominated era through to late times in the $\Lambda$-dominated era. In creating these plots we have interpolated between gridpoints, avoiding points where our solver produced large numerical errors.

It can be seen from Figure \ref{fig_kfail3d_ham} that the simple interpolation works well for both $\mu$ and $\xi$ on most spatial scales (with errors below $\sim 1\%$), but with the oscillations around the Hubble scale leading to larger errors (up to $\sim 5\%$). The detail of these plots shows some interesting structure, if we look at how the error evolves with time in each of the two cases. For the coupling $\mu$ the situation is relatively simple: there is primarily a single peak that does not change substantially with time, and which tracks the position of the Hubble scale. The coupling $\xi$ is more complicated: in this case there are several peaks that move position relative to the Hubble scale. In a cosmology with $\Lambda=0$ this more complicated behaviour does not occur, and one recovers a plot that is qualitatively similar to the case involving $\mu$.

\section{Evolution using Parameterised Equations}
\label{sec:paramevolution}

Part of the purpose of having a parameterised framework for gravity in cosmology is to be able to model gravitational physics {\it without} having to specify a theory or class of theories {\it a priori}. Practical implementation of such a framework therefore requires us to be able to evolve perturbations using only the parameterised equations, which is what we will consider in this section. We will perform this evolution using the zero-parameter interpolating function from Eq. (\ref{eqn_timinterp0}), which we will call the `simple parameterised' model, and evaluate the accuracy of the results by comparing to our class of example theories only after the evolution has been calculated. The quantities we will consider when using this approach will be the density contrast, as well as some example observables.

With regard to implementation, we will evolve the density perturbation and the velocity using the conservation equations for these quantities, which do not depend on the theory of gravity for their functional form (assuming the underlying theory admits mass and momentum conservation at leading order). 
The parameterised Hamiltonian constraint equation (\ref{pert1}), together with the slip $\eta$, will then be used to evolve the leading-order scalar metric potentials. For the functional form of the parameters $\{ \alpha , \, \gamma , \, \alpha_c \, , \gamma_c\}$ we will use the evolution determined by the background, as discussed in section \ref{sec:bg}. For the slip, the small-scale limit will be calculated from the parameters $\alpha$ and $\gamma$, while the large-scale limit will be extracted directly from the evolution of our example theory\footnote{We do this as we do not currently have a way of specifying this quantity in terms of the PPNC parameters, as discussed previously.}. We start the evolution of our perturbations at redshift $z=1\,100$, and end at the present day $\tau_0$, so that we cover both the matter and $\Lambda$-dominated eras of cosmic history. We use the same initial conditions for evolving the parameterised equations as in our example theories, for ease of direct comparison.

\subsection{Density Contrasts}

\begin{figure}
\centering
\vspace{-0.5cm}
\includegraphics[height =4.7cm]{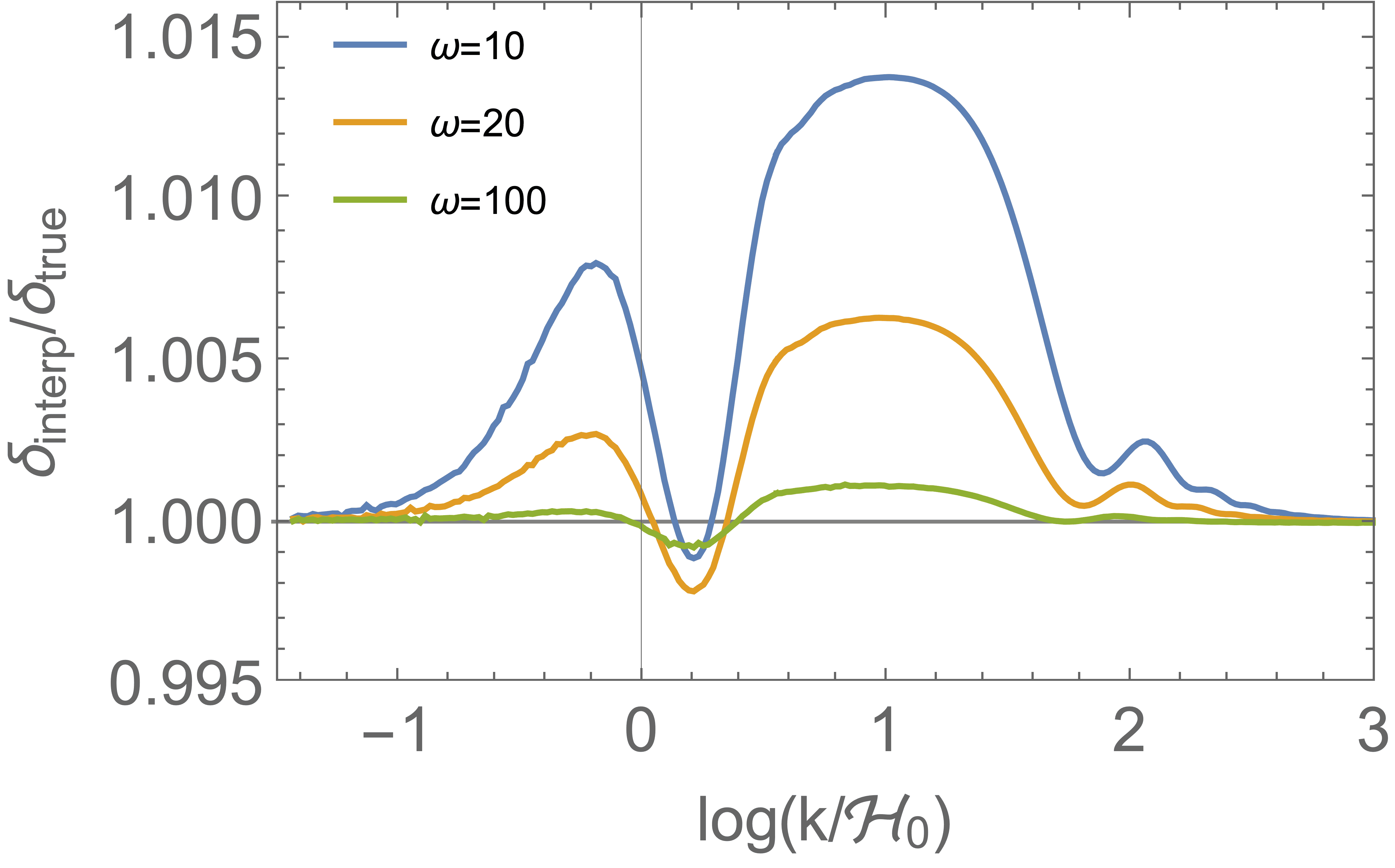}
\includegraphics[height =4.7cm]{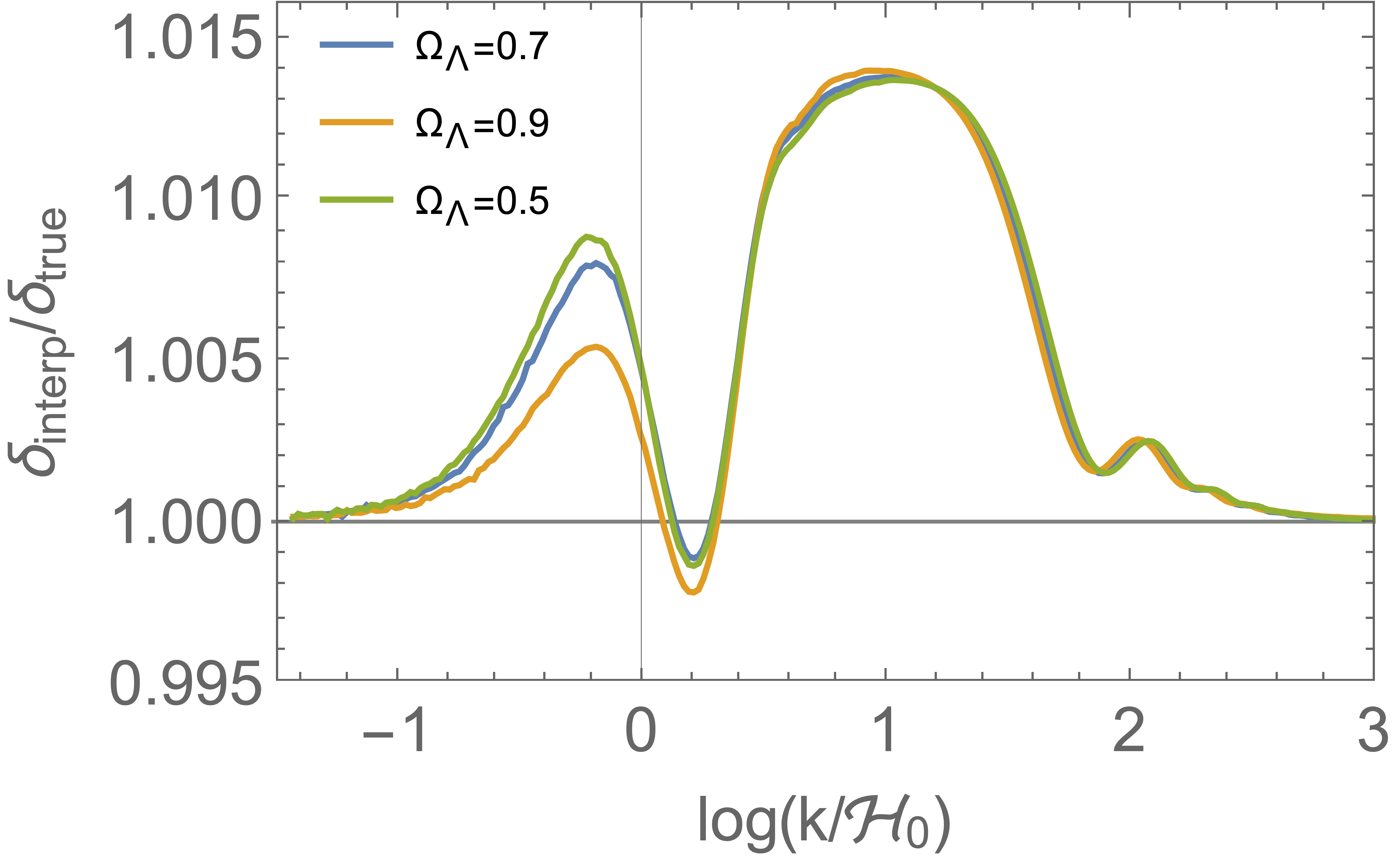}
\vspace{-0.25cm}
\caption{The ratio of the density contrasts at $z=0$, calculated using the parameterised equations and the example theories directly. Left: differing values of $\omega$ with $\Omega_\Lambda=0.7$. Right: differing values of $\Omega_\Lambda$ with $\omega=10$. 
}
\label{fig_deltaratio}
\vspace{0.25cm}

\centering
\includegraphics[height =4.9cm]{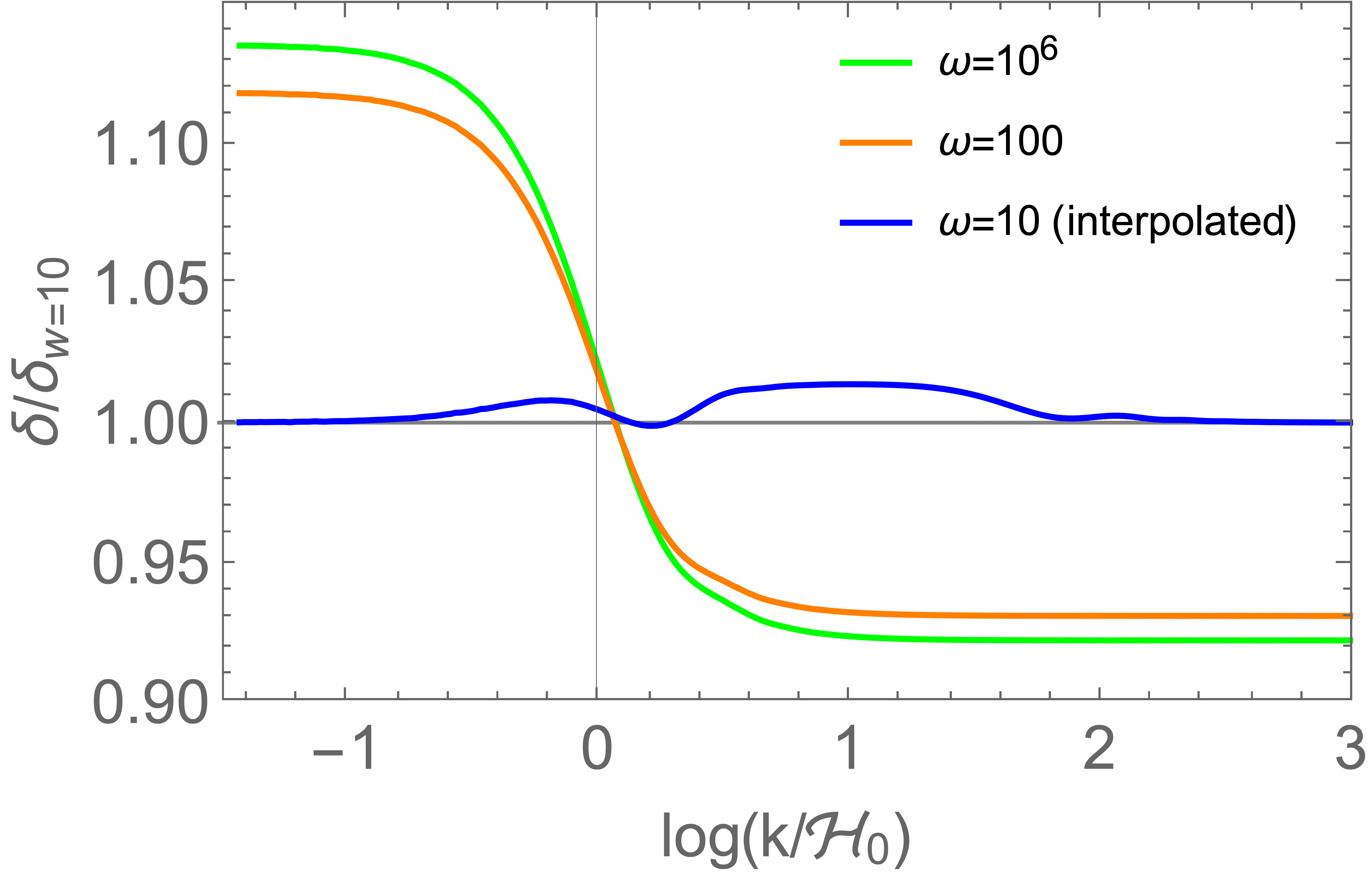}
\includegraphics[height =4.9cm]{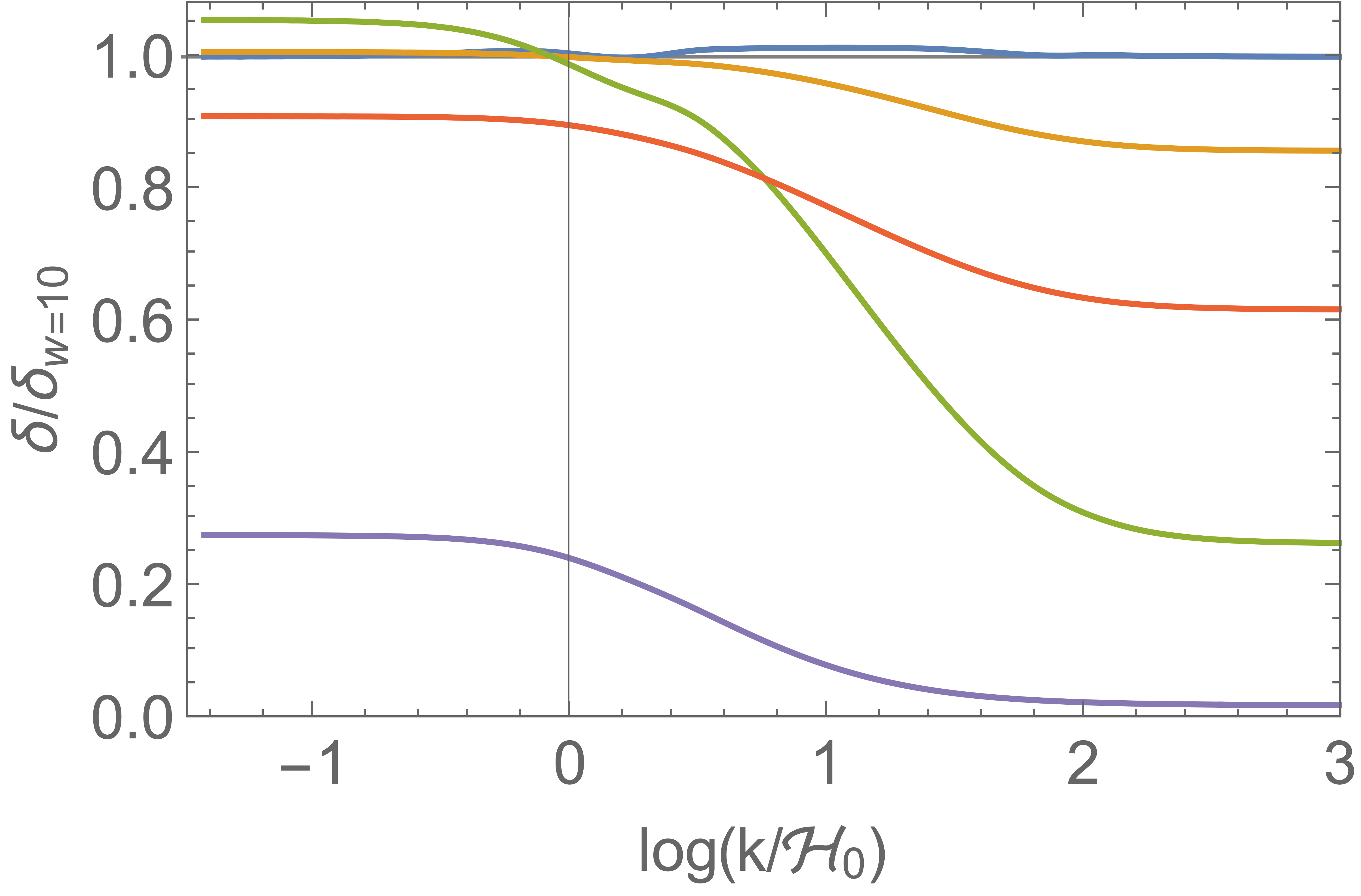}
\vspace{-0.25cm}
\caption{
Left: { Accuracy} of the density contrast at $z=0$ from evolving the simple parameterised equations (blue), compared to the difference induced by changing $\omega$ to $100$ (orange) and $10^6$ (green). Right: { Accuracy} of the density contrast today from evolving the simple parameterised equations (blue), compared to the error introduced by assuming a $\Lambda$CDM background (red for $\omega=100$; purple for $\omega=10$) or $\Lambda$CDM perturbation equations (orange for $\omega=10$; green for $\omega=100$). In each plot we take $\Omega_{\Lambda}=0.7$ and $\omega=10$, unless otherwise stated.
}
\label{fig_deltaratio_theories}
\end{figure}

Calculating the density contrast at $\tau_0$ using the parameterised equations, and the method described above, leads to results that are in good agreement with the results we obtain directly from our example theories across a range of spatial scales. There are, however, errors introduced due to the inadequacy of the simple interpolating function (\ref{eqn_timinterp0}) to model all of the detail around the horizon scale (as was found in section \ref{sec:acc}). The accuracy of using the simple interpolation is displayed in the plots in Figures \ref{fig_deltaratio}, where we show the ratio of the density contrast computed by evolving the parameterised equations with the simple interpolation to the value obtained from our example theories directly. The inaccuracy of the interpolation can be seen to create an error of order $\sim 1\%$ around the horizon scale. Decreasing $\omega$ (and therefore increasing the deviation from GR) increases this error, whereas it appears to be mostly insensitive to the value of $\Omega_\Lambda$. These errors are smaller than those in $\mu$ and $\xi$, as displayed in Figure \ref{fig_kfail3d_ham}.

As well as considering the fractional error introduced from evolving the parameterised equations with the simple interpolation, it is also of interest to consider the relative size of this error compared to the difference in density contrast that occurs due to changing the underlying theory of gravity (which is, after all, the signal that we would ultimately hope to detect). These two things are displayed at $\tau_0$, and over a range of spatial scales, in the left-hand plot of Figure \ref{fig_deltaratio_theories}. Here we have chosen to plot the fractional error in $\delta \rho$ for an example theory with $\omega=10$ and $\Omega_\Lambda=0.7$ (the blue curve in each of the panels in Figure \ref{fig_deltaratio}), with the change in density contrast that one gets from increasing the value of $\omega$ to $100$ and $10^6$ (keeping $\Omega_\Lambda=0.7$ for all curves). It can be seen that the effects from changing the value of $\omega$ is much larger than the error introduced from the imperfect interpolating function, suggesting that {  the bias on constraints on the parameters will be small compared to any detected deviation from General Relativity. Of course this bias should still be understood or modelled, which would presumably be at the expense of adding further parameters to the framework.}

In the right-hand plot of Figure \ref{fig_deltaratio_theories}, we compare the inaccuracies introduced from evolving the simple parameterised equations with the inaccuracies induced by applying the modified gravity parameterisation to either the background only or the perturbations only (instead of both). The inaccuracy introduced by only modifying the background or the perturbations is much larger than that introduced by evolving the simple parameterised equations directly (this is true even for theories that are much closer to GR). Interestingly, the inaccuracy introduced by not including the correct background expansion appears to be even larger than that from not evolving the perturbations with the correct equations. This result is significant for the PPNC approach because the PPNC approach provides a way to consistently parameterise both the background {\it and} the perturbations in terms of the {\it same} underlying parameters. This is not true of any other attempts at creating a theory-independent framework for testing gravity in cosmology of which we are aware. {  These results suggest that when constraining modified gravity in a model independent or phenomenological way, the background should generally be allowed to deviate from $\Lambda$CDM (for example using a dark energy equation of state $w(a)$) in order to prevent biasing the constraints.}

\subsection{Lensing of Light}

We now wish to consider the effect of evolving the perturbations using the simple parameterised equations on some cosmological observables. This is not intended to be a detailed observability study, nor is it carried out in detail for specific experimental setups. Instead we wish to make a preliminary examination of how well the parameterised equations fare (in comparison to the ``true'' values), when calculating observables. For the lensing of light we simply have to integrate $(\Phi+\Psi)^2$ over the comoving distance $\chi$, from $\chi=0$ out to some source $\chi_*$. Doing this, we see that the parameterised equations reproduce the numerical values that one would get using the specific equations of any of our example theories with errors that are entirely negligible. This is expected, as lensing is primarily operating well within the horizon, and within this regime the parameterised equations give very good results. We also note that we reproduce the same results if we include the lensing function $\left(1-{\chi}/{\chi_*} \right)/\chi^{4}$ in the integral over $\chi$. This simple approach to weak-lensing calculations will of course break down on the largest angular scales; we leave a detailed examination of the effect of using the parameterised equations in this regime for future work.
 
\subsection{Integrated Sachs-Wolfe}
\label{sec_isw}

\begin{figure}
\centering
\includegraphics[height =4.9cm]{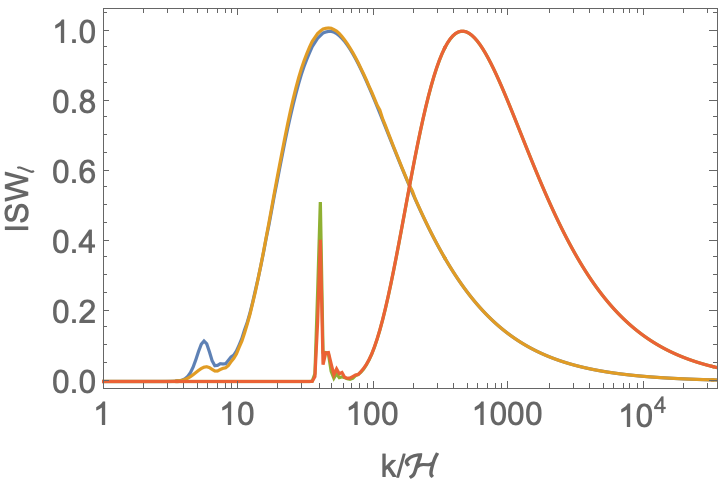}
\includegraphics[height =4.9cm]{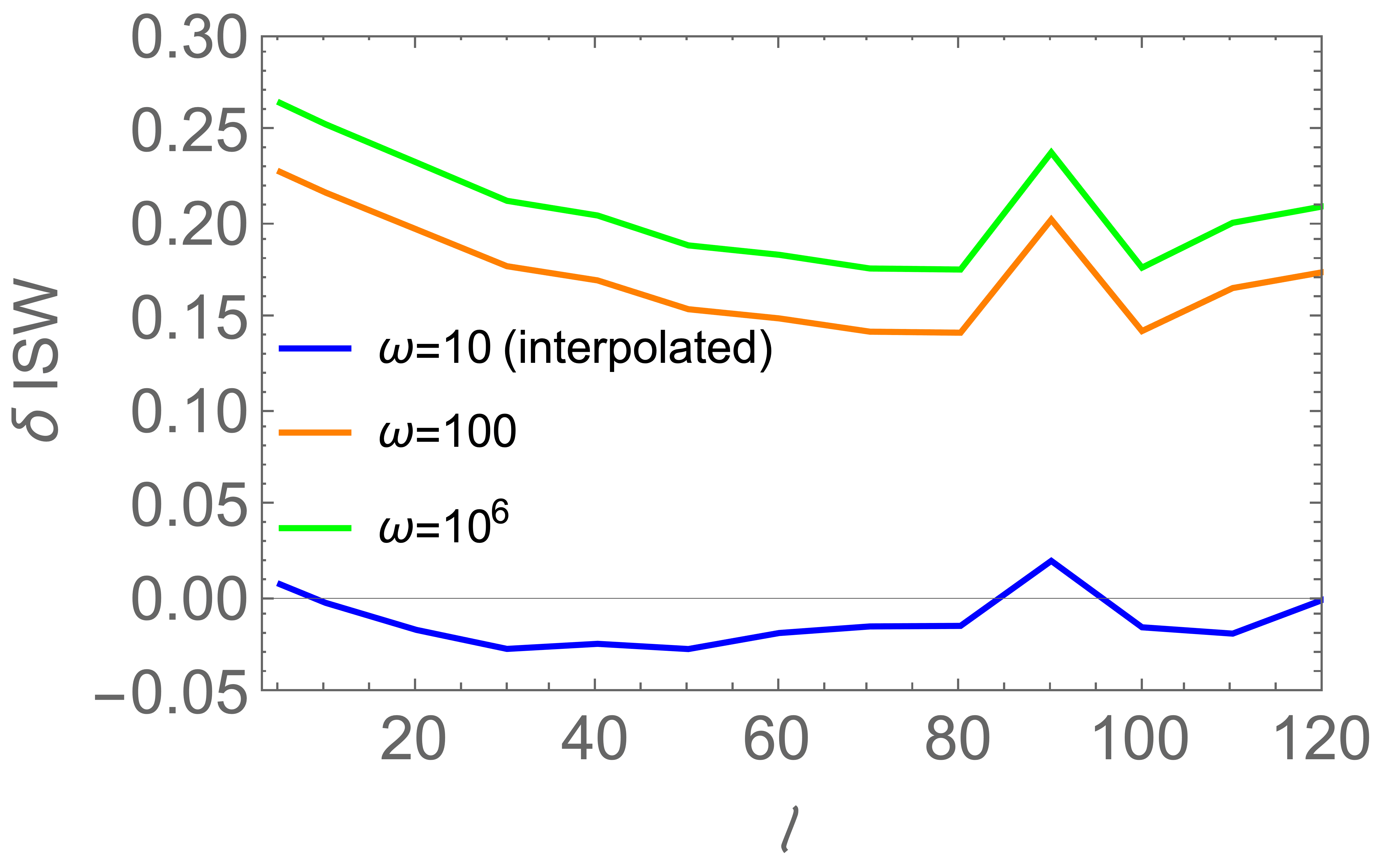}
\caption{Left: The contribution of different $k$ modes to the ISW effect, for $\ell=10$ and $100$. For $\ell=10$, the value obtained from the theory with $\omega=10$ is blue, and the value from evolving with the simple parameterised equations is orange. For $\ell=100$, the theory with $\omega=10$ is green, and the parameterised version is red. Each pair of curves has been normalised to the highest point, to allow them to be shown on the same axes. In all cases $\Omega_{\Lambda}=0.7$. Right: Errors induced in the ISW observable at $\tau_0$ from using the simple parameterised equations with $\omega=10$, compared to different theories. Each curve is the ISW observable normalized by its value in the fiducial $\omega=10$ theory. Blue is obtained by evolving the parameterised equations, orange is $\omega=100$, and green is $\omega=10^6$.
}
\label{fig_combinedisw}
\end{figure}

As the Integrated Sachs-Wolfe (ISW) effect primarily operates over larger scales, we expect that this observable will be more sensitive to the inaccuracy of our simple interpolation between small and large scales. To investigate this, for a given $\ell$, we look at the integral giving the contribution of each $k$-mode to that $\ell$ mode:
\begin{equation}
{\rm ISW}_l = \int^{\chi_\text{max}}_0 d\chi \left(j_l \left(\dot{\Psi}+\dot{\Phi} \right) \right)^2\text{,}
\end{equation}
where $j_l$ is the Bessel function. We calculated this quantity with both the simple parameterised equations as well as with the correct evolution for specific example theories at $\ell=10$ and $\ell=100$. Our results are shown in the left-hand panel of Figure \ref{fig_combinedisw}. As expected, there is a noticeable difference between the curves for some $k$ values. 

We can also integrate our results over $k$, with the usual factor of $k^{-1}$ to account for a scale-invariant primordial spectrum, in order to examine the accuracy of the ISW $C_\ell$s. This is shown by the blue curve in the right-hand plot of Figure \ref{fig_combinedisw}. The error peaks around a couple of percent, similar to (but a little worse than) the inaccuracy we found in the density contrast. We note that these errors are never greater than $20\%$ of cosmic variance over this range of $\ell$. Finally, we again consider the consequence of changing $\omega$, and show the results in Figure \ref{fig_combinedisw}. Again, each curve is plotted as a ratio of the results obtained in a theory with $\omega=10$, and again we see that the errors introduced by integrating the simple parameterised equations directly is substantially smaller than the change in the observable due to changing the value of $\omega$. This indicates that the errors introduced when integrating the simple parameterised equations directly is not likely to strongly bias any parameter constraints based on this observable.

\newpage

\section{Discussion}
\label{sec:conc}

We have investigated the scale-dependence of gravitational couplings in the PPNC approach to cosmological modelling, in which the coupling functions that appear in the cosmological perturbation equations can be directly linked to the parameters of the PPN framework (the workhorse of astrophysical tests of gravity). As well as just describing perturbations, however, the PPNC approach also specifies the evolution of the cosmological background in terms of the same set of underlying parameters. This link between gravitational physics on small and large cosmological scales has allowed the limits of the relevant gravitational couplings to be deduced, but (before the present work) without any information about how to interpolate between these different regimes.

Our study has included an investigation of the application of the PPNC framework within the class of canonical scalar-tensor theories of gravity given in Eq. (\ref{action}). Within these example theories we have numerically determined the evolution of the PPNC parameters $\{ \alpha ,\, \gamma ,\, \alpha_c ,\, \gamma_c \}$, and used their values to determine the large and small-scale limits of the coupling parameters that govern the evolution of perturbations. These results agree perfectly with existing theoretical predictions in every case we considered, and therefore verify the ability of this framework to be applied to the cosmological background, the small-scale quasi-static linear and quasi-non-linear regime, as well as the large-scale super-horizon regime.

We have investigated the way in which the gravitational coupling parameters from the perturbation equations transition from small to large cosmological scales. This is the first time that this transition region has been investigated in the PPNC approach. We have found that, within our class of example theories, there is an underlying smooth transition between the small and large scales, which exists together with oscillations in these couplings around the Hubble scale (due to the existence of the scalar field). The smooth transition is reasonably well modelled by a simple $\tanh$ function, as proposed in Ref. \cite{Sanghai_2019}, and the oscillations would appear to be linked to the Gravity Acoustic Oscillations (GAOs) from Ref. \cite{gao}.

{  We have considered the inaccuracy introduced into some basic observational predictions by using the simple $\tanh$ function. For flat-sky lensing calculations we find this inaccuracy to be small, as expected since the modes that contribute most are well below the horizon scale thus outside of the transition regime. This is not the case for the ISW effect, where we have shown (see Section \ref{sec_isw}) that using the simple $\tanh$ function can introduce errors at the level of about a percent. However, this inaccuracy is significantly below the level of uncertainty in the ISW effect caused by cosmic variance, as well as the effect of modifying the theory of gravity itself. Nonetheless this inaccuracy represents a potential source of uncertainty in applying this formalism to interpret real observations, and may well also be larger for other observables we have not considered here. It should therefore be recognised and understood, or itself modelled (which would presumably be at the expense of adding further parameters to the framework).} We find that these conclusions remain qualitatively unchanged when $\omega$ and $\Lambda$ are made smooth functions of the scalar field $\phi$.

Our study lays the groundwork for sensibly implementing the PPNC approach in Boltzmann codes and N-body simulations, and thus constraining gravity using all cosmological scales, and in a way that can be combined with gravitational constraints from other regimes and scales (due to its explicit links with PPN). Of course, much work remains to be done in working through the practicalities of such an application, including how to model unspecified functions of time, and whether methods such as binning or reconstruction from observations would be possible or appropriate. One might also consider applying this approach to constraining vector gravitational fields in cosmology, which could be achieved using results and methods in Refs \cite{Anton_2021} and \cite{v1, v2, v3, v4}, or higher-order gravitational fields \cite{adamek1, adamek2, adamek3}. Investigations of these questions, as well as remaining theoretical issues, will be the subjects of future work.

\section*{Acknowledgements} We acknowledge helpful conversations and comments from Tessa Baker and Alkistis Pourtsidou, and support from the Science and Technology Facilities Council (STFC) grant ST/P000592/1.

\appendix

\newpage

\section*{Appendices}

\section{Time-Dependence of Parameters}
\label{app:params}

\begin{figure}[b!]
\centering
\includegraphics[height =4.9cm]{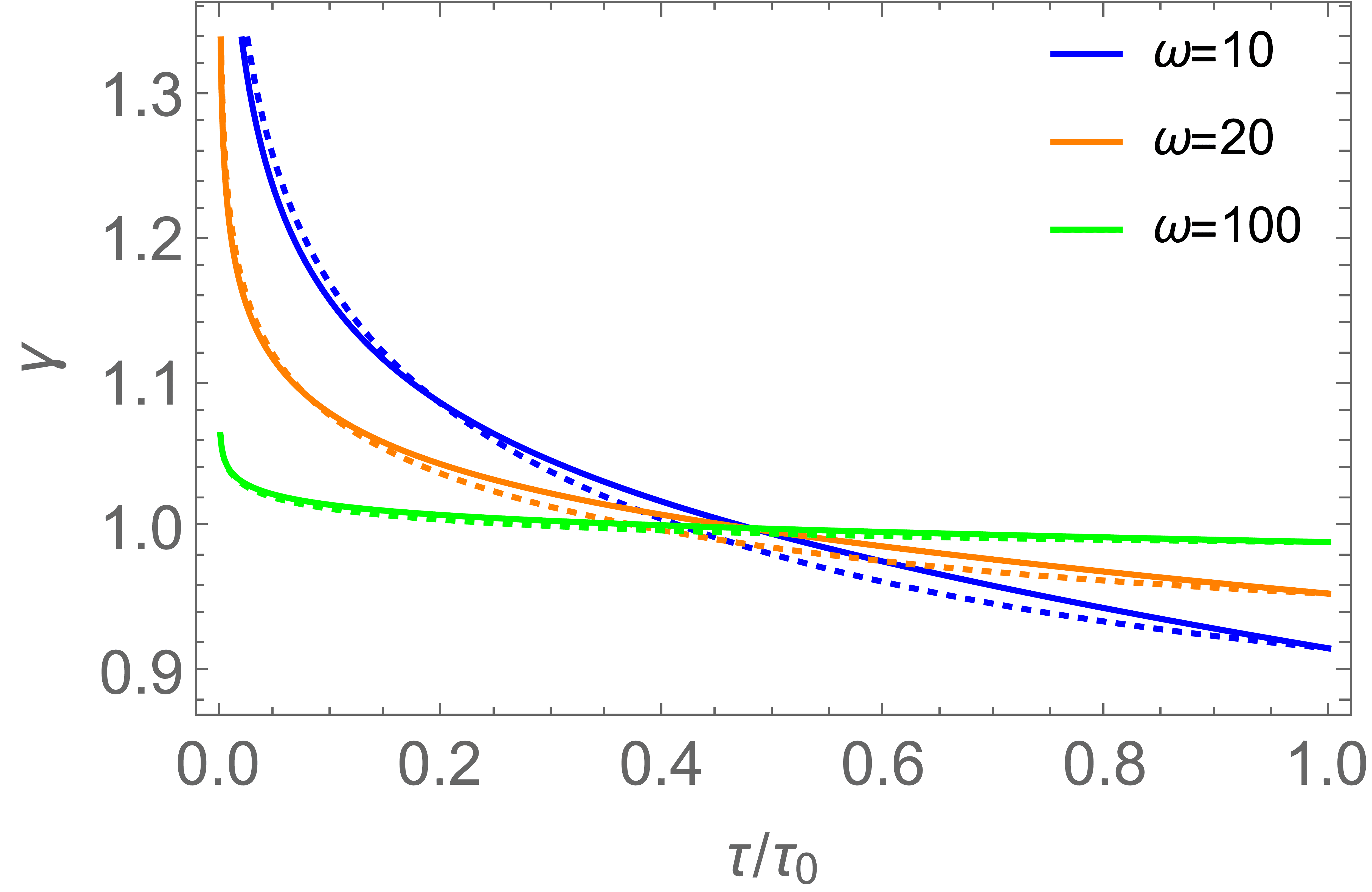}
\includegraphics[height =4.9cm]{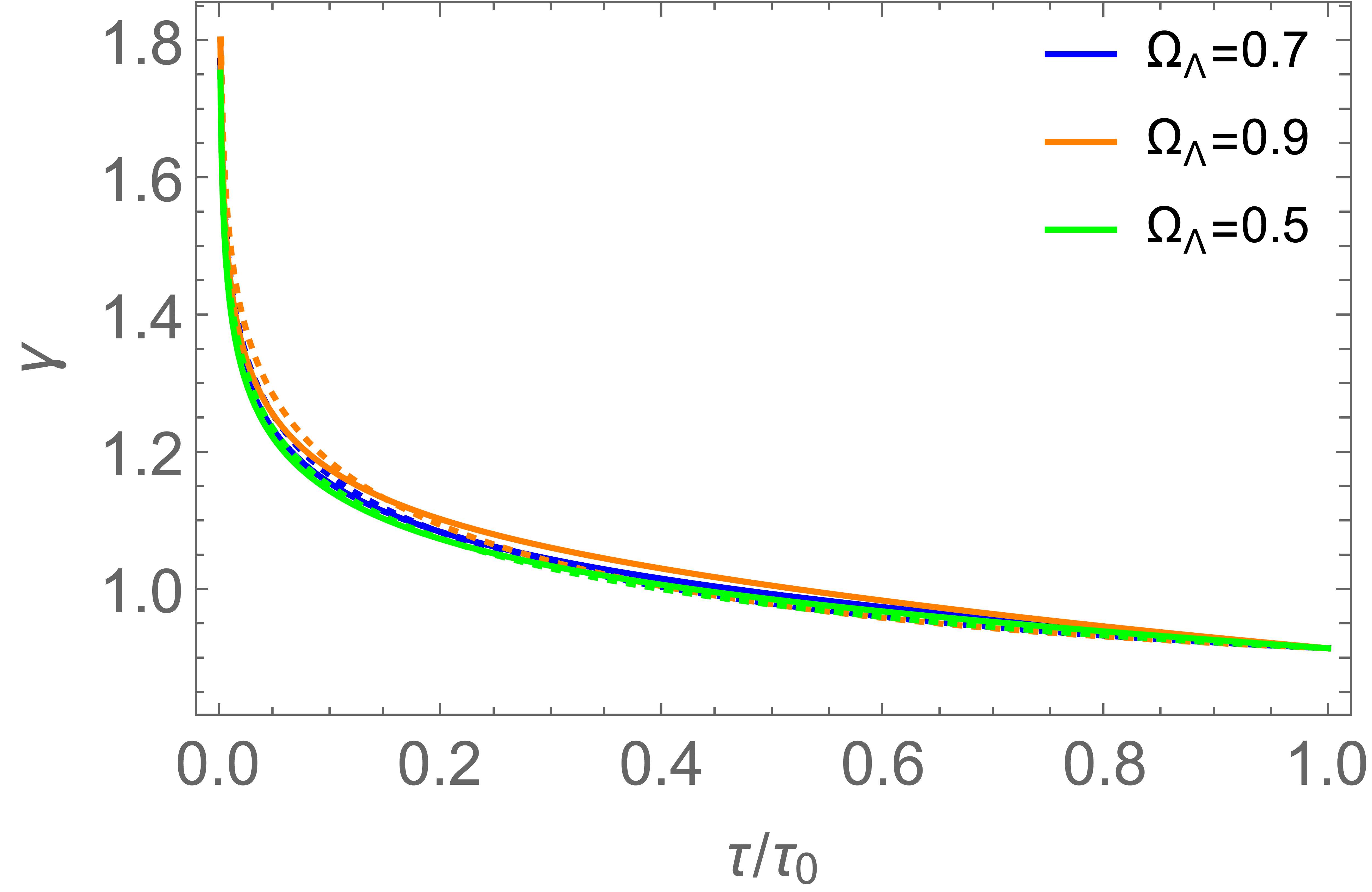}
\caption{Behaviour of $\gamma$ (solid lines) as a function of time $\tau$, along with simple power-law approximations (dashed lines). Left: differing values of $\omega$, with $\Omega_\Lambda=0.7$. Right: differing values of $\Omega_\Lambda$, with $\omega=10$.
}
\label{fig_gammaguess}
\vspace{0.5cm}

\centering
\includegraphics[height =5cm]{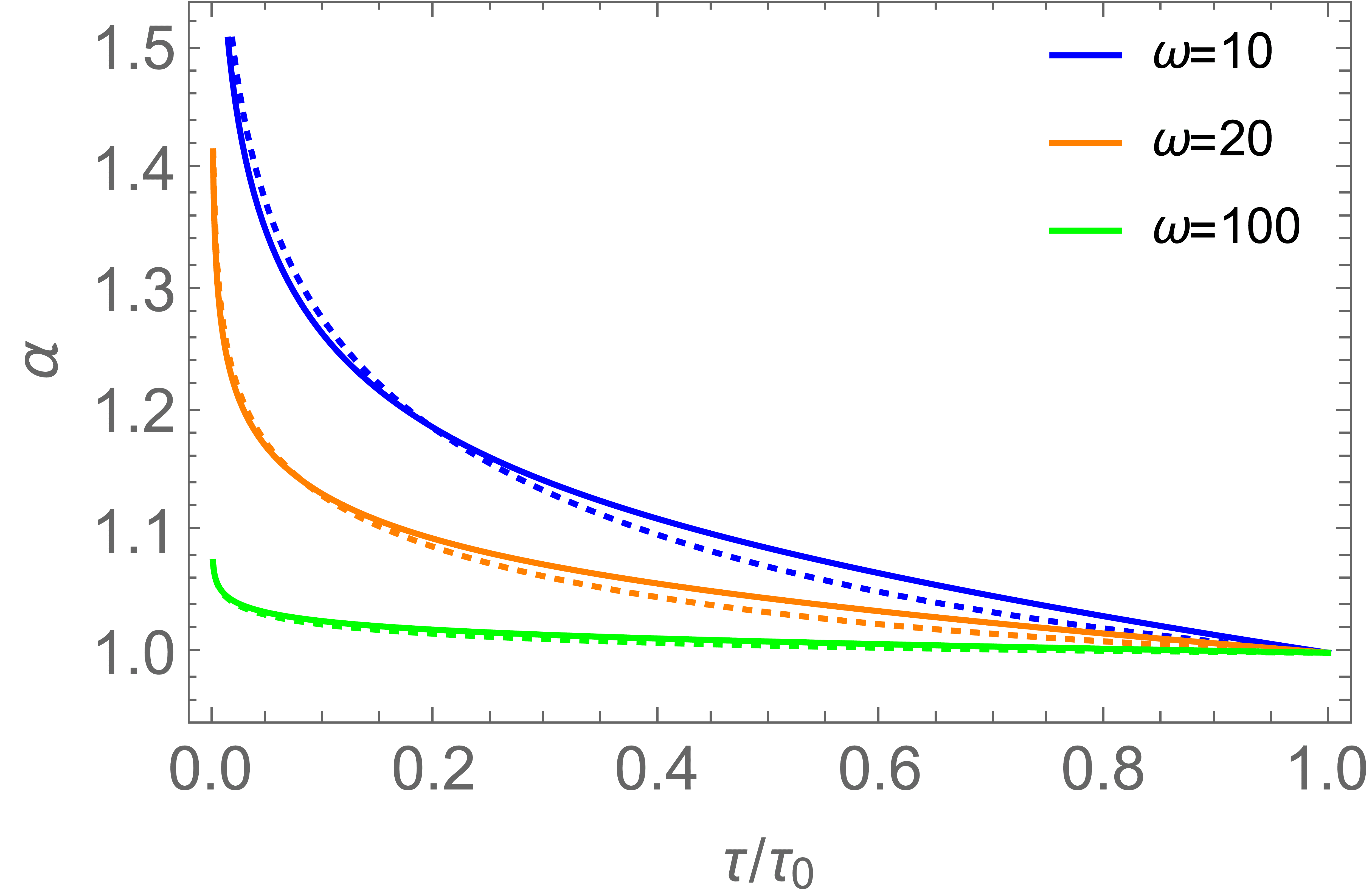}
\includegraphics[height =5cm]{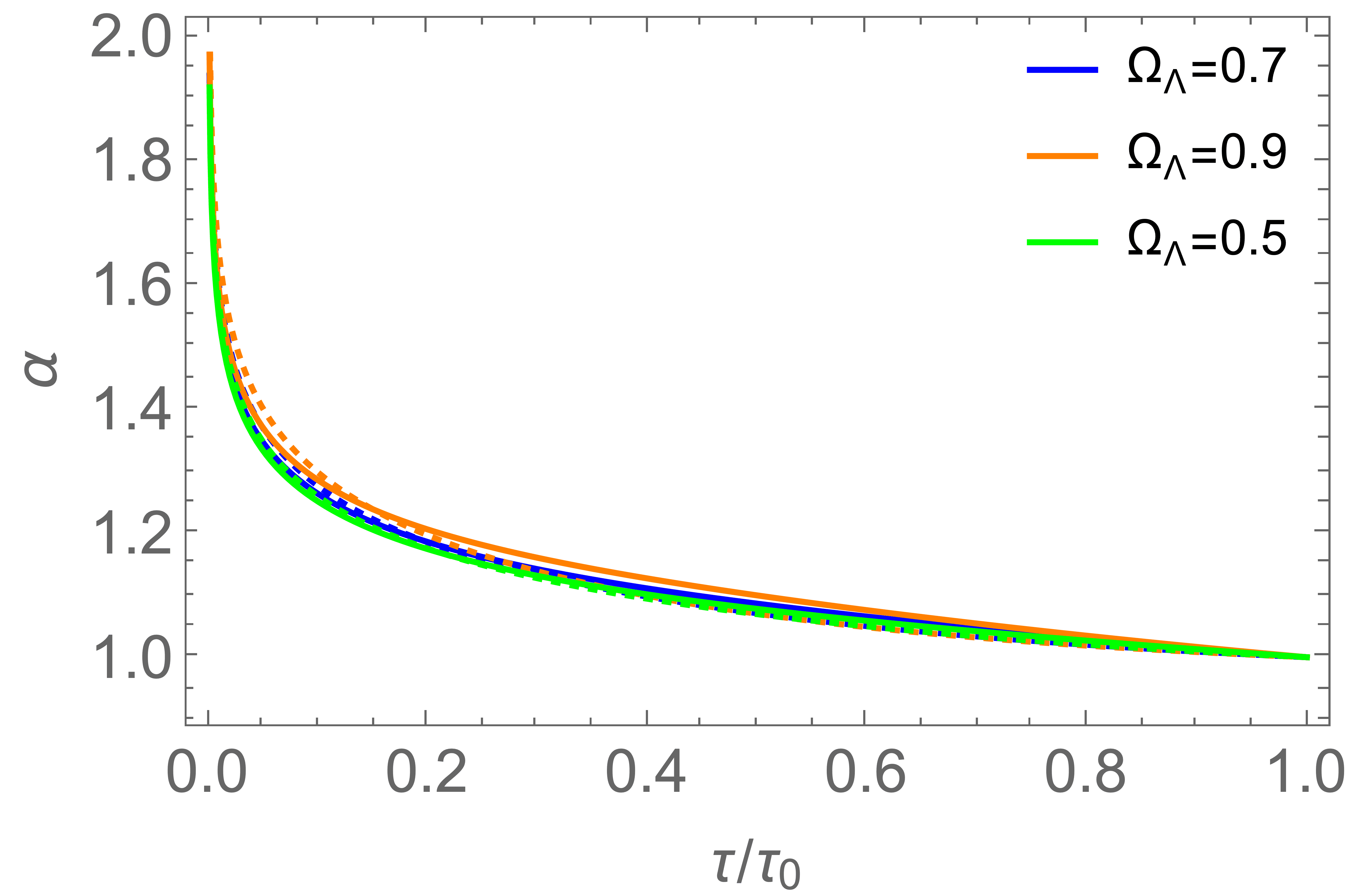}
\caption{Behaviour of $\alpha$ (solid lines) as a function of time $\tau$, along with simple power-law approximations (dashed lines). Left: differing values of $\omega$, with $\Omega_\Lambda=0.7$. Right: differing values of $\Omega_\Lambda$, with $\omega=10$.
}
\label{fig_alphaguess}
\end{figure}

Figures \ref{fig_gamma} and \ref{fig_alpha} showed the behaviour of the PPN parameters $\alpha$ and $\gamma$ as functions of conformal time $\tau$, within some of our example theories of gravity. In general, however, the time dependence of these parameters is not known, and indeed is likely to vary considerably between different theories of gravity. It is therefore useful to pause and consider how well we might have reproduced the calculated evolutions of $\alpha$ and $\gamma$ if we had instead posited them as simple functions, with degrees of freedom to be fixed. This situation is illustrated in Figures \ref{fig_gammaguess} and \ref{fig_alphaguess}, where we again reproduce the behaviour of $\gamma$ and $\alpha$ from our example theories with different values of $\omega$ and $\Omega_\Lambda$, but this time also with a simple fitting function $f(t)=At^{-n}+B$ where $A$, $B$ and $n$ are constants. Fixing the end points, as would be achieved by (for example) using astrophysical data of gravitational experiments at the present time, and adjusting the remaining freedom we can see it is possible to achieve reasonable fits. In reality, of course, we would not fit functions to pre-specified curves (as these would not be known), but instead would use data to constrain constants.

\begin{figure}[b!]
\centering
\includegraphics[height=4.8cm]{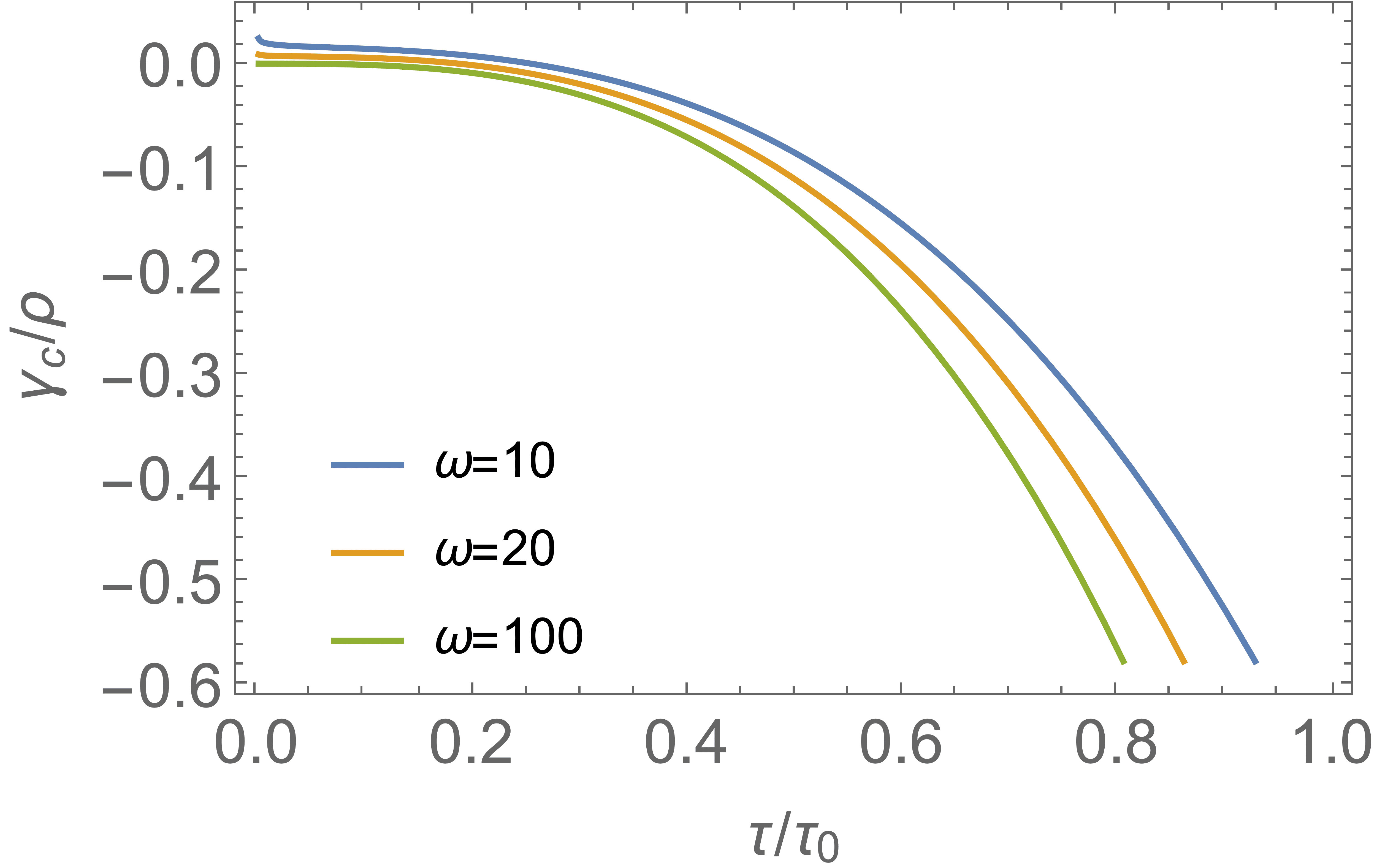}
\includegraphics[height=4.8cm]{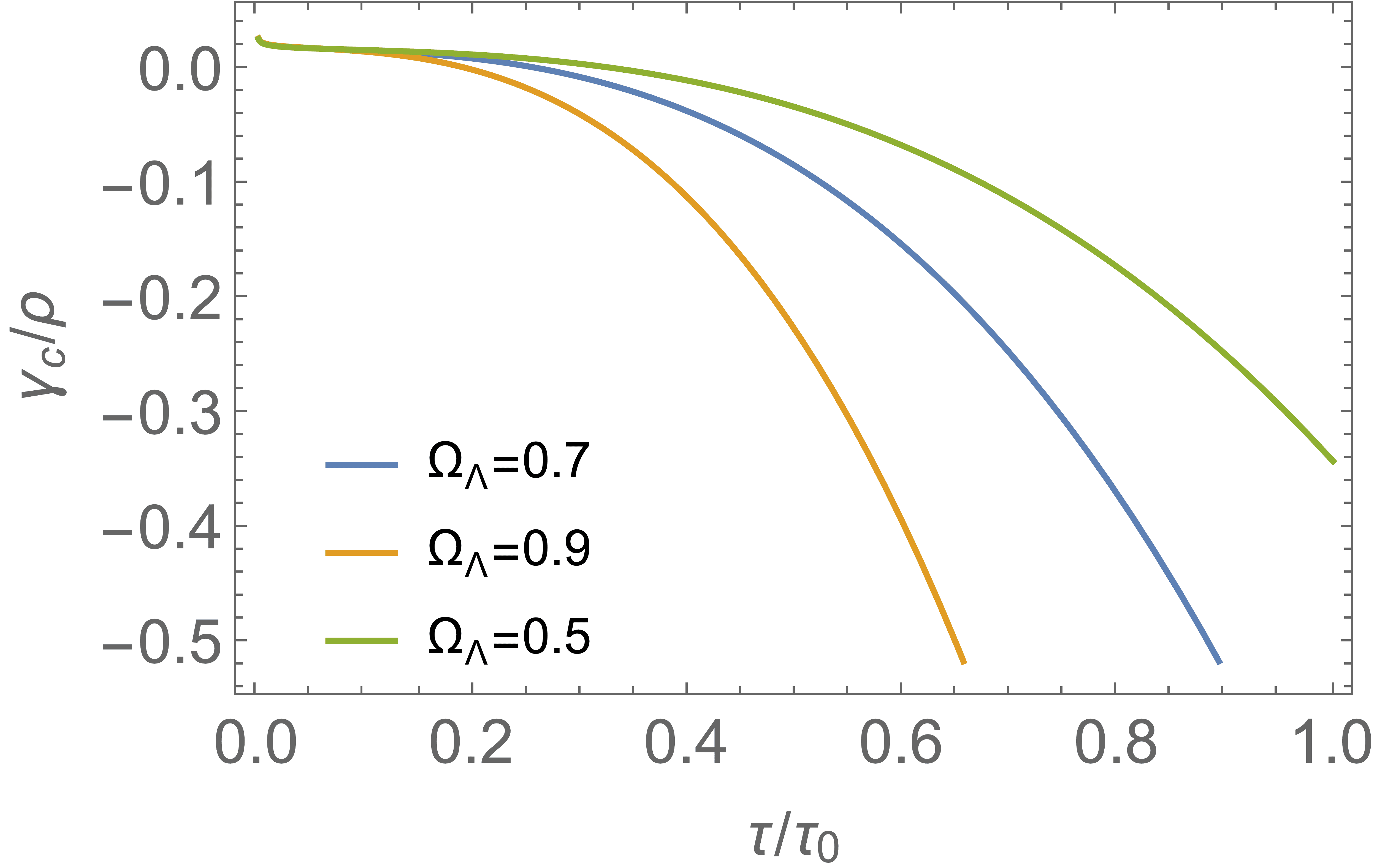}
\newline

\vspace{-1cm}
\includegraphics[height =5cm]{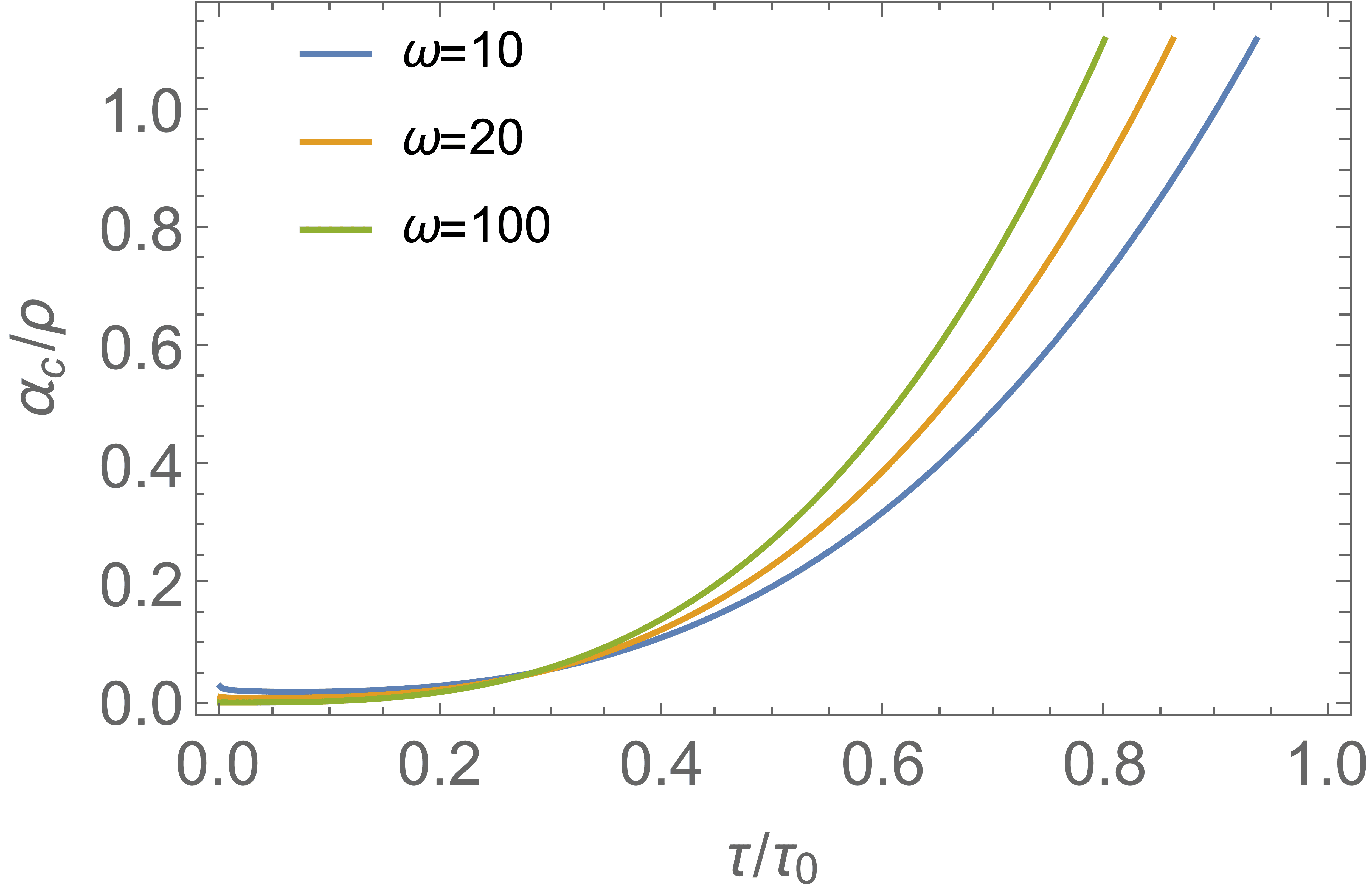}
\includegraphics[height =5cm]{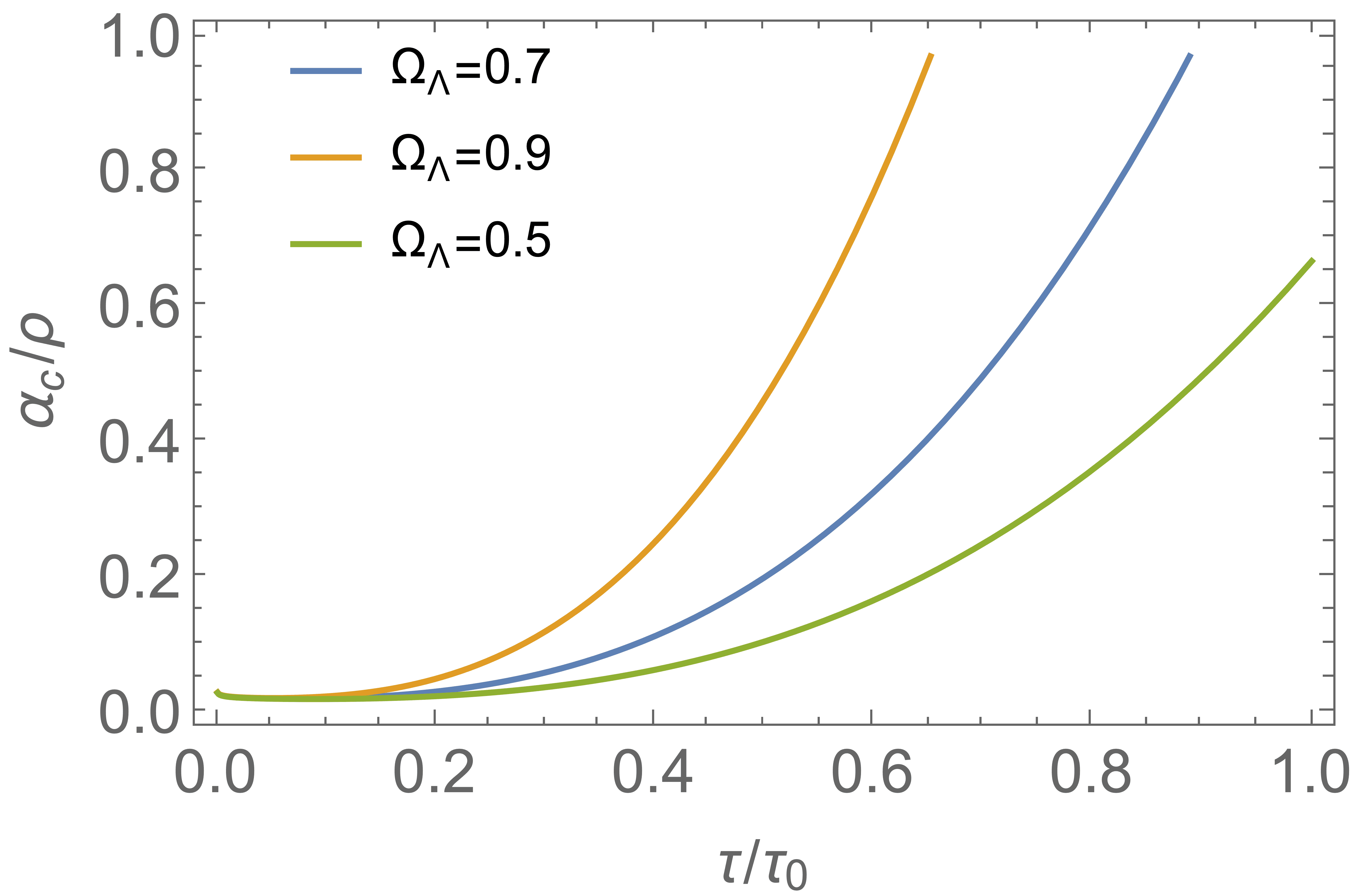}
\newline

\vspace{-1cm}
\includegraphics[height =4.9cm]{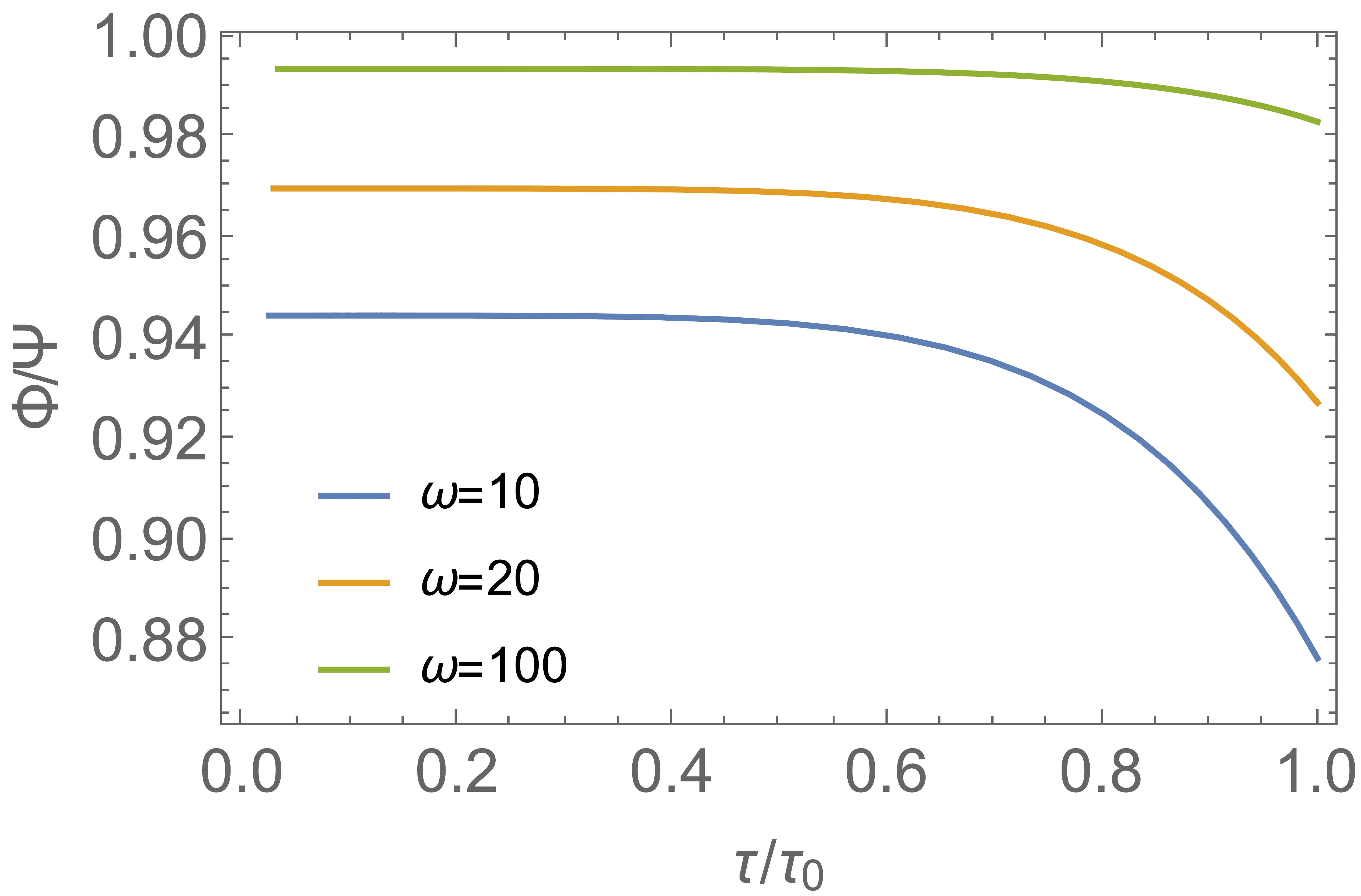}
\includegraphics[height =4.9cm]{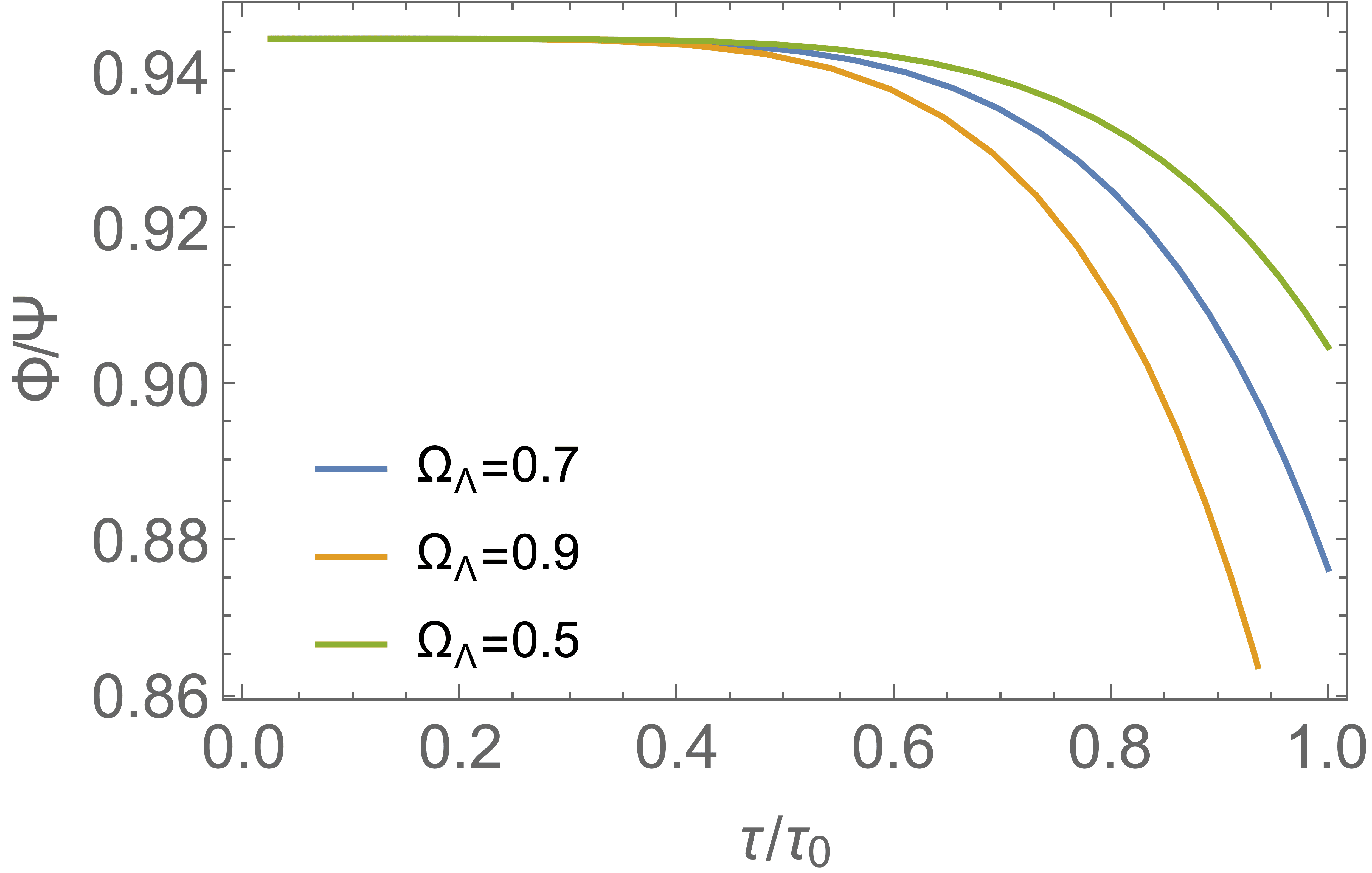}
\caption{The behaviour of $\alpha_c$, $\gamma_c$ and the large-scale limit of $\eta$ as a function of time $\tau$. Left: differing values of $\omega$, with $\Omega_\Lambda=0.7$. Right: differing values of $\Omega_\Lambda$, with $\omega=10$.}
\label{fig_largescaleslip}
\end{figure}

For completeness, let us also present the behaviour of the parameters $\gamma_c$, $\alpha_c$ and the large-scale limit of $\eta=\Phi/\Psi$ as functions of $\tau$ in Figure \ref{fig_largescaleslip}, for several different choices of $\omega$ and $\Omega_\Lambda$. The reader may note that we have chosen to plot the ratio of $\gamma_c$ and $\alpha_c$ to the background matter density $\bar{\rho}(\tau)$, in order to make them dimensionless. This choice also more accurately reflects the importance of these parameters at different stages in the cosmological evolution, and is thus more comparable with Figures \ref{fig_gamma} and \ref{fig_alpha}. The signs of $\alpha_c$ and $\gamma_c$ are determined by the signs of the $\Lambda$ term in Eq. (\ref{eqn_gammac}) and (\ref{eqn_alphac}), as this term dominates when these parameters are cosmologically relevant. Increasing $\Omega_\Lambda$ results in the amplitude of these parameters increasing, while decreased $\omega$ (making the theory further from GR) decreases their amplitude. The behaviour of the large-scale limit of $\eta$ can be seen to have a different behaviour to the other quantities. In all cases, the large-scale slip $\eta$ is below the GR value, and becomes further from GR with decreasing $\omega$ (with this effect increasing at late times). Increasing $\Omega_\Lambda$ reduces the value of $\eta$ compared to its value in GR at late times.

\section{Other Interpolating Functions}
\label{app:func}

\begin{figure}[h!]
\centering
\includegraphics[height =7.5cm]{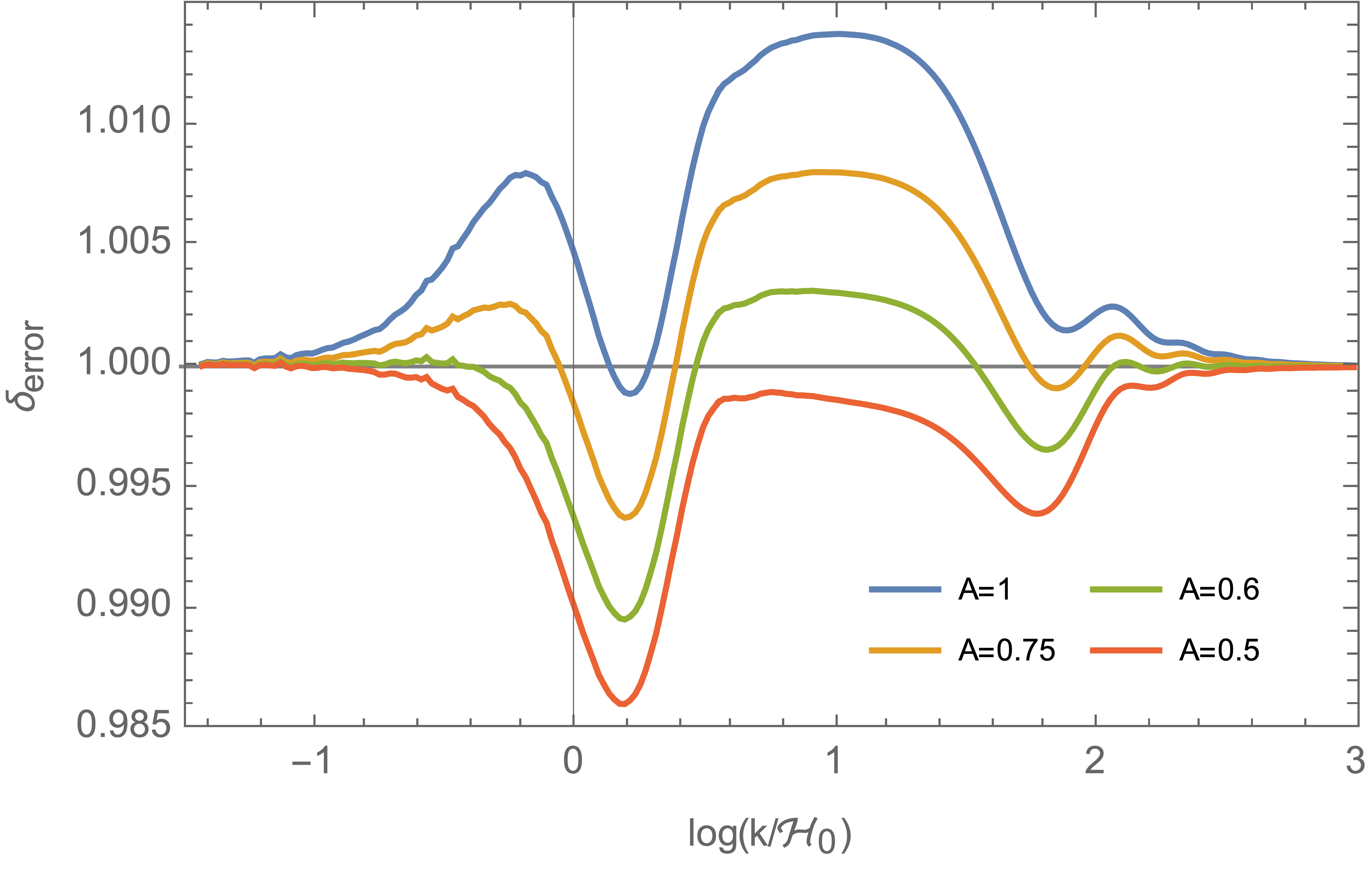}
\caption{The error in the density contrast today, calculated using the generalized interpolating function from Eq. (\ref{eqn_timinterp2}), with $\omega=10$ and $\Omega_{\Lambda}=0.7$.
Curves corresponds to different values of $A$. }
\label{fig_varytimfac}
\end{figure}

One possible generalization of the zero-parameter interpolation function introduced in Eq. (\ref{eqn_timinterp0}) is to expand it to include a single additional constant $A$, as
\begin{equation}
\label{eqn_timinterp2}
f(k)=\frac{1}{2}\left( S+L\right)+\frac{1}{2}\left(S-L \right)\frac{1-A\frac{k^2}{\mathcal{H}^2}}{1+A\frac{k^2}{\mathcal{H}^2}} \, \text{.}
\end{equation}
The fit of the interpolation can be improved by tailoring the value of $A$, as shown in Figure \ref{fig_varytimfac}. However, the optimal $A$ will certainly vary between different theories of gravity, so if we wish to remain as model-independent as possible then the zero-parameter interpolation seems most appealing. Should a more complicated interpolation function be required then the parameter $A$ (or its equivalent in other possible generalizations) would need to be included in the data analysis as a nuisance parameter, and marginalised over. Whether or not this would be beneficial would require a study of relevant statistics, such as Bayesian information.

\end{document}